 \documentclass[12pt,preprint]{aastex}

\newcommand{\kms}{~km s$^{-1}$~} 
\newcommand{\snr}{~SNR~1987A~}
\newcommand{\myemail}{zhekovs@colorado.edu}

\shorttitle {LETG and HETG Observations of \snr}
\shortauthors{Zhekov et al.}

\begin{document}


\title{High-Resolution X-ray Spectroscopy of \snr: Chandra LETG
and HETG Observations in 2007 }


\author{Svetozar A. Zhekov\altaffilmark{1,5}, 
Richard McCray\altaffilmark{1}, 
Daniel Dewey\altaffilmark{2}, 
Claude R.  Canizares\altaffilmark{2}, 
Kazimierz J. Borkowski\altaffilmark{3},
David N. Burrows \altaffilmark{4}, and 
Sangwook Park\altaffilmark{4} }

\altaffiltext{1}{JILA, University of Colorado, Boulder, CO
80309-0440; \myemail, dick@jila.colorado.edu}
\altaffiltext{2}{MIT, Kavli Institute, Cambridge, MA 02139;
dd@space.mit.edu, crc@space.mit.edu} 
\altaffiltext{3}{Department
of Physics, NCSU, Raleigh, NC 27695-8202; kborkow@unity.ncsu.edu}
\altaffiltext{4}{Department of Astronomy and Astrophysics,
Pennsylvania State University, 525 Davey Laboratory, University
Park, PA 16802; burrows@astro.psu.edu; park@astro.psu.edu}
\altaffiltext{5}{On leave from Space Research Institute, Sofia,
Bulgaria}


\begin{abstract} 
We present an extended analysis of the deep {\it Chandra} LETG and
HETG observations of the supernova remnant 1987A (\snr) carried out
in 2007.
The global fits to the grating spectra show that
the temperature of the X-ray emitting plasma in
the slower shocks in this system has remained stable for the last
three years, while that in the faster shocks has decreased. 
This temperature evolution is confirmed by
the first light curves of strong X-ray emission lines
and their ratios.
On the other hand,
bulk gas velocities inferred from the X-ray line
profiles are too low to account for the post-shock plasma
temperatures inferred from spectral fits.  This suggests that
the X-ray emission comes from gas that has been shocked twice,
first by the blast wave and again by shocks reflected from the
inner ring of \snr.  
A new model that takes these considerations into account gives 
support to this physical picture.
\end{abstract}





\section{Introduction} 
\label{sec:intro} 

Since its appearance on
February 23, 1987, SN~1987A has passed through various stages of
its evolution and it is now well into the supernova remnant (SNR) phase
\citep{Dick07}. Its reappearance in radio \citep{SS92, SS93} and
in X-rays \citep{Beuer94, Gor94, Ha96} at $\sim 1200$~days after
the explosion marked the beginning of this evolutionary phase.
Since then it has been continuously brightening in X-rays, at an
accelerating rate in the past few years \citep{P05, P06, P07}.
On day 6067, Bouchet et al. (2004 ; see also Bouchet et al. 2006) 
obtained a 10 micron image of the remnant which has similar 
morphology to the X-ray image.
X-ray observations are crucial for understanding the underlying
physics of this phase because they manifest
most of the energetics at present.  Detailed X-ray
spectral information deduced from grating observations 
provides us with essential information about the shocked plasma:
its temperature, bulk gas velocity, chemical composition, etc.

Continuous brightening of \snr in the last decade has enabled
spectral observations to be carried out with steadily improving
photon statistics.  The first such observations were performed
with the High Energy Transmission Grating (HETG) aboard
 {\it Chandra} in October 1999 \citep{Eli02}.  Although
the dispersed spectra contained relatively few source counts 
(602 in MEG, and 255 in HEG), emission
lines of thermal origin were clearly detected.  Since then,
dispersed spectra of \snr have been obtained in several
observations: May 2003 ({\it XMM-Newton}; Haberl et al. 2006);
Aug-Sep 2004 ({\it Chandra} LETG - Low Energy Transmission 
Grating; Zhekov et al. 2005, 2006 - hereafter Z05, Z06); 
January 2007 ({\it XMM-Newton}; Heng et al.
2008); March-April 2007 ({\it Chandra} HETG; Dewey et al. 2008 -
hereafter D08); and September 2007 ({\it Chandra} LETG; this work).

The superb spatial and spectral resolution of the {\it Chandra}
observatory opens new possibilities for studying \snr.  The
grating observations provide us not only with details about the
thermodynamics and kinetics of the X-ray emitting gas but also
about its geometry.  For example, the analysis of the first LETG
observations with good photon statistics (Z05) demonstrated that
the X-ray emission region is an expanding ring, having geometry
similar to optical image of the inner ring in \snr as seen with
the Hubble Space Telescope.  This conclusion was confirmed by data
from more recent {\it Chandra} HETG observations (D08).  In fact,
this conclusion was suggested by the similarity between the
optical and X-ray images (e.g., Burrows et al. 2000;
Park et al. 2004, 2006) but one
could not rule out a more nearly spherical geometry from the X-ray
images alone.

Here we present an extended analysis of the HETG and LETG grating
observations of \snr that were carried out with {\it Chandra} in
2007.  D08 have already reported the first part of our analysis,
which focused mainly on the HETG data.
For the current analysis,
we assume that the X-ray emitting plasma distribution is
symmetric about the axis of the inner circumstellar ring.
This assumption is clearly not true, as can be seen from the X-ray
images
(Park et al. 2004, 2006), but we adopt it here as a first
approximation
as we did with the LETG 2004 data in order to see how the inferred
parameters change with time.
In this study, we also carry on a detailed comparison between the
techniques used so far for deriving parameters of the strong
lines in the X-ray spectrum of \snr. This relates in a
more transparent way the results obtained with
different instrumentions.
The better constrained observational facts are a good basis for a
more elaborated interpretation as well.

This paper is organized as
follows. In \S~\ref{sec:obs} we briefly review the X-ray
observations. In \S~\ref{sec:global_fits} we present models for
the emitting gas that are based on global fits to the entire
spectrum of \snr. In \S~\ref{sec:line_prof} we analyze the
profiles and ratios of strong X-ray emission lines.
In \S~\ref{sec:2D} we report results from the two-dimensional
analysis of the X-ray observations.
In \S~\ref{sec:disc} we discuss the physical picture that
emerges from analysis of current and previous data.
Finally,
in \S~\ref{sec:conclusions} we list our conclusions.

\section{Observations and Data Reduction} 
\label{sec:obs} 

In 2007,
SNR~1987A was observed with both grating spectrometers
aboard {\it Chandra}: HETG and LETG. The observations  in
configuration LETG-ACIS-S  were carried out in nine consecutive
sequences (\dataset[ADS/Sa.CXO\#obs/07620]{Chandra ObsIDs: 7620,}
\dataset[ADS/Sa.CXO\#obs/07621]{ 7621,}
\dataset[ADS/Sa.CXO\#obs/09580]{ 9580,}
\dataset[ADS/Sa.CXO\#obs/09581]{ 9581,}
\dataset[ADS/Sa.CXO\#obs/09582]{ 9582,}
\dataset[ADS/Sa.CXO\#obs/09589]{ 9589,}
\dataset[ADS/Sa.CXO\#obs/09590]{ 9590,}
\dataset[ADS/Sa.CXO\#obs/09591]{ 9591 and}
\dataset[ADS/Sa.CXO\#obs/09592]{ 9592}) that took place between
Sep 4 and Sep 16, 2007 ($7498- 7510$~days after the explosion) and
provided a total effective exposure of 285 ksec.  The roll angle
was between $95^\circ$ and $98^\circ$, therefore, the dispersion
axis was aligned approximately with the minor axis (North-South)
of the inner ring in \snr (P.A. $\approx 354^\circ$ ~Sugerman et
al. 2002).

The positive and negative first-order LETG spectra for each of the
nine observations were extracted following the Science Threads for
Grating Spectroscopy in the CIAO 3.4 \,\footnote{Chandra
Interactive Analysis of Observations (CIAO),
http://cxc.harvard.edu/ciao/} data analysis software.  The
resultant spectra were merged into one spectrum each for the
positive and negative LETG arms with respective total counts of
23,797 (LETG+1) and 15,358 (LETG-1).  The ancillary response functions 
for all spectra were generated using the Chandra calibration database
CALDB v.4.0. (The terms LETG and LEG will be used
throughout the text to refer to the LETG data and spectra.)

We discussed the 2007 HETG data of \snr in D08; here we only
recall that the first-order spectra contain the following numbers
of counts: 15,200 (MEG-1), 14,000 (MEG+1), 5,800 (HEG-1) and 4,700
(HEG+1).

Finally, the instrument configurations were such that the {\it
negative} first-order LEG arm was pointing to North, and the same
was true for the {\it positive} first-order MEG/HEG arm.
Anticipating the discussion of the spatial-spectral effects, we
note that the narrowed (`compressed') spectral lines are thus
found in their corresponding spectra (LEG$-1$; MEG/HEG$+1$).  

\begin{figure}[ht]
\begin{center}
\includegraphics[width=1.75in, height=1.25in]{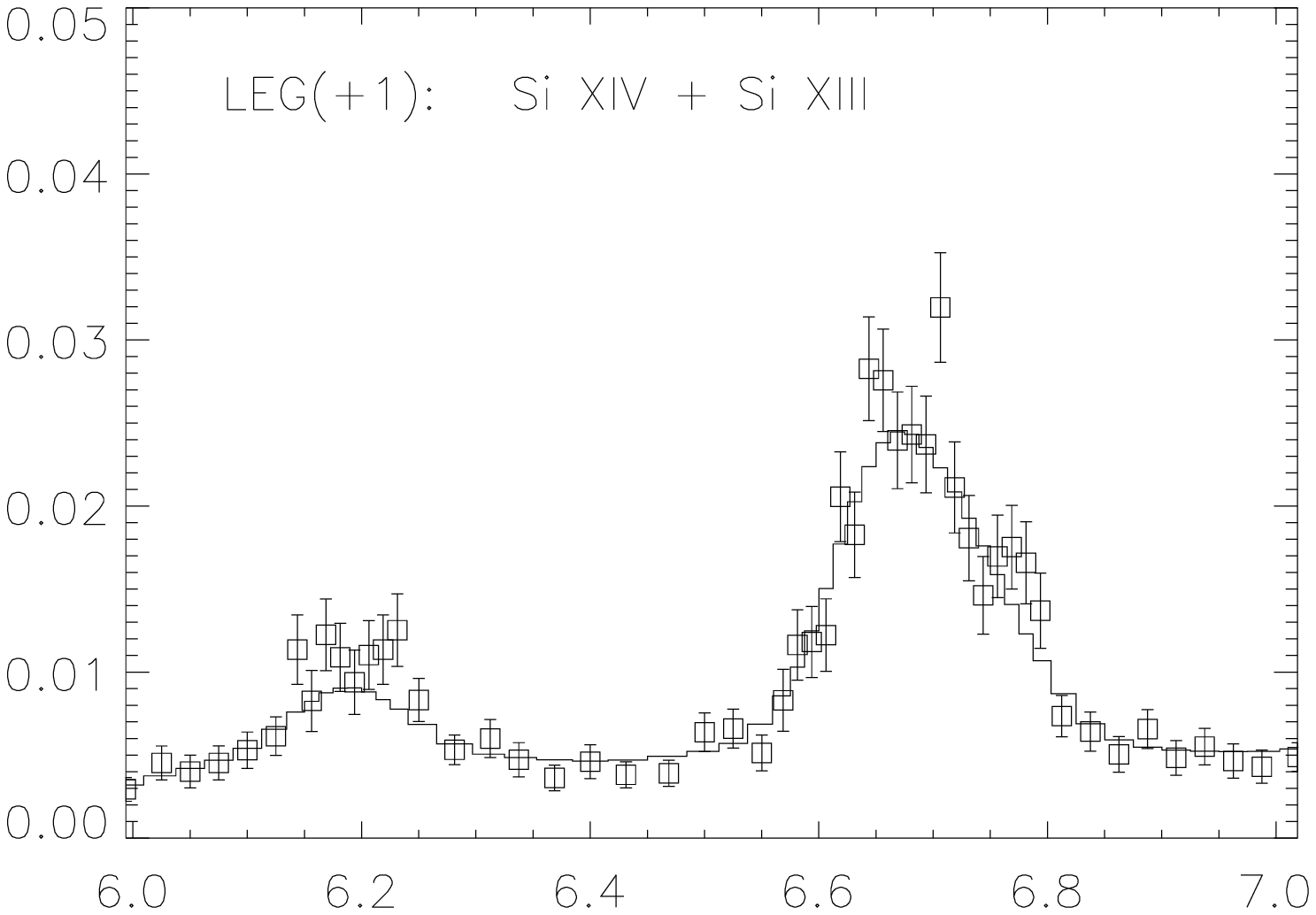}
\includegraphics[width=1.75in, height=1.25in]{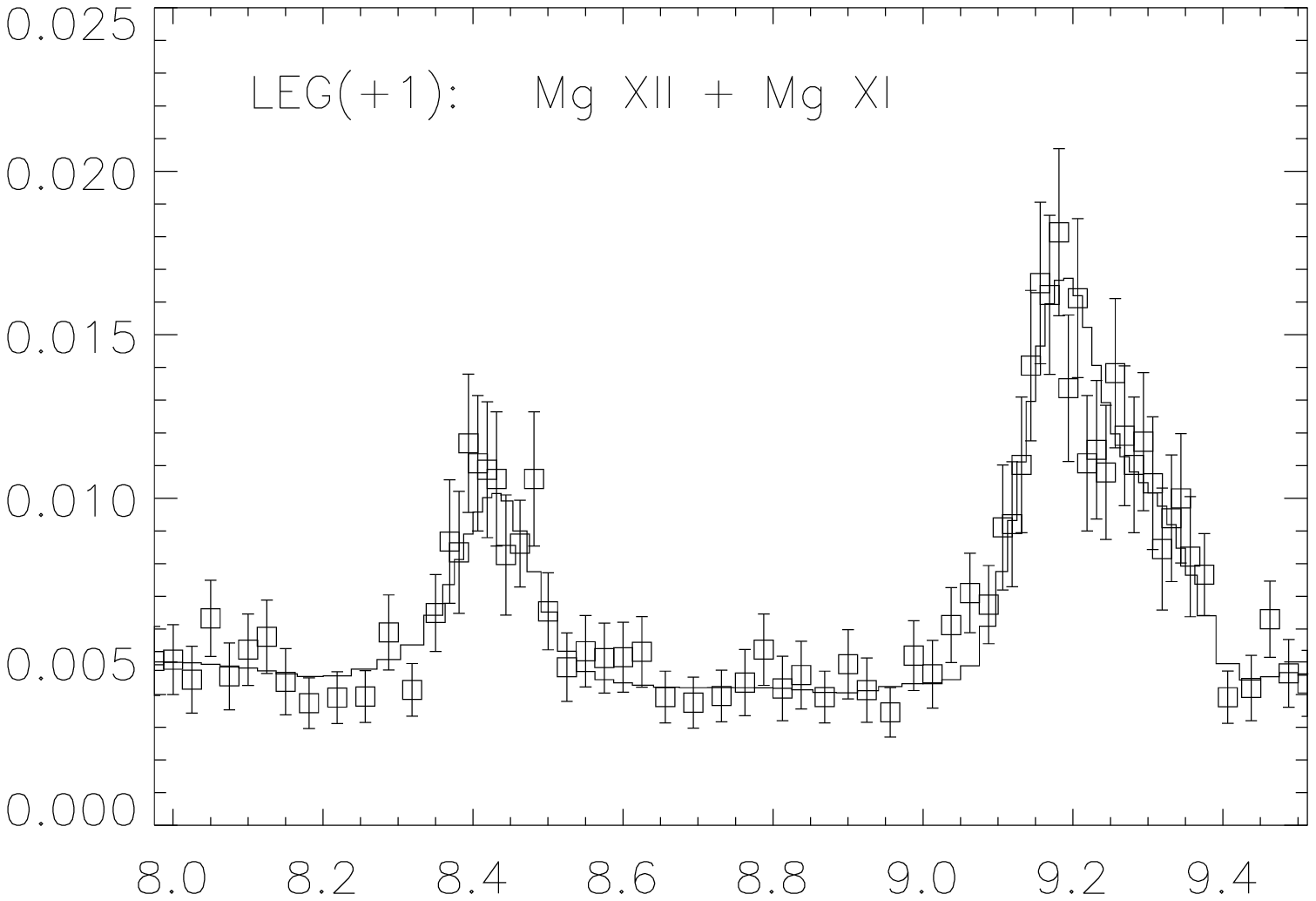}
\includegraphics[width=1.75in, height=1.25in]{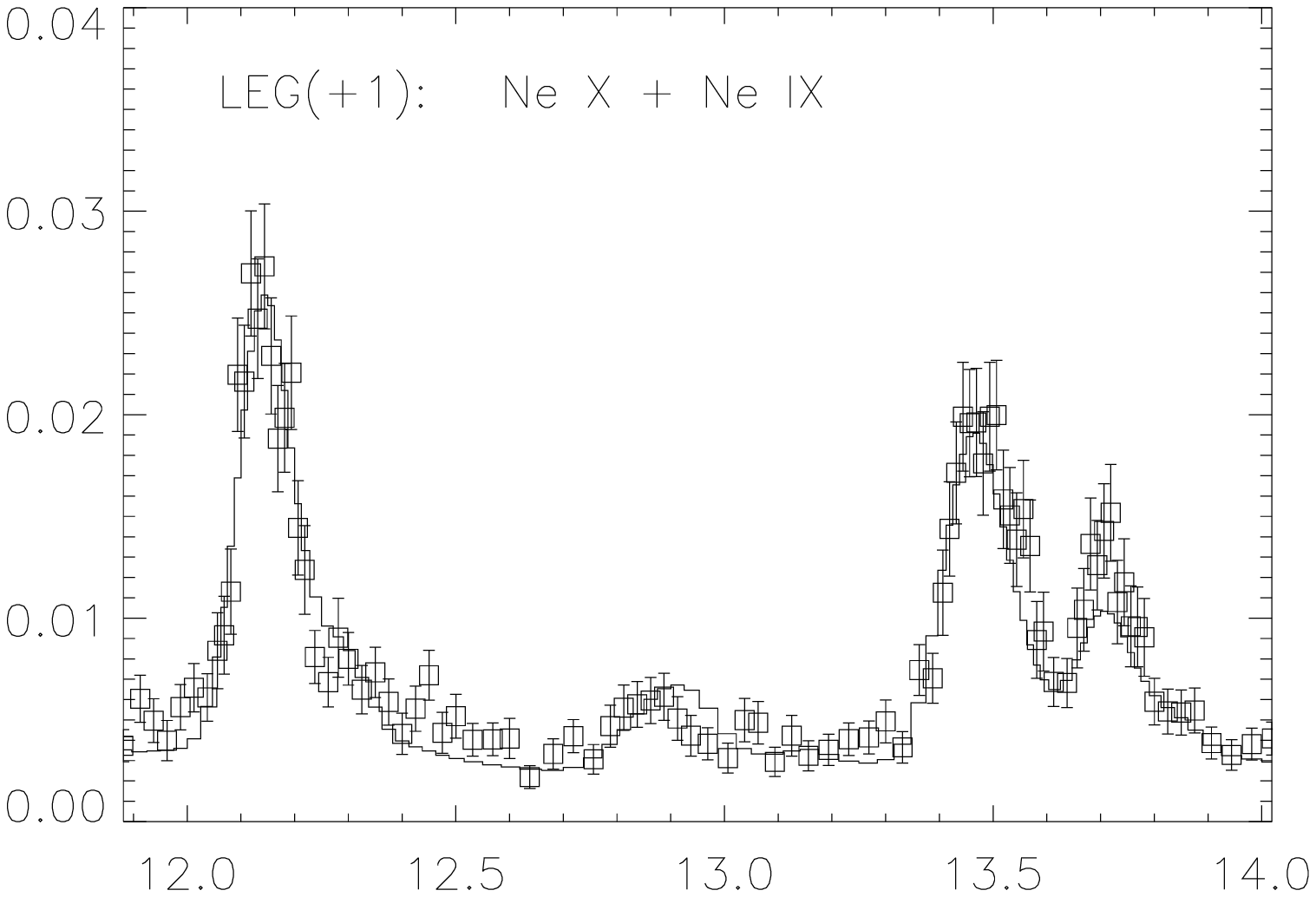}
\includegraphics[width=1.00in, height=1.25in]{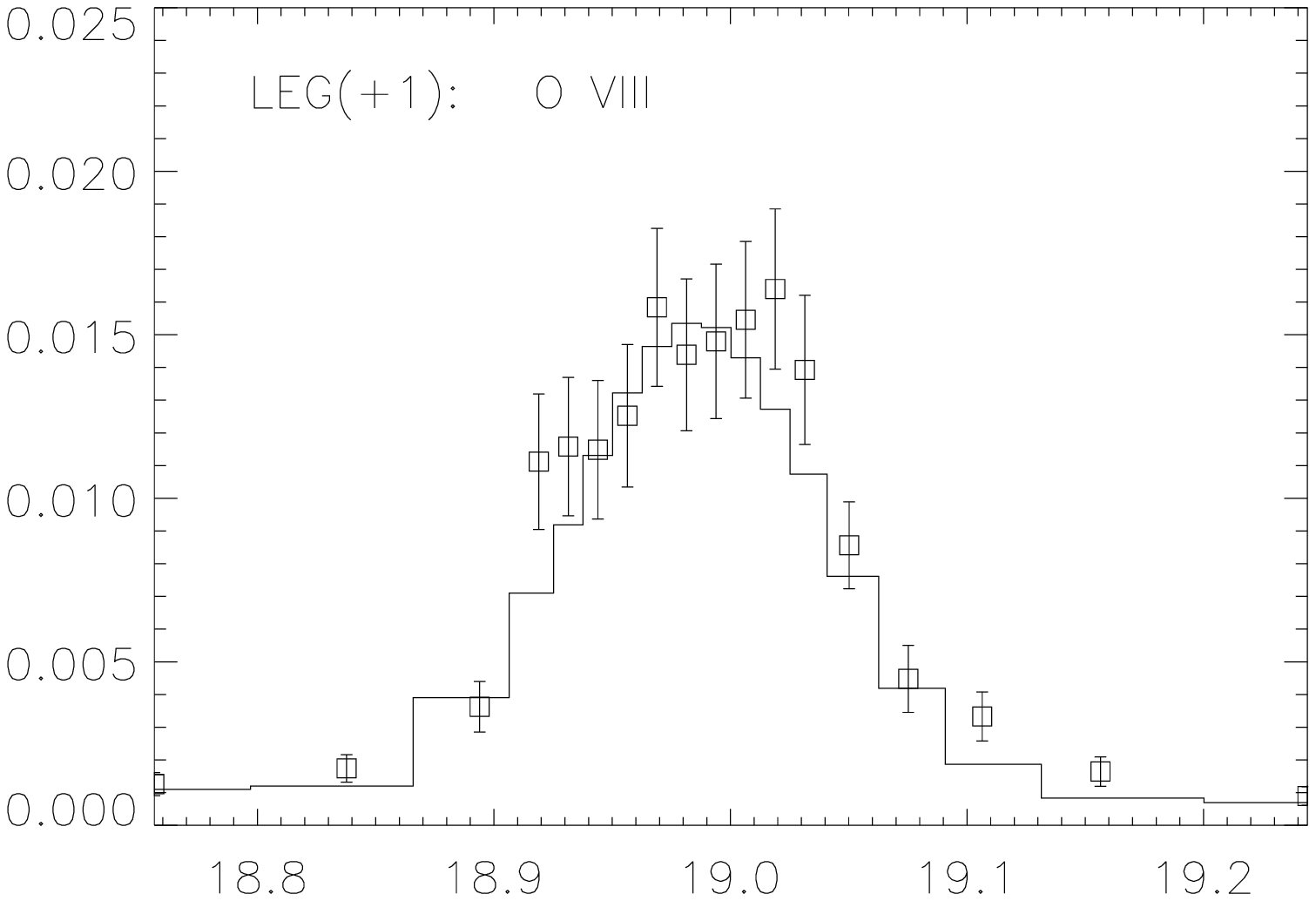}
\includegraphics[width=1.75in, height=1.25in]{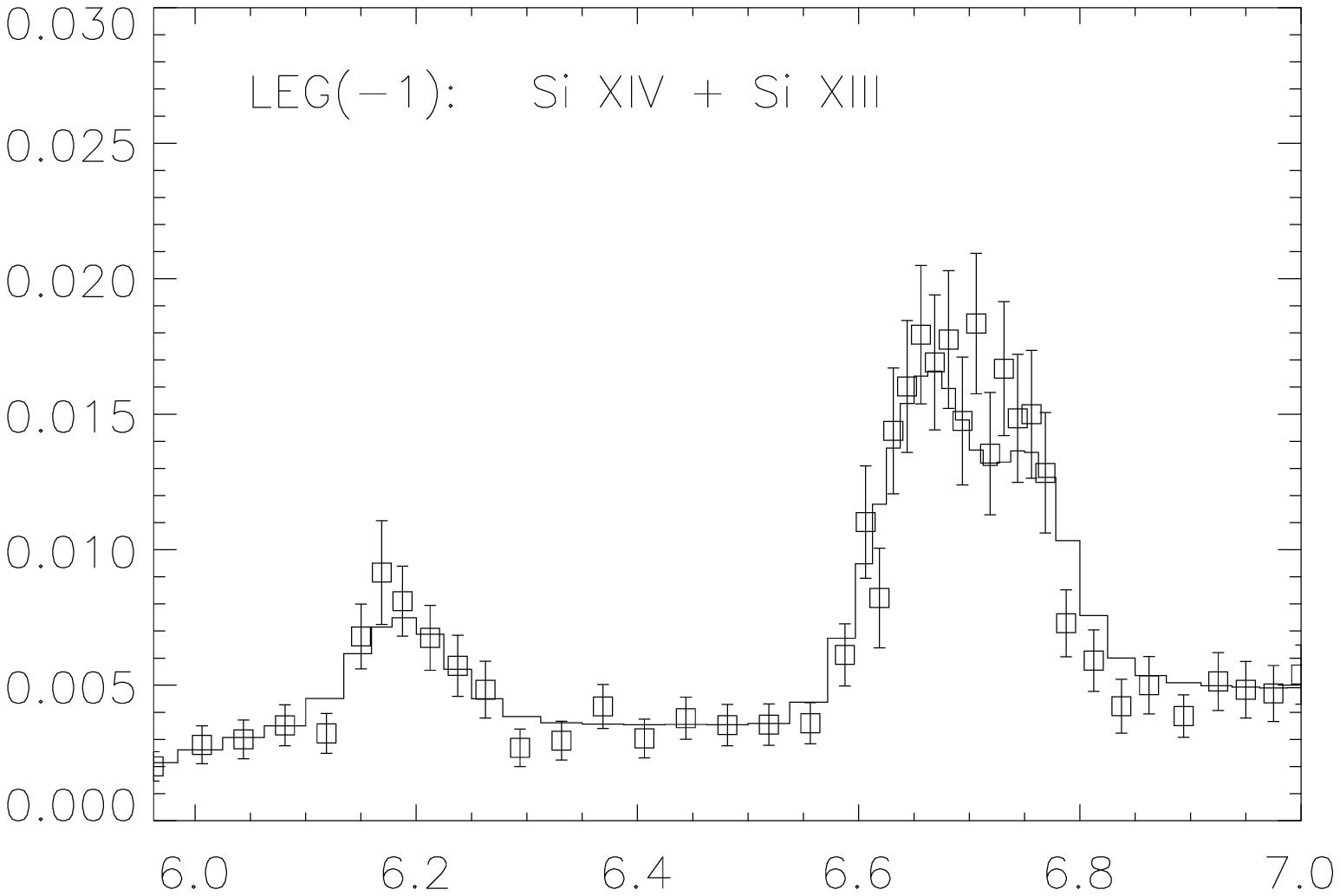}
\includegraphics[width=1.75in, height=1.25in]{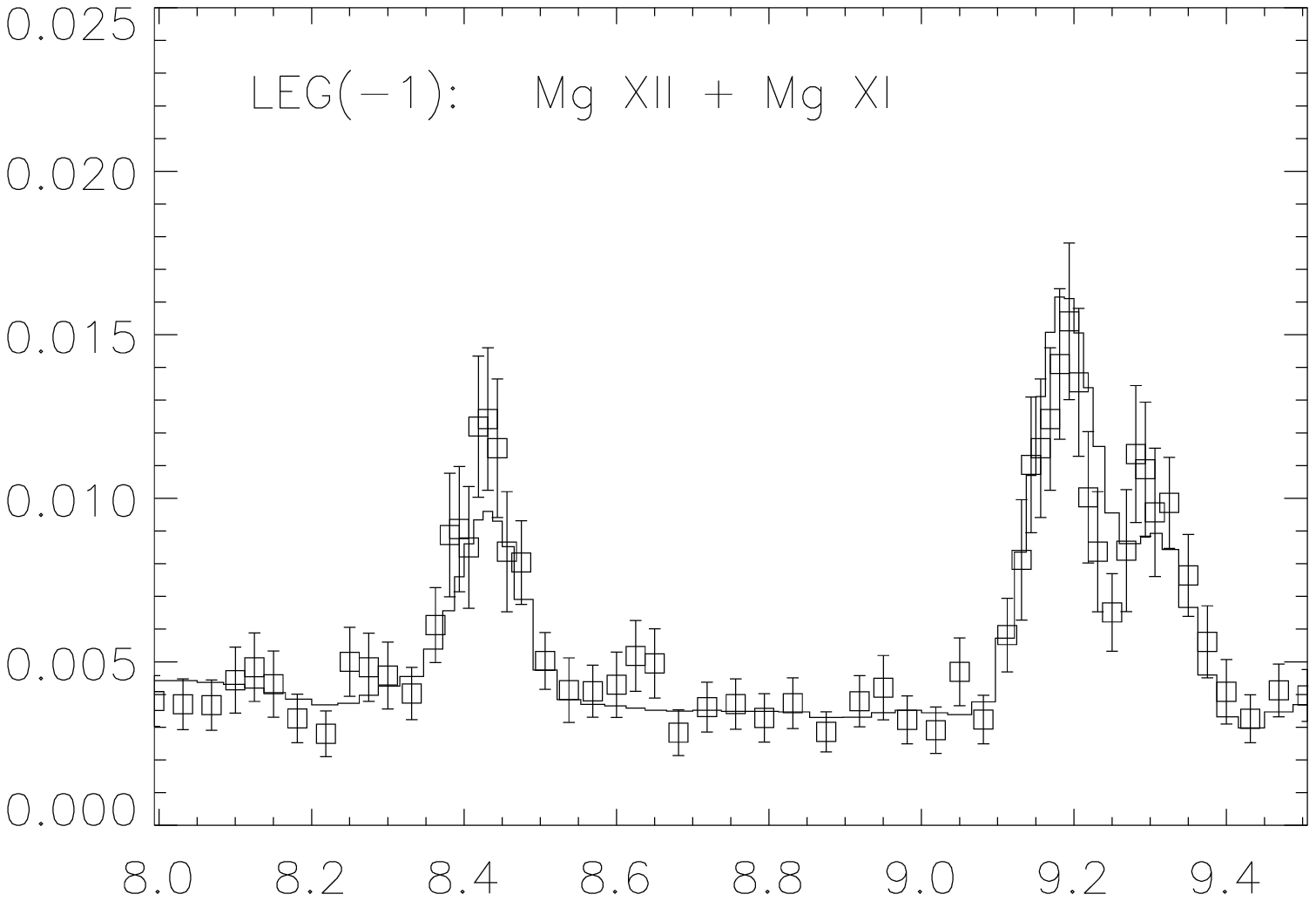}
\includegraphics[width=1.75in, height=1.25in]{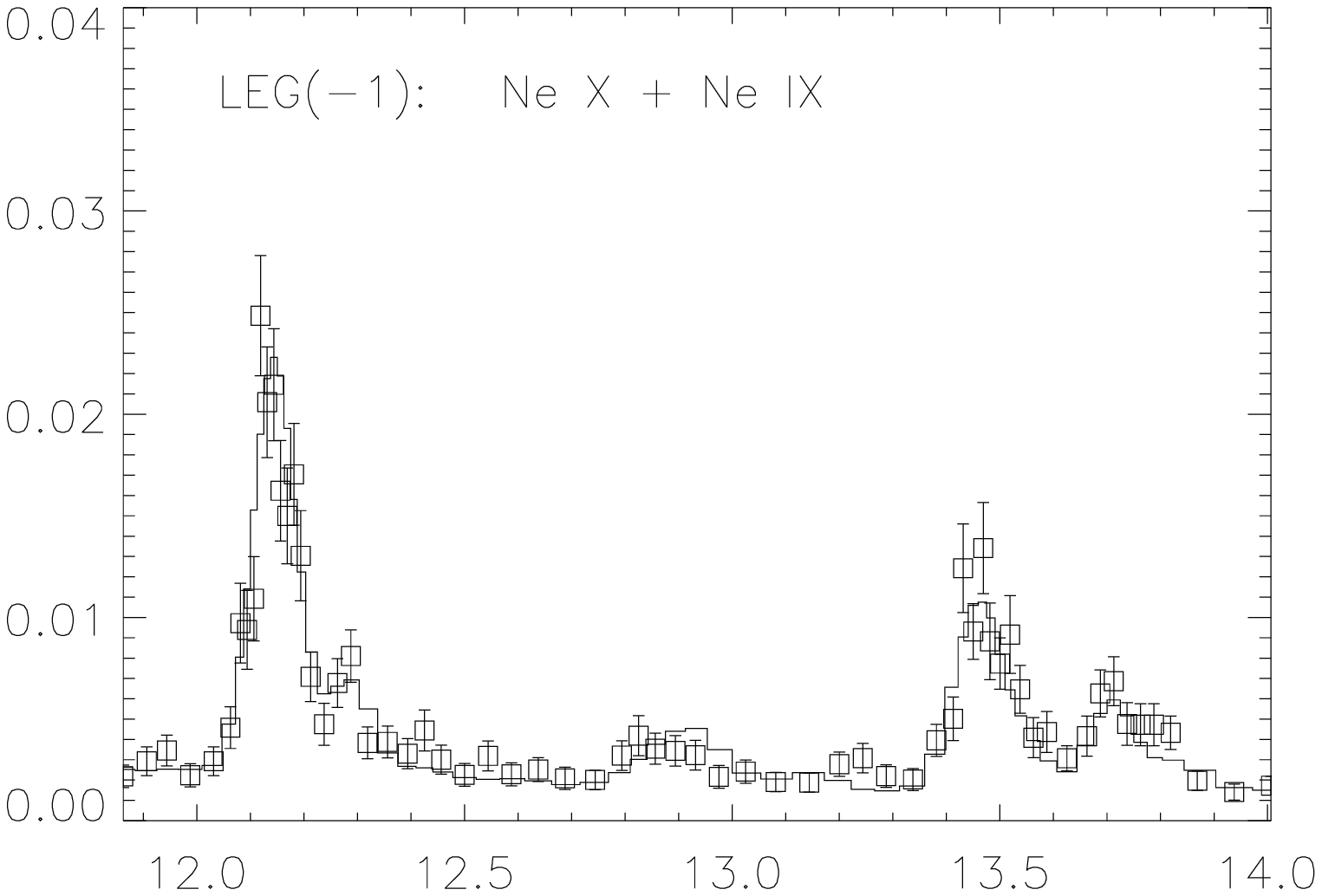}
\includegraphics[width=1.00in, height=1.25in]{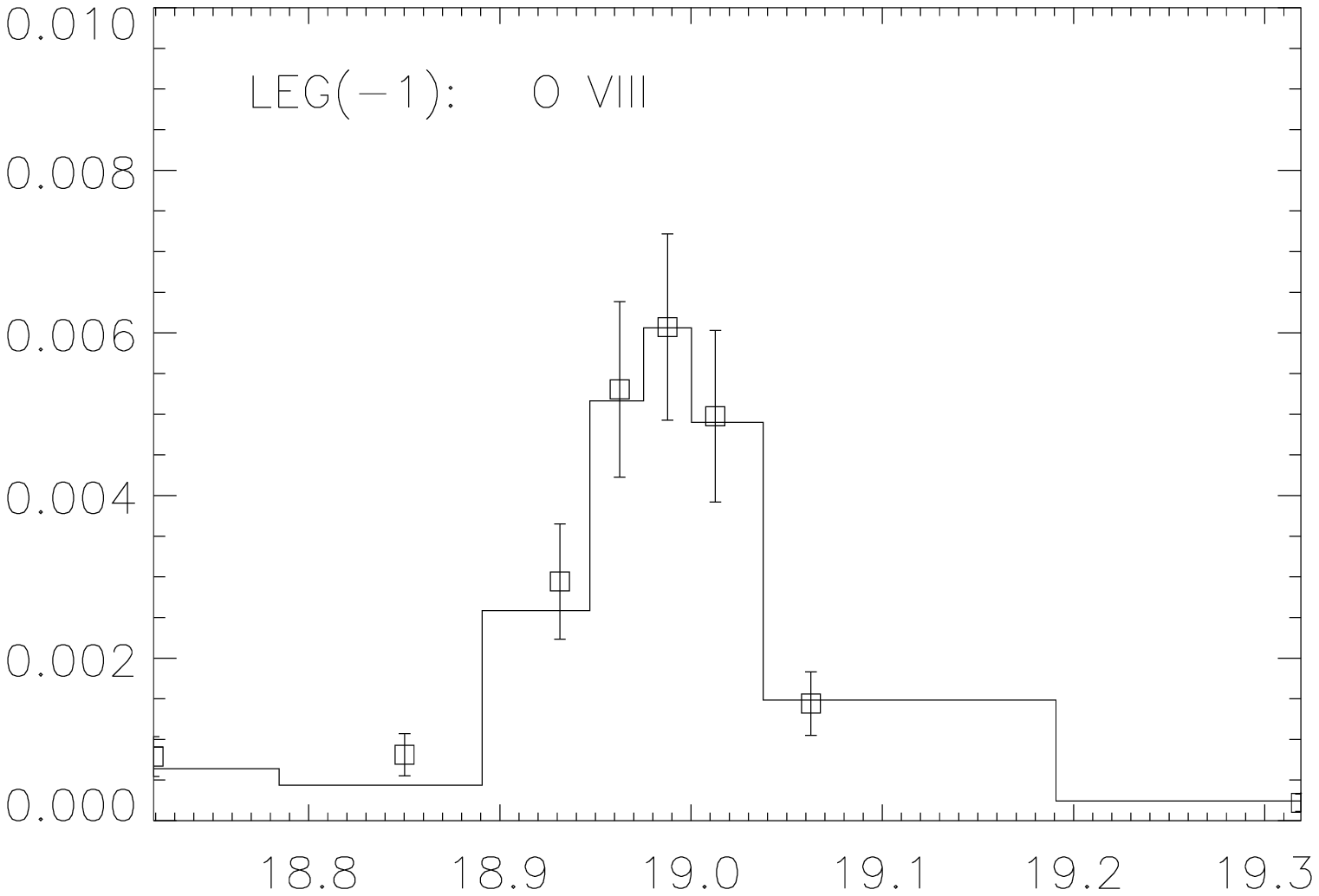}
\includegraphics[width=1.75in, height=1.25in]{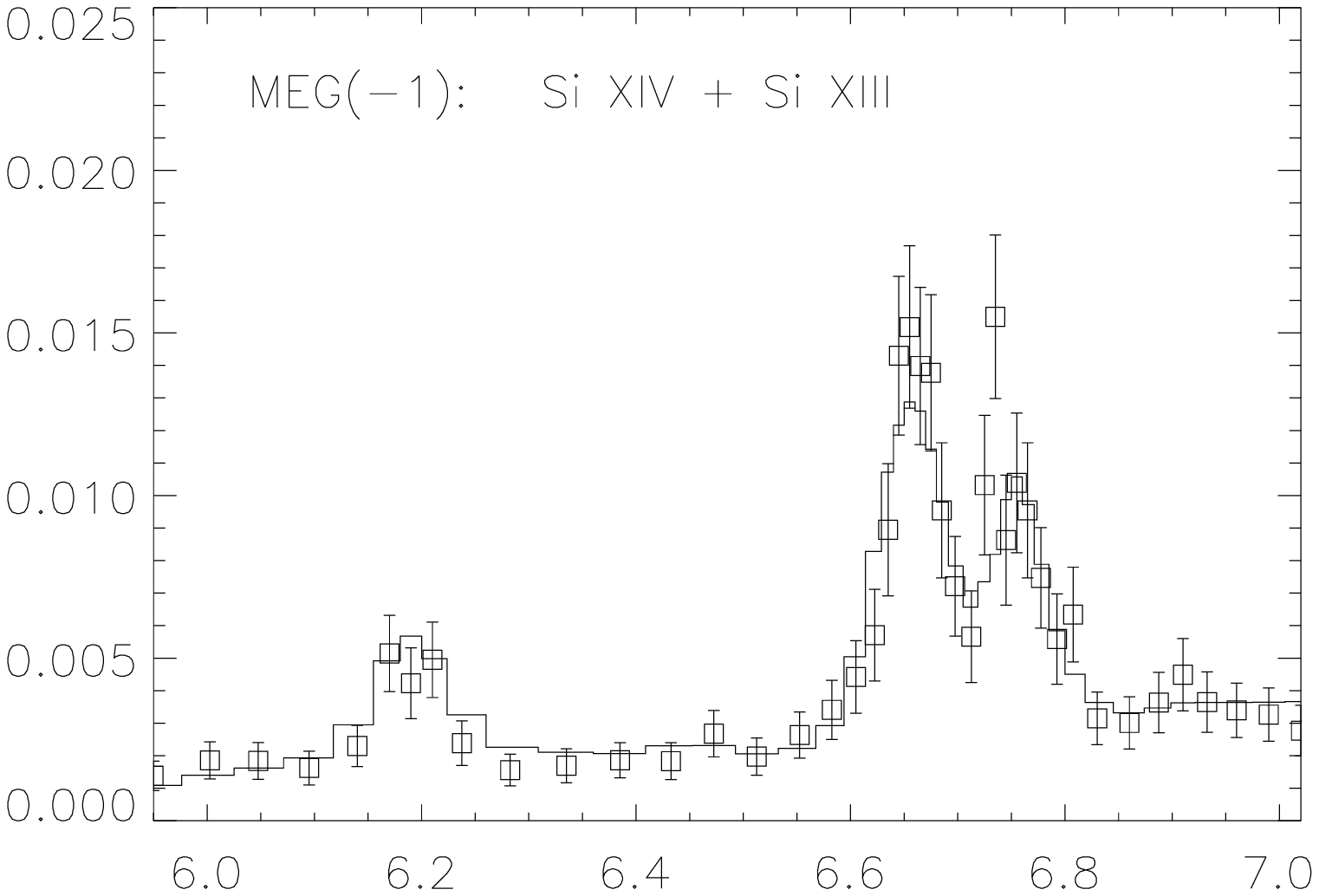}
\includegraphics[width=1.75in, height=1.25in]{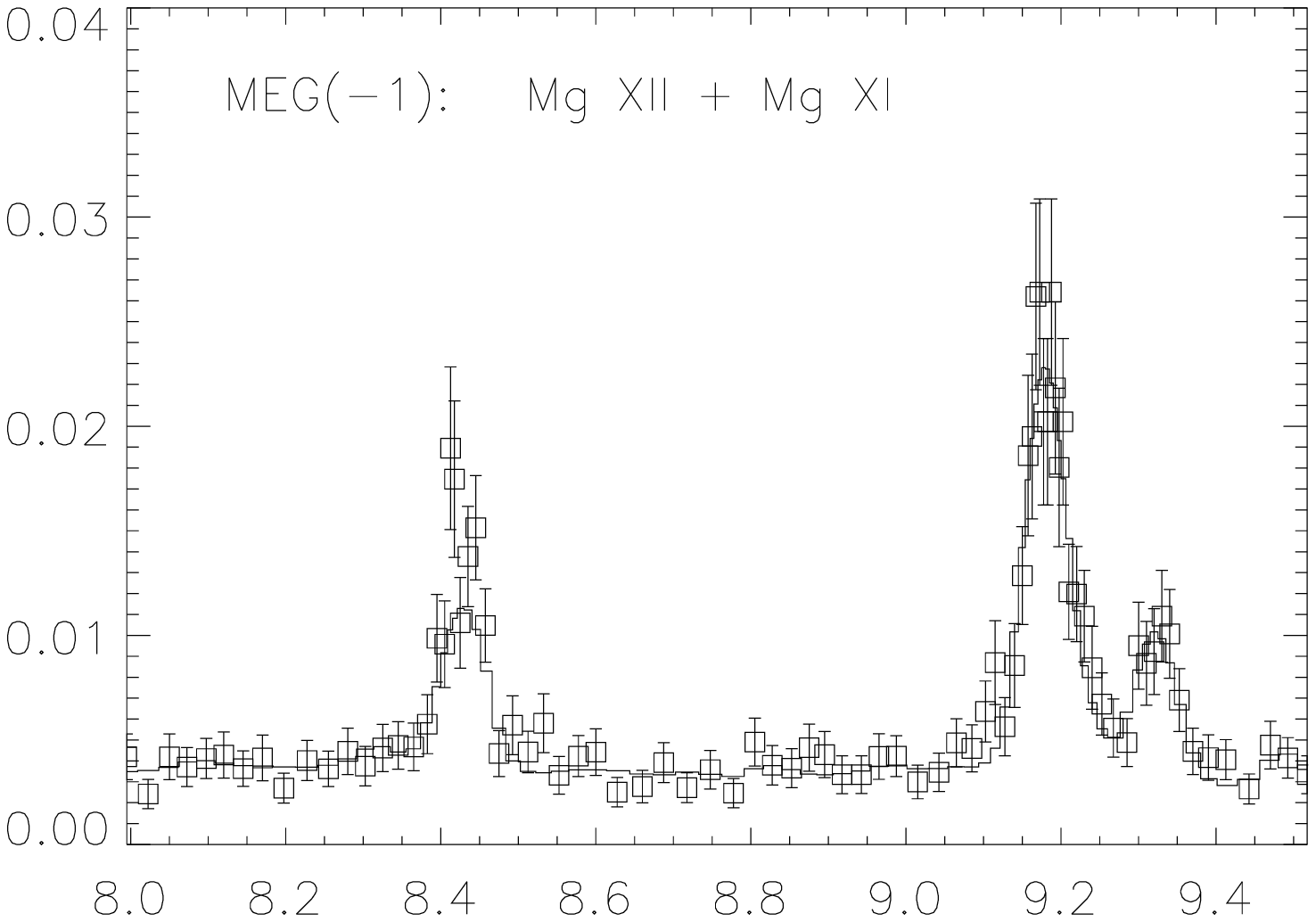}
\includegraphics[width=1.75in, height=1.25in]{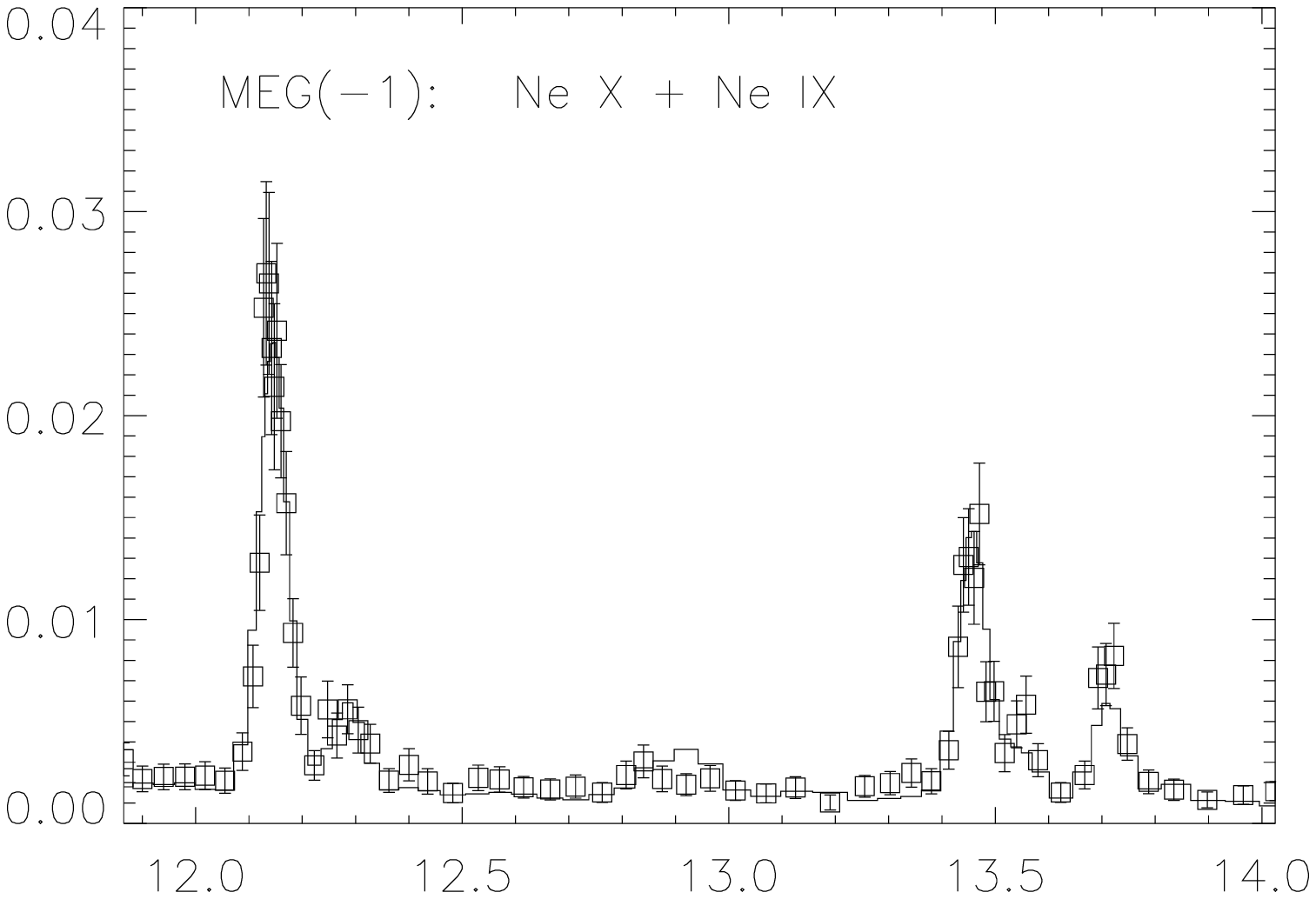}
\includegraphics[width=1.00in, height=1.25in]{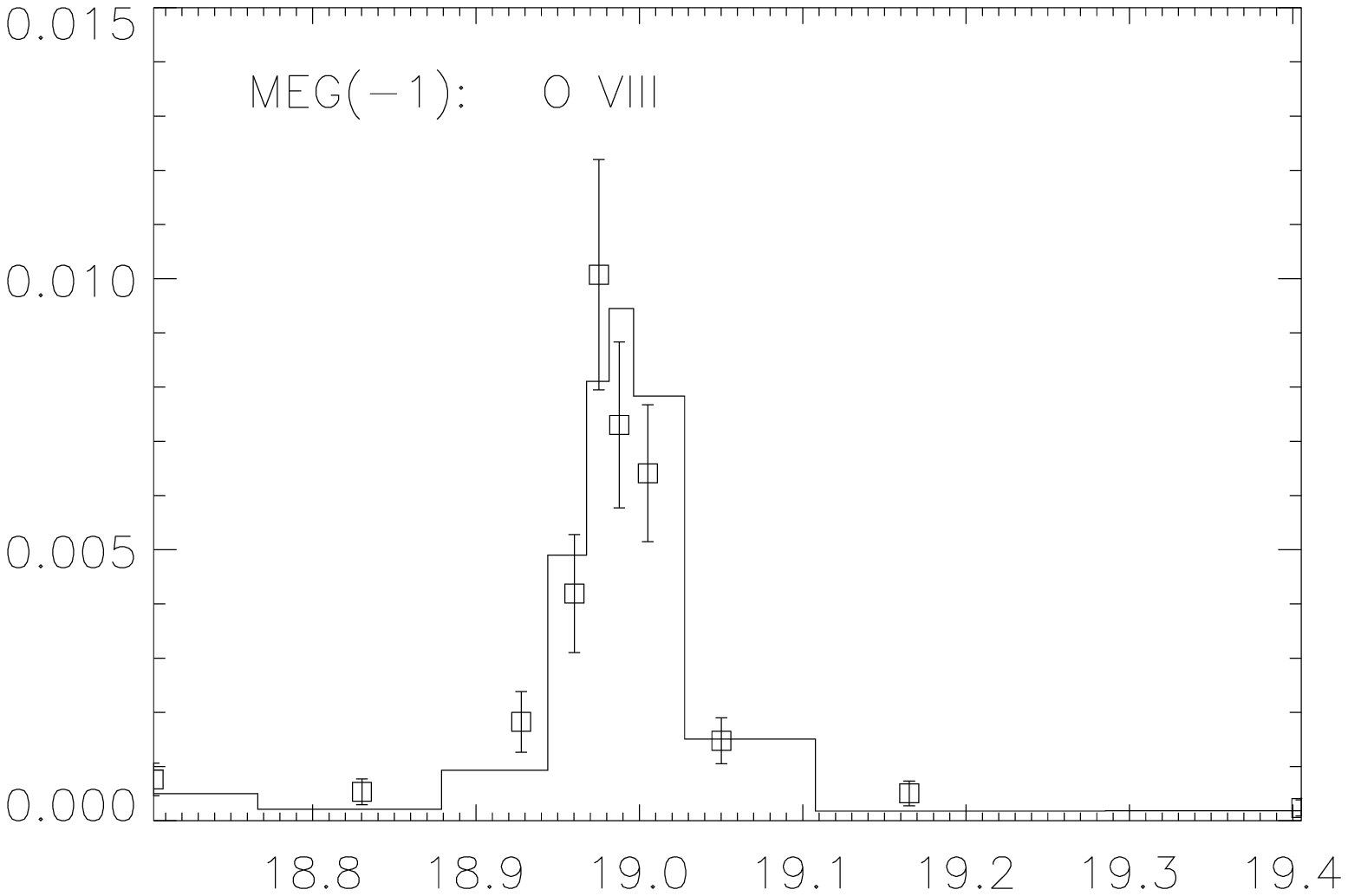}
\includegraphics[width=1.75in, height=1.25in]{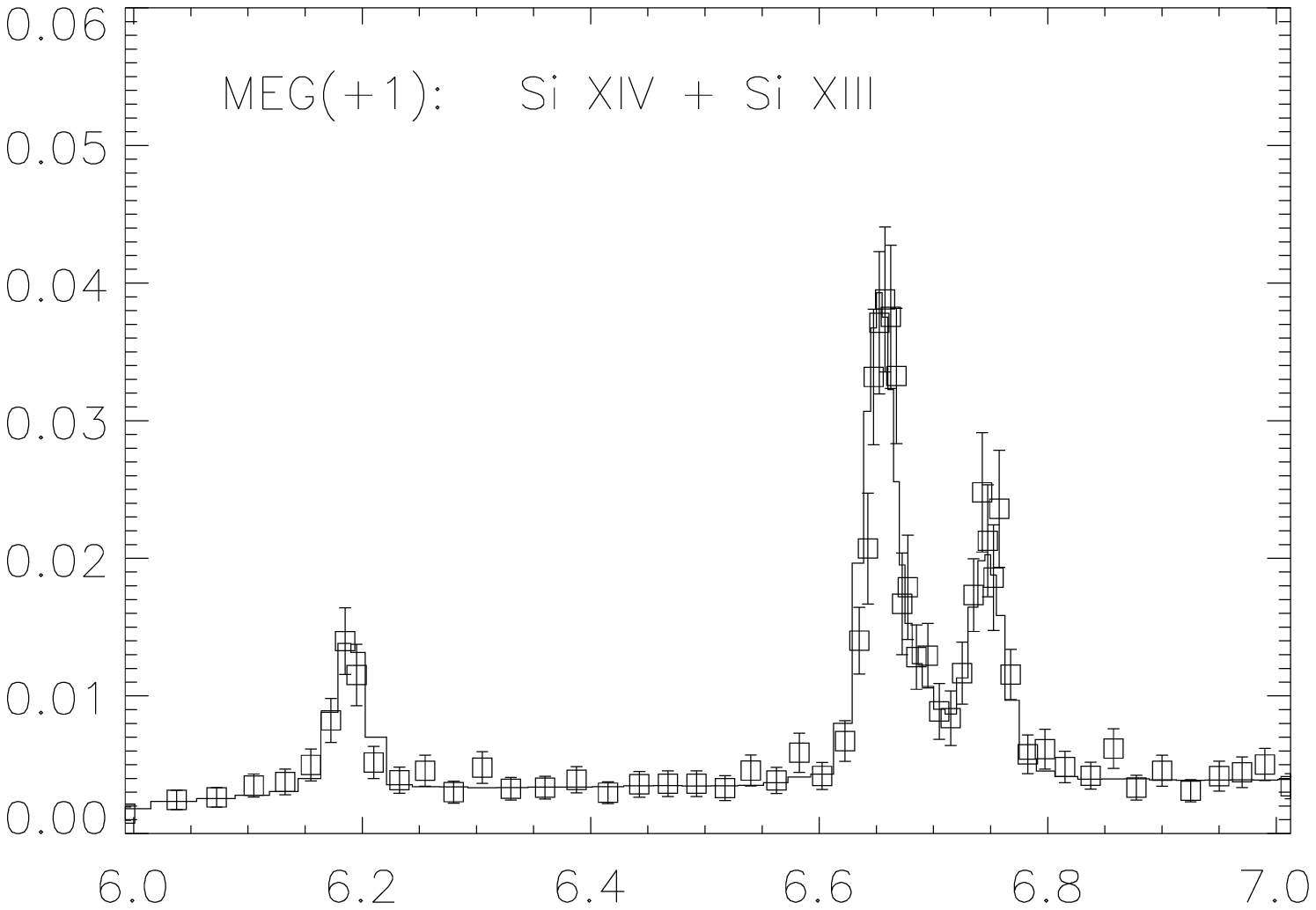}
\includegraphics[width=1.75in, height=1.25in]{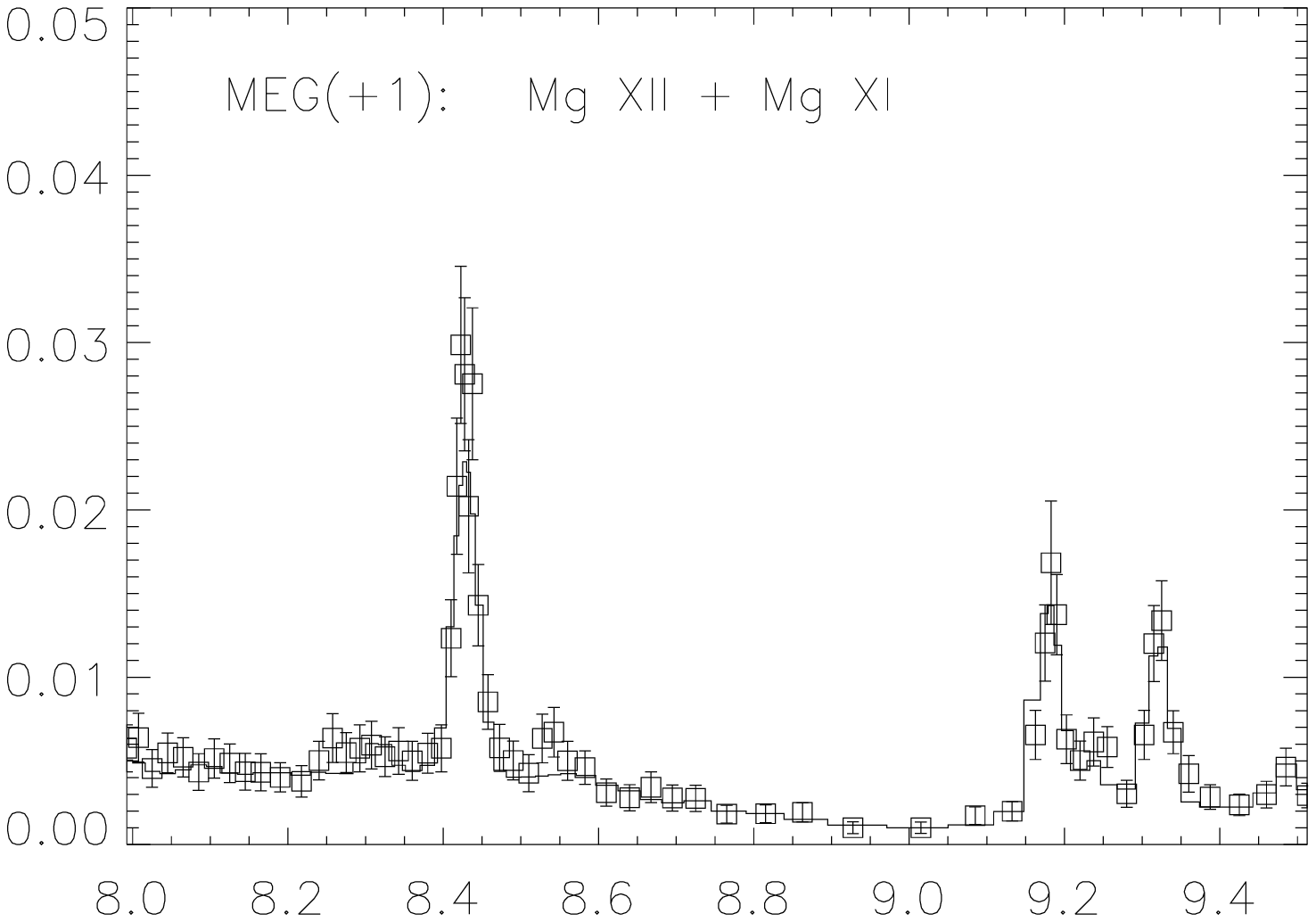}
\includegraphics[width=1.75in, height=1.25in]{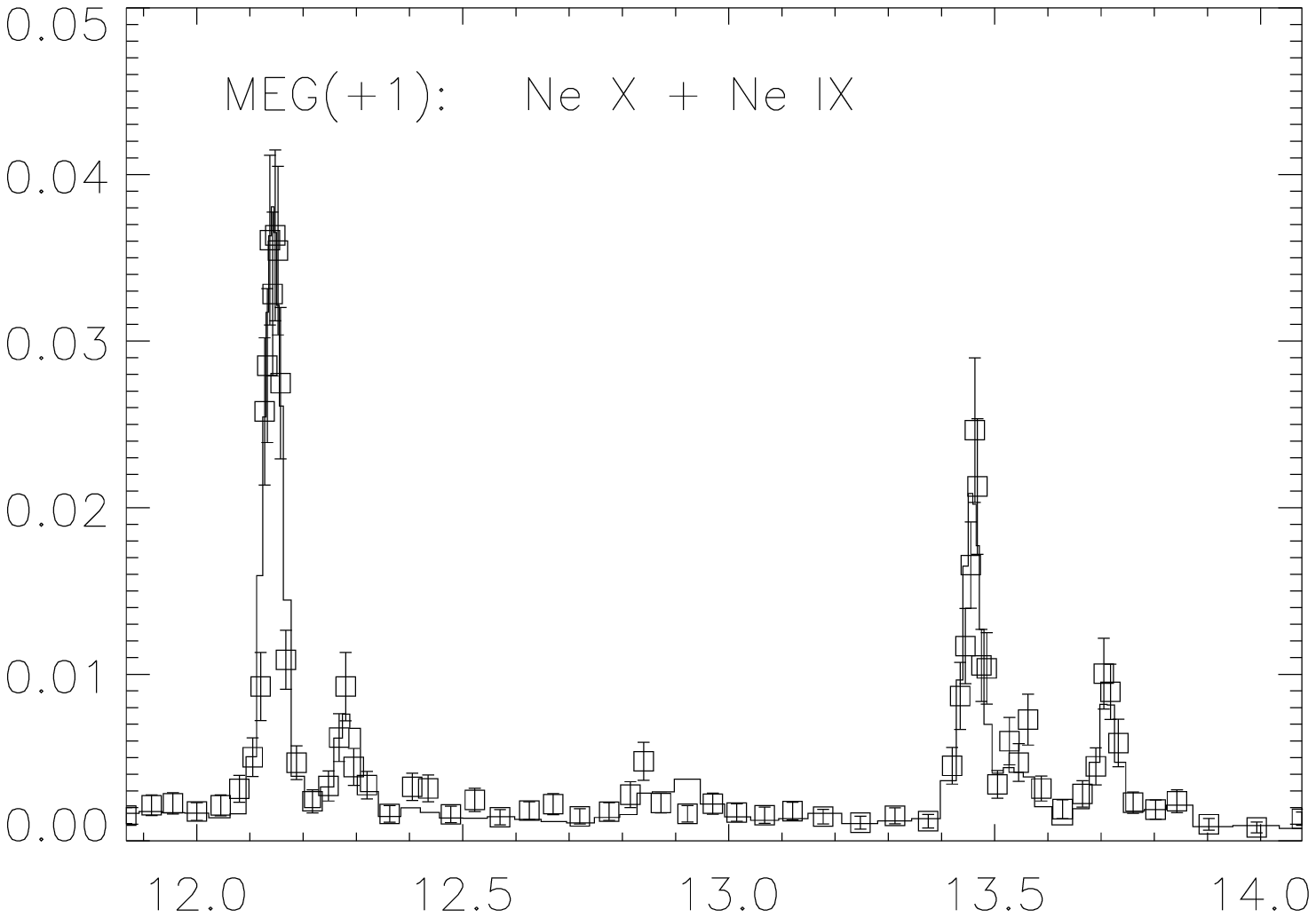}
\includegraphics[width=1.00in, height=1.25in]{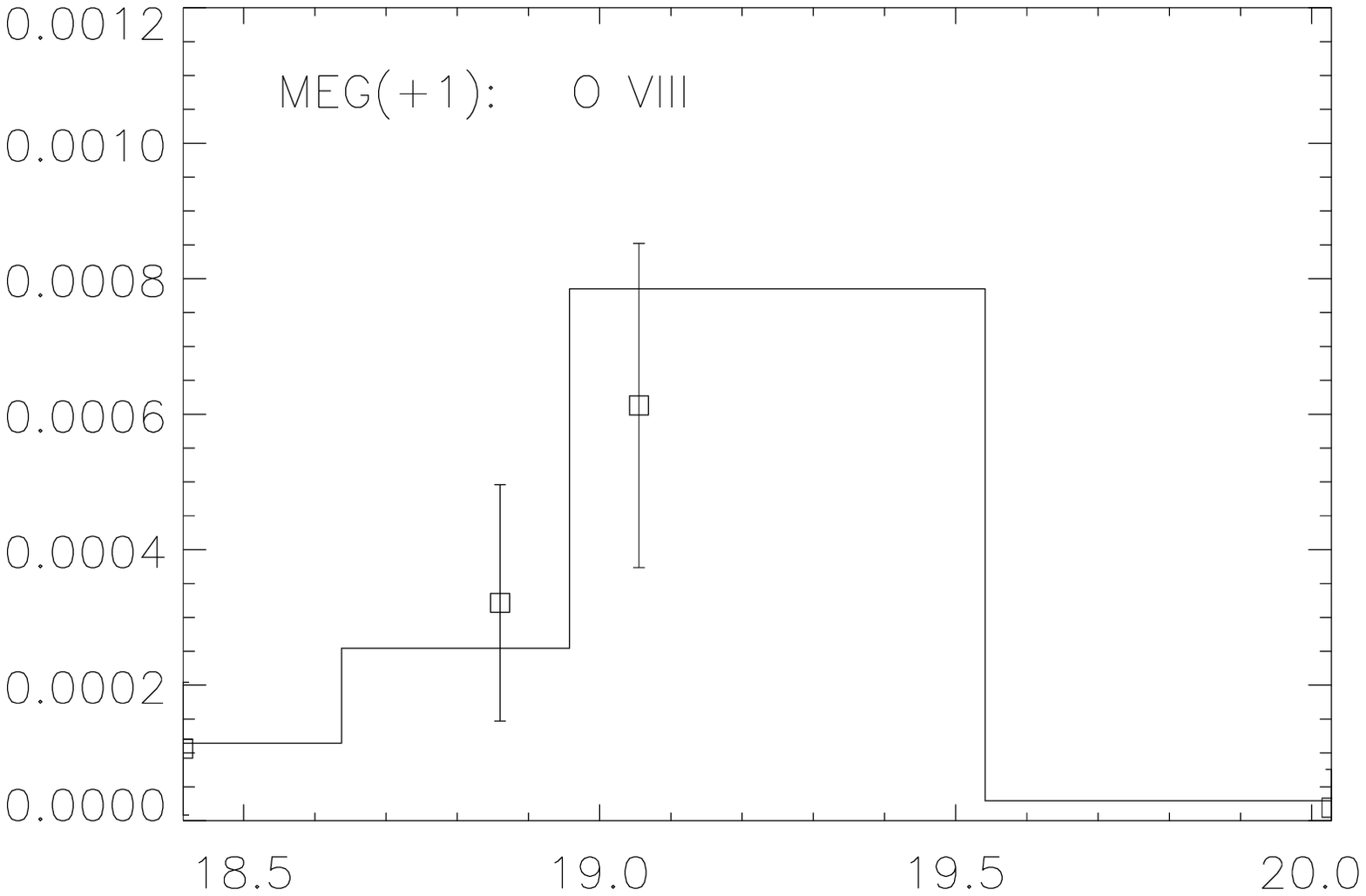}
\end{center}
\caption{
X-ray spectra of SNR 1987A in 2007 near the strong spectral
lines of the ions Si XIV, Si XIII, Mg XII, Mg XI, Ne X, NeIX and
O VIII: (i) empty squares with $1\sigma$-error bars -- observed
spectra; (ii) solid curves -- the theoretical spectra from
the two-shock model.  Horizontal axes -- observed wavelength
(\AA); vertical axes -- flux density (photons s$^{-1}$
\AA$^{-1}$).
The positive and negative LEG first-order data are shown in the first
and second row, respectively.
The negative and positive MEG first-order data are shown in the
third and fourth row, respectively. All observed spectra are
rebinned to have a minimum of 30 counts per bin.
For the 2004 LEG spectra see Fig.2 in Z06.
}
\label{fig:spectra}
\end{figure}

Fig.~\ref{fig:spectra} displays portions of these spectra in the
vicinity of strong emission lines.  The superior spectral
resolution of the MEG is evident in the emission line profiles.

\section{Global fits} 
\label{sec:global_fits} 

We performed global
fits following the procedure we used to analyze the 2004 LETG
spectra (Z06; details of the global models are found therein).  
We considered two basic models: (i) a two-shock
model; (ii) a DS-shock model (distribution of shocks). In all the
cases, we constructed simultaneous model fits to six
background-subtracted (positive and negative first-order) spectra
of LETG 2004, HETG 2007 (MEG) and LETG 2007.  We rebinned all
spectra to ensure a minimum of 30 counts per bin. The basic
assumptions of our models are outlined in \S~\ref{subsec:basic}
and the results are given in \S~\ref{subsec:res}.

\subsection{Basic assumptions} 
\label{subsec:basic} 

In our global
analysis we fit our data to a model consisting of a distribution
of plane-parallel strong adiabatic (index $\gamma = 5/3$) shocks
with velocity $V_{sh}$ that enter stationary gas.  Therefore, the
post-shock gas velocity is $V_{psh} = 3/4 V_{sh}$. Given the
adopted set of abundances (see Z06 and below) the mean molecular
weight per particle is $\mu = 0.72$, thus the post-shock
temperature is 
\begin{equation} kT = 1.4(V_{sh}/1000~\mathrm{km~s^{-1}})^2 ~keV.  
\label{eq:kT}
\end{equation}  
We
model the X-ray spectrum of the shocked plasma taking into account
the effects of non-equilibrium ionization (NEI), by using either
the corresponding models in XSPEC ({\it vpshock} in version 11.3.2)
\citep{Arnaud96} that are based on \citet{Bor01} or custom XSPEC
models constructed on the same basis (Z06).  
In fitting models consisting of a distribution of shocks, we
take as the basic `vectors' plane-parallel strong adiabatic shocks
characterized by their post-shock temperature and emission
measure. The NEI effects are characterized by the `ionization
age,' ($n_e t$), i.e., the product of the post-shock electron
number density and time since shock passage.

In our global fits we assume that all spectra share the same
value, $N_H$, of the interstellar absorption, which does not
change with time.

Likewise, we assume that the chemical abundances of the hot
(shock-heated) plasma are constant over time.  To determine
them, we adopted the same procedure as in our previous spectral
analyses (Michael et al. 2002; Park et al.  2002, 2004, 2006; Z06;
D08). That is, we only varied the abundances of elements having
strong emission lines in the observed (0.5 - 4 keV) energy range:
N, O, Ne, Mg, Si, S and Fe.  We held the abundances of H,
He, C, Ar, Ca and Ni fixed to their corresponding values as discussed
originally in \citet{Eli02}.  Some evolution of the
chemical composition might take place due to destruction of dust
grains when the shocks engulf the denser parts of the equatorial
disk. 
But we find at most modest depletions of refractory
elements in the X-ray emitting gas, and neglect dust destruction
effects in the present work (see also \S~\ref{subsec:res}).

\subsection{Fit Results} 
\label{subsec:res} 

Table~\ref{tab:fit} displays results from our simultaneous two-shock 
model fit to the LEG 2004, MEG
2007 and LEG 2007 spectra of \snr. 
We note that the shock parameters in 2004 from the simultaneous fit
are very close to those derived from the LETG 2004 analysis alone
(Z06) which
makes us confident about the model assumptions listed
in \S~\ref{subsec:basic}.
Also, we see a gradual increase in the shock ionization ages which
is natural to expect if the blast wave  encounters gas of
increasing density as it propagates into the circumstellar ring.

We also constructed a simultaneous fit ($\chi^2/dof = 1789/2242$)
of all the data with the distribution of shocks model
and inferred the following values (with $1~\sigma$~errors) for
absorption column density and elemental abundances (as ratios to
their solar values, Anders \& Grevesse 1989): 
$N_H = 1.44\pm0.01\times10^{21}$~cm$^{-2}$, N$=0.73\pm0.07$,
O$=0.10\pm0.01$, Ne$=0.34\pm0.01$, Mg$=0.31\pm0.01$,
Si$=0.37\pm0.01$, S$=0.33\pm0.04$, and Fe$=0.20\pm0.01$.  Note
that these results are close both to the values given in
Table~\ref{tab:fit} and to those from the LETG 2004 DS analysis.

\begin{figure}[ht]
\begin{center}
\includegraphics[width=3.2in, height=2.25in]{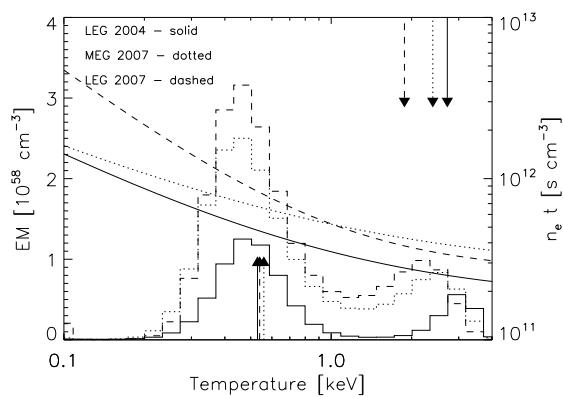}
\end{center}
\caption{
Emission measure (EM) of the distribution of shocks in SNR~1987A
as derived from the DS model with 25 points logarithmically spaced
in the (0.1 - 4 keV) post-shock  temperature range.
The arrows indicate the post-shock  temperature from the two-shock
model
fits (see Table~\ref{tab:fit}): low (high)-temperature components are
plotted from the lower (upper) x-axis.
The smooth lines show the derived ionization age of each shock
($n_e t$).
}
\label{fig:dem}
\end{figure}

Fig.~\ref{fig:dem} displays the distributions of emission measures
and ionization ages
inferred from DS model fits to the LEG 2004, MEG 2007 and LEG 2007
spectra.  The figure shows that the shock distribution is
quite stable over a period of three years, especially in its soft
part.
On the other hand, we do see noticeable
evidence of a change in the high-temperature tail.
This result is consistent with the results we infer from a
two-shock model fit to the spectra (Table~\ref{tab:fit}).
The
fact that the shape of the shock distribution is almost the same
for the HETG (MEG) 2007 and LETG 2007 data sets gives us
confidence that our DS analysis procedure is robust with regard to
the instrument spectral resolution.

Unfortunately, it is not feasible for us to carry out a detailed
error analysis for the DS shape.  The parameter space is too large
and the CPU time requirements would be prohibitive.  Instead, we
performed the standard F-test that shows whether the derived 
distributions have significantly different variances. We note
that the variance is a conventional measure of the `width', that 
is of the shape, of a given distribution. The results of the
F-test show that the probability of the null hypothesis (or of
`identical' shapes) is: 0.006 for LEG 2004 and LEG 2007 data;
0.01 for LEG 2004 and MEG 2004; and 0.85 for LEG 2007 and MEG
2007. 
Therefore, we conclude that the slight change in the shape of the
shock distribution from MEG 2007 to LEG 2007 is visually suggestive 
but not statistically significant whereas there is clear significance 
to the change in the three-year interval between the LEG 2004 
observation and the LEG 2007 observation.

Inspecting the results from our shock fits (Table~\ref{tab:fit}
and Fig.~\ref{fig:dem}), we note the following points.

The
abundances of \snr derived from analysis of the entire set of
grating spectra taken so far with {\it Chandra} confirm the
conclusions from our previous analysis (Z06). Namely, Ne, Mg, Si,
S and Fe have abundance values typical for the LMC \citep{RD92},
while N and O abundances derived from the X-ray spectra are a
factor $\sim 2$ lower than those found by
\citet{LF96} from optical and UV spectra of the inner ring.  The
lower nitrogen abundance found here is in accord with the value
inferred by \citet{Pun02} from their analysis of the optical 
and UV spectra of Spot 1 on the inner circumstellar ring. 

The gas-phase abundance of silicon measured from the X-ray spectrum is of
particular interest because it places constraints on the depletion of
silicon onto dust grains in the circumstellar matter and on the
destruction of the dust grains in the X-ray emitting gas (Dwek et al
2008).  As seen from Table~\ref{tab:fit}, the silicon abundance in the
X-ray emitting region has a value typical for LMC.  This fact implies
either  that there was little silicon depletion in the circumstellar ring
or that most of the dust in the X-ray emitting gas has been vaporized.
This result is consistent with an earlier analysis of the first mid-IR
spectrum of \snr \citep{Bou06} which concluded that the dust-to-gas ratio
in the inner ring of \snr is considerably lower than that for the
interstellar medium in the LMC (i.e., no Si depletion in the gas phase).
But in a recent and more elaborate analysis, \citet{Dwek08} found that the
mid-IR emission of the circumstellar ring can be explained if a dust
abundance typical for the interstellar medium in LMC is adopted.  
Dwek et al. also found that the ratio of infrared to X-ray emission
decreased from day 6190 to 7137 and interpreted this evolution as evidence
that dust grains were being destroyed by sputtering in the X-ray
emitting gas.  If so,
one might expect to see a secular increase of the abundance of silicon in
the X-ray spectrum; but we do not see evidence for such an increase.  The
values of silicon abundance derived from analysis of different X-ray
spectra of \snr with good photon statistics are consistent with each other
and with the value typical for LMC (Si $= 0.31$; Russel \& Dopita 1992):
Si $= 0.30\pm0.1$~ (1$\sigma$-error from CCD spectra; Michael et al.
2002); Si $= 0.32\pm0.07$~ (1$\sigma$-error from CCD spectra; Park et al.
2004); Si $= 0.28~[0.22-0.32]$~ (90\% confidence interval from grating
spectra; Z06 ); Si $= 0.33~[0.32-0.35]$~ (90\% confidence interval from
grating spectra; this work). Thus, we can rule out an increase of  the
gas-phase silicon abundance by a factor greater than $\sim 30\%$ in the
period 2002 - 2007.  These constraints on the 
physical processes governing the evolution of the shocked dust
deserve a more detailed analysis than we intend to provide
in this paper.  Such an analysis should take account of uncertainties in
the grain size distribution; temperature stratification of the X-ray
emitting plasma (see Fig.~\ref{fig:dem});  uncertainties in the unabsorbed
X-ray flux etc.  Obviously, these results are tantalizing and suggest that
future observations both in X-rays and mid-IR and more elaborate, joint
modeling of these spectra will tell us much about dust
destruction in SNR 1987A.

The cooling trend for the shocked-heated plasma in \snr that we
infer from both the two-temperature model and the DS model fits to
the grating data is consistent with the trend inferred from fits to 
the pulse height spectra of the imaging data \citep{P06}.  We
believe that the trend is a result of the fact that the blast wave
is encountering gas of increasing density as it enters the inner
ring.  The inferred evolution of the shock ionization age,
although not a very tightly constrained
parameter, is consistent with this conclusion.
It is worth noting that the X-ray plasma models assumed a complete
adiabaticy. Such an assumption is likely to break when the blast wave
reaches deep in the inner ring and a more elaborate modeling will be
required that also takes into account the efficient hot gas cooling
due to dust emission.

The actual distributions of densities, velocities, and
temperatures, and ionization ages of the X-ray emitting gas in
\snr are likely to be very complex.  The model fits to global
spectra
presented here are surely oversimplified and miss many details;
but we believe that they provide insight into the actual
situation.  The two-shock models give us an idea of the
temperature trends of the shock distribution, while the DS models
provide a bit more detail of the temperature distribution
function.  The main advantage of these models is that they take
into account the non-equilibrium ionization.  But to begin to
understand how the temperature distribution of the shocked gas is
related to the actual hydrodynamics and geometry of the shock
interaction region, we need to analyze line profiles.

\section{Spectral Lines: profiles, ratios, and light curves}
\label{sec:line_prof}

The widths and ratios of spectral lines can provide us with
additional diagnostics of the shocked gas. For example, the
evolution of the ratios of lines from particular ions indicates
the temperature evolution of the plasma that emits these lines and
the importance of non-equilibrium ionization. On the other hand,
the line profiles give us a measure of the bulk velocity,
$V_{bulk}$, of the X-ray emitting plasma.  It is interesting to
compare the velocities inferred from line profiles with the
velocity of the shocked gas inferred from its temperature
according to equation (1).  Two cases are worth noting:
\begin{itemize} 
\item 
$V_{bulk} > 3/4 V_{sh}$,~in which case the
explanation could be: (i) the ion and electron temperatures behind
the shock have not equilibrated yet; or (ii) the shocks are
oblique.

\item 
$V_{bulk} < 3/4 V_{sh}$,~in which case the most likely
explanation is that reflected shocks are the dominant source of
heat for the X-ray  emitting plasma.

\end{itemize}

\subsection{Line profile fitting}

A general and very important feature of the line profile analysis
of the {\it Chandra} X-ray grating spectra of \snr is that due to the
complex
nature of the spatial-spectral effects we cannot measure directly
the bulk gas velocities in the hot plasma. In fact, we can measure
the line widths; but to infer the bulk gas velocities, we must
carry on some modeling as well. Two approaches have been
adopted so far: (i) Z05 derived the widths of individual spectral
lines in the  positive and negative first-order spectra which were
then fitted simultaneously to deduce the gas velocities;
(ii) D08 used an expanding ring model to fit the line profiles and
the gas velocities were derived for each spectral line.
Both techniques were used in the present analysis of the 2007 HETG and
LETG spectra.  This approach served two purposes: first, to ensure
compatibility with the 2004 LETG analysis; and, second, to better
constrain the derived line parameters.

For the first technique, we follow the procedure employed by Z05
and we used a Gaussian fit to the line profiles.
Having derived the line widths for the
first-order spectra, we fit them by the formula that assumes that
the net line width (FWHM) is:
\begin{equation} \Delta
\lambda_{tot} = 2 \Delta \lambda_0 \pm 2 z_0
(\lambda/\lambda_0)^\alpha \lambda\,,~
\label{eqn:stratified}
\end{equation}
where the plus (minus) sign refers to the ($+1/-1$)
order of the LEG spectrum and to the ($-1/+1$) order of the MEG
spectrum, respectively. In this simplified representation of the
spatial-spectral effects, the first term accounts for the spatial
broadening and the second term accounts for the bulk gas motion.
The parameter $z_0$~ determines the line broadening at some fiducial
wavelength $\lambda_0$.
The coefficient $\alpha$ accounts for velocity stratification of
the shocked gas.  If all spectral lines have the same bulk gas
velocity, $\alpha = 0$; otherwise, $\alpha \neq 0$.

\begin{figure}[ht]
\begin{center}
\includegraphics[width=1.55in, height=1.25in]{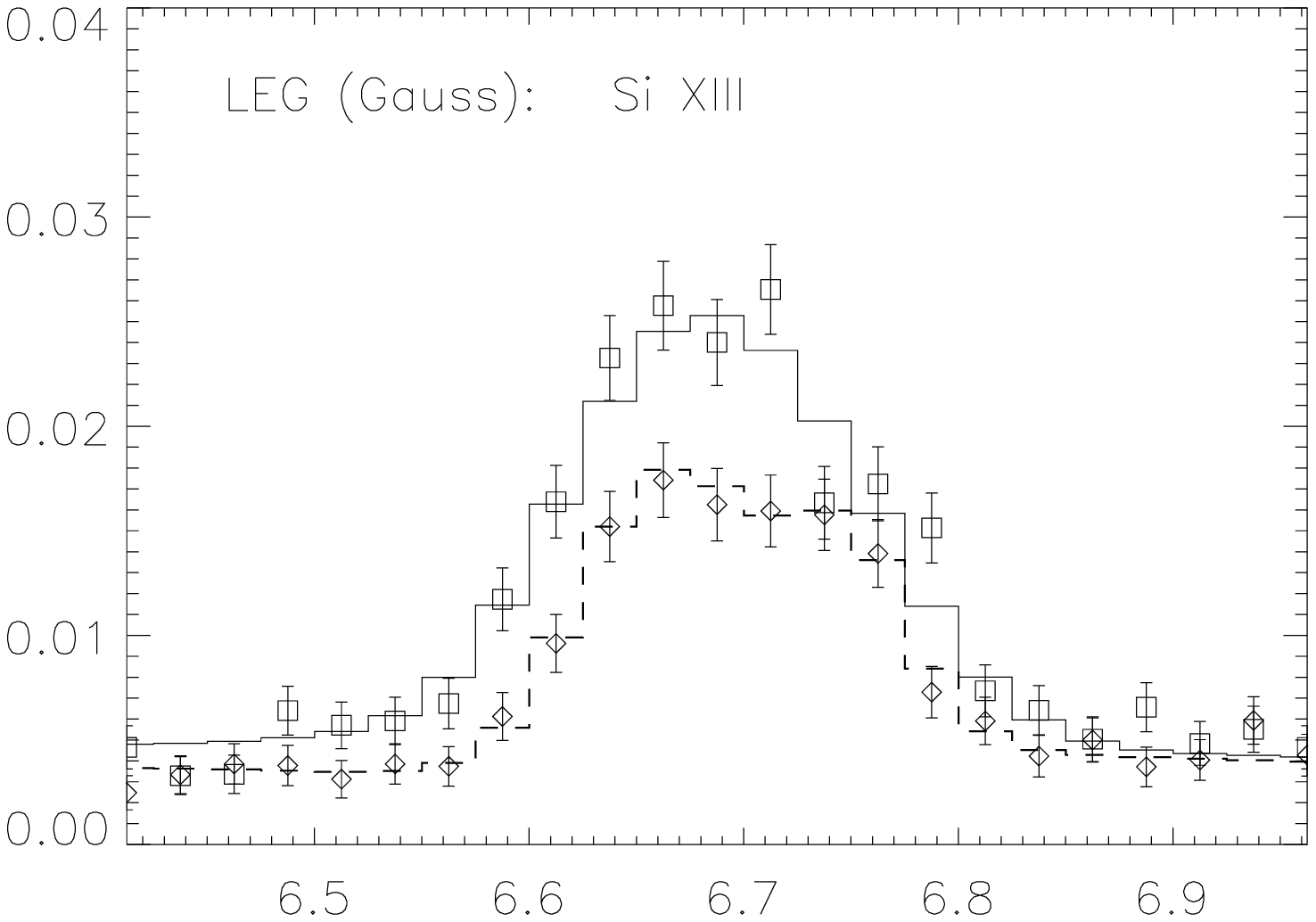}
\includegraphics[width=1.55in, height=1.25in]{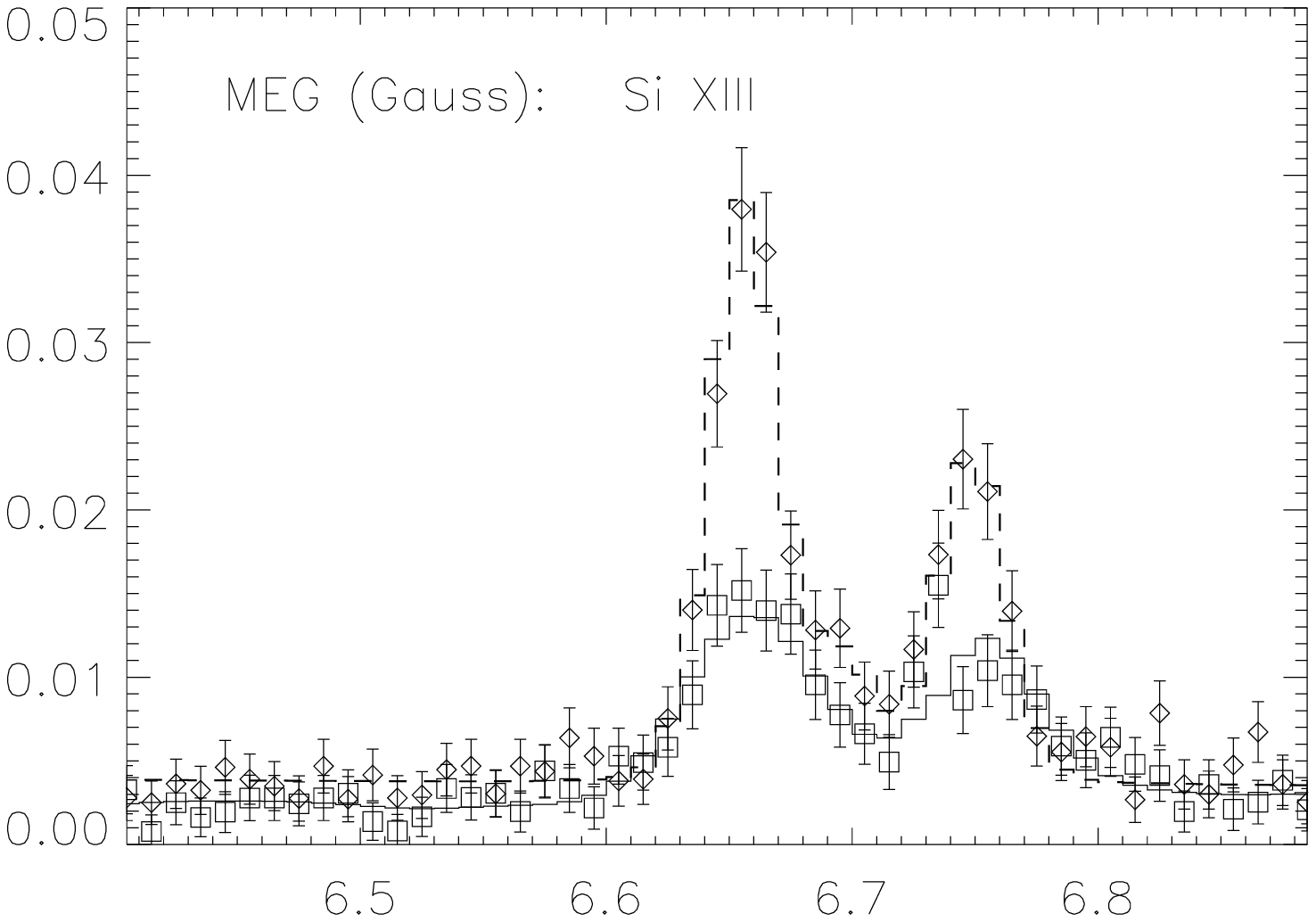}
\includegraphics[width=1.55in, height=1.25in]{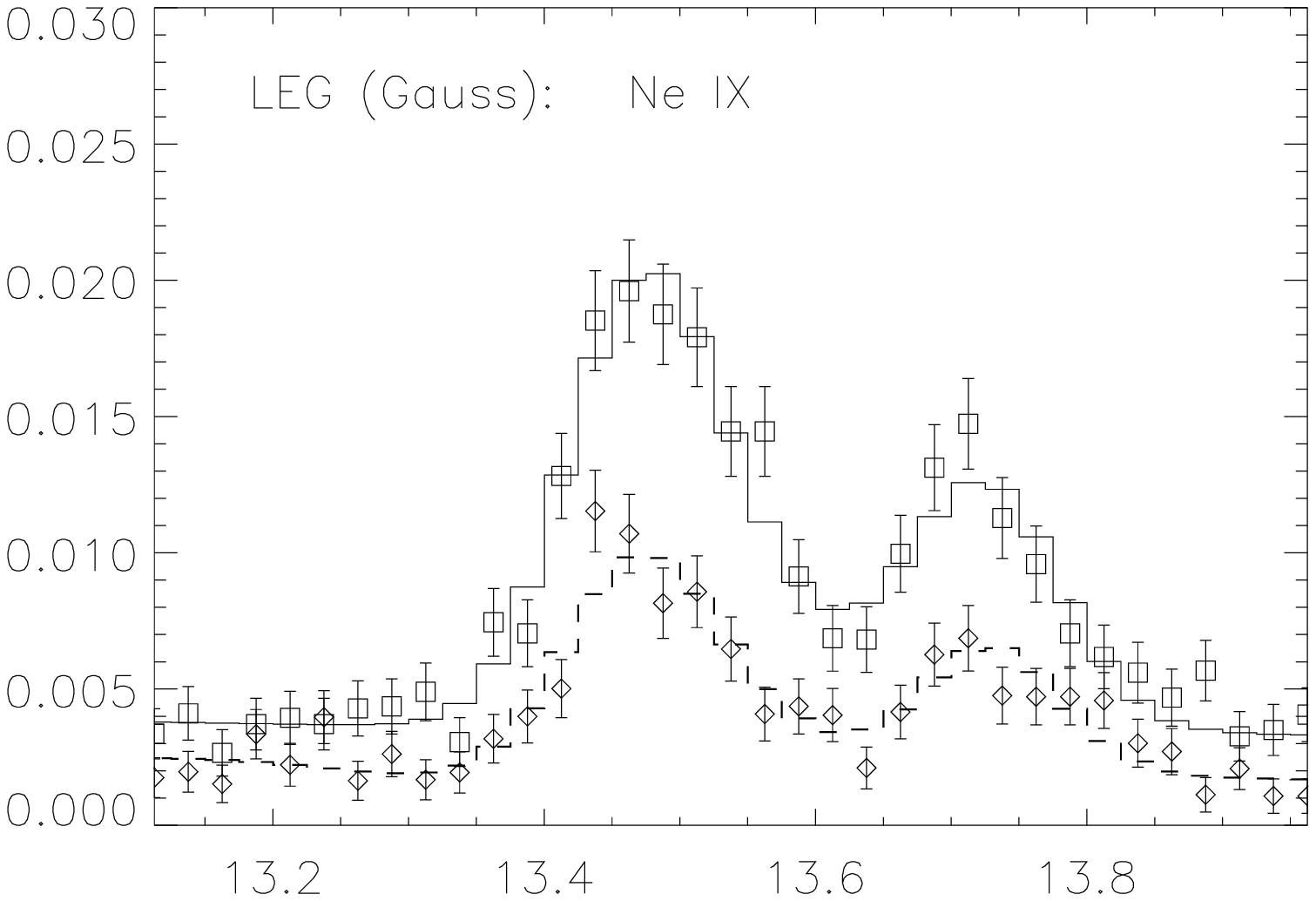}
\includegraphics[width=1.55in, height=1.25in]{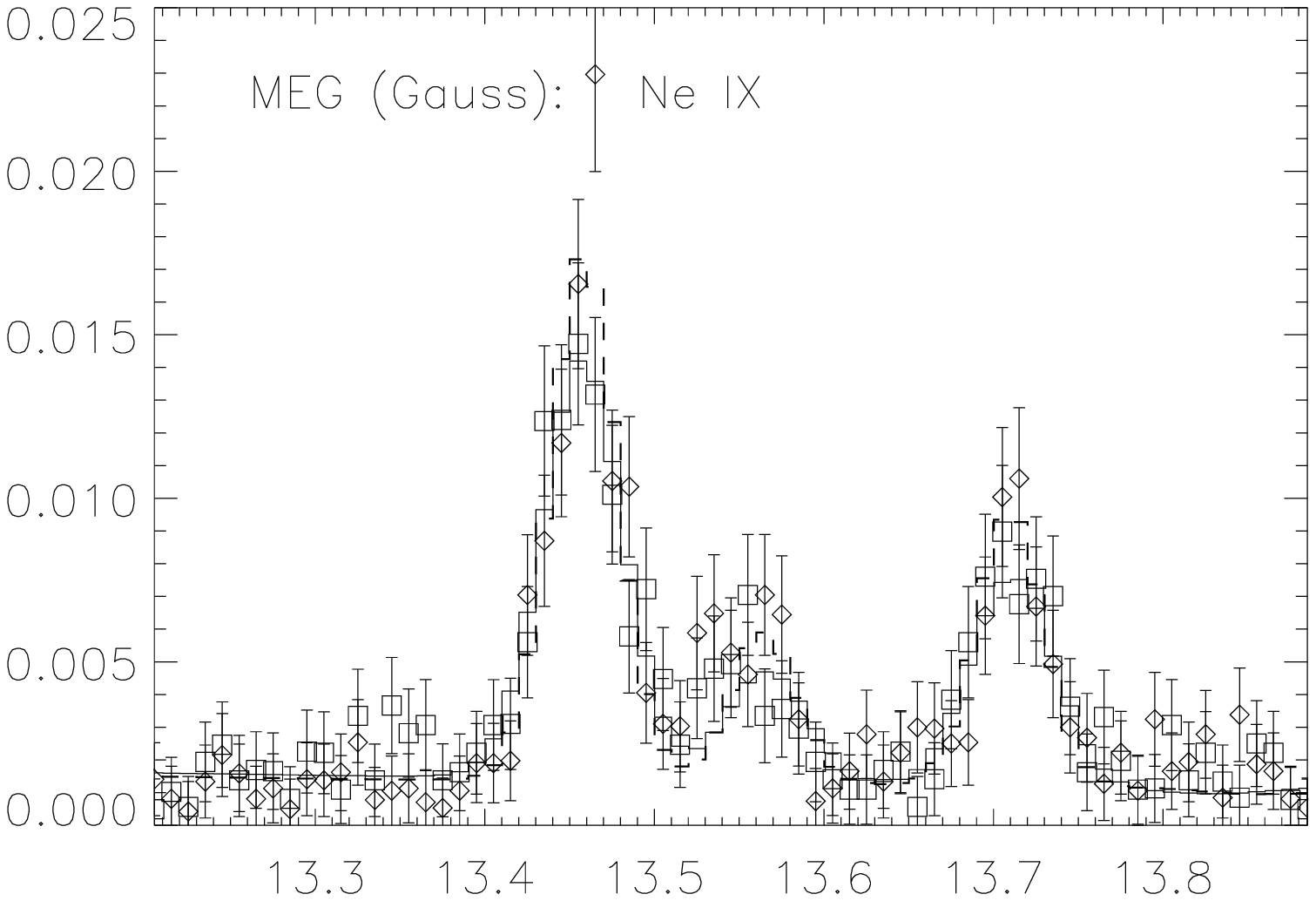}
\includegraphics[width=1.55in, height=1.25in]{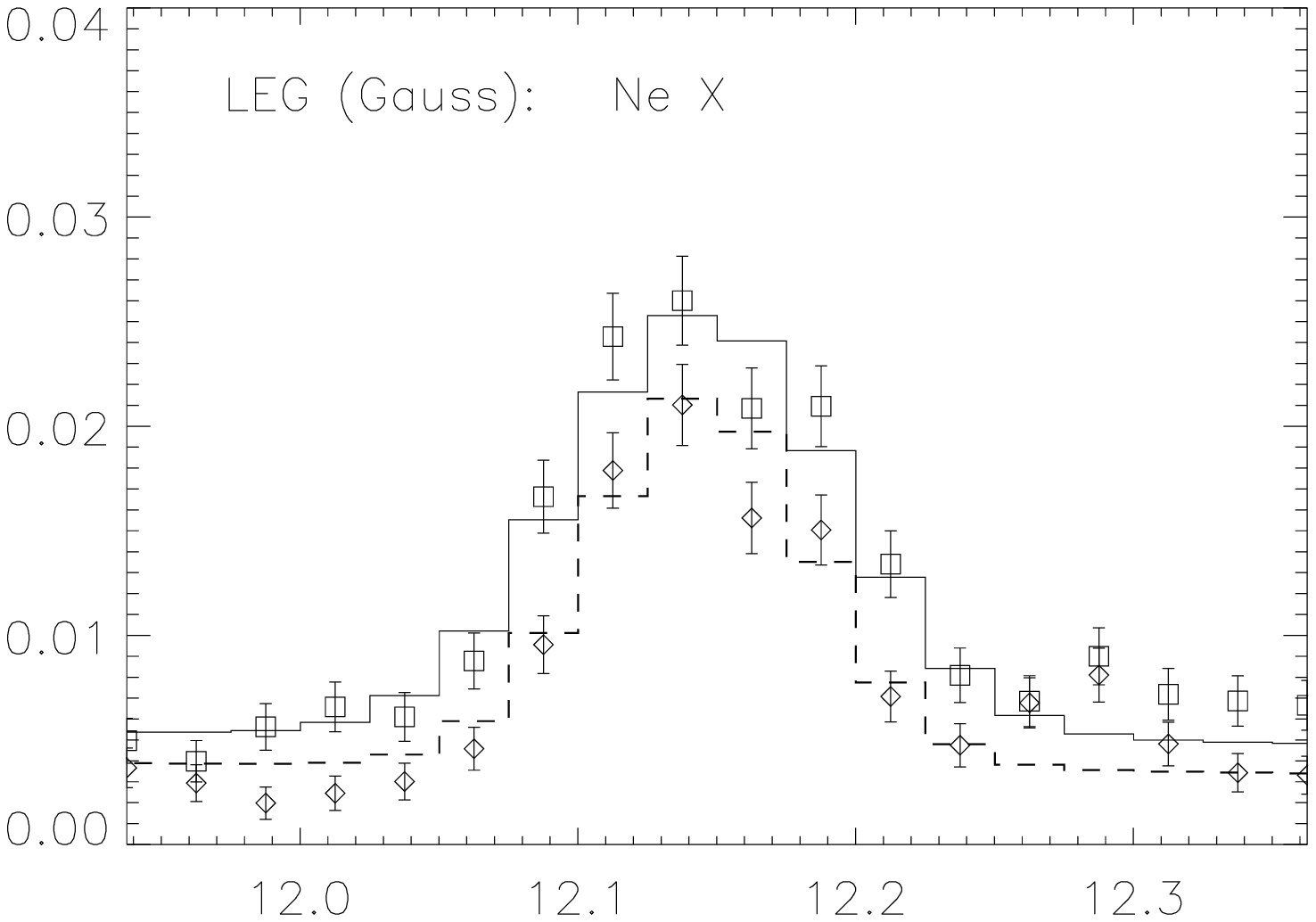}
\includegraphics[width=1.55in, height=1.25in]{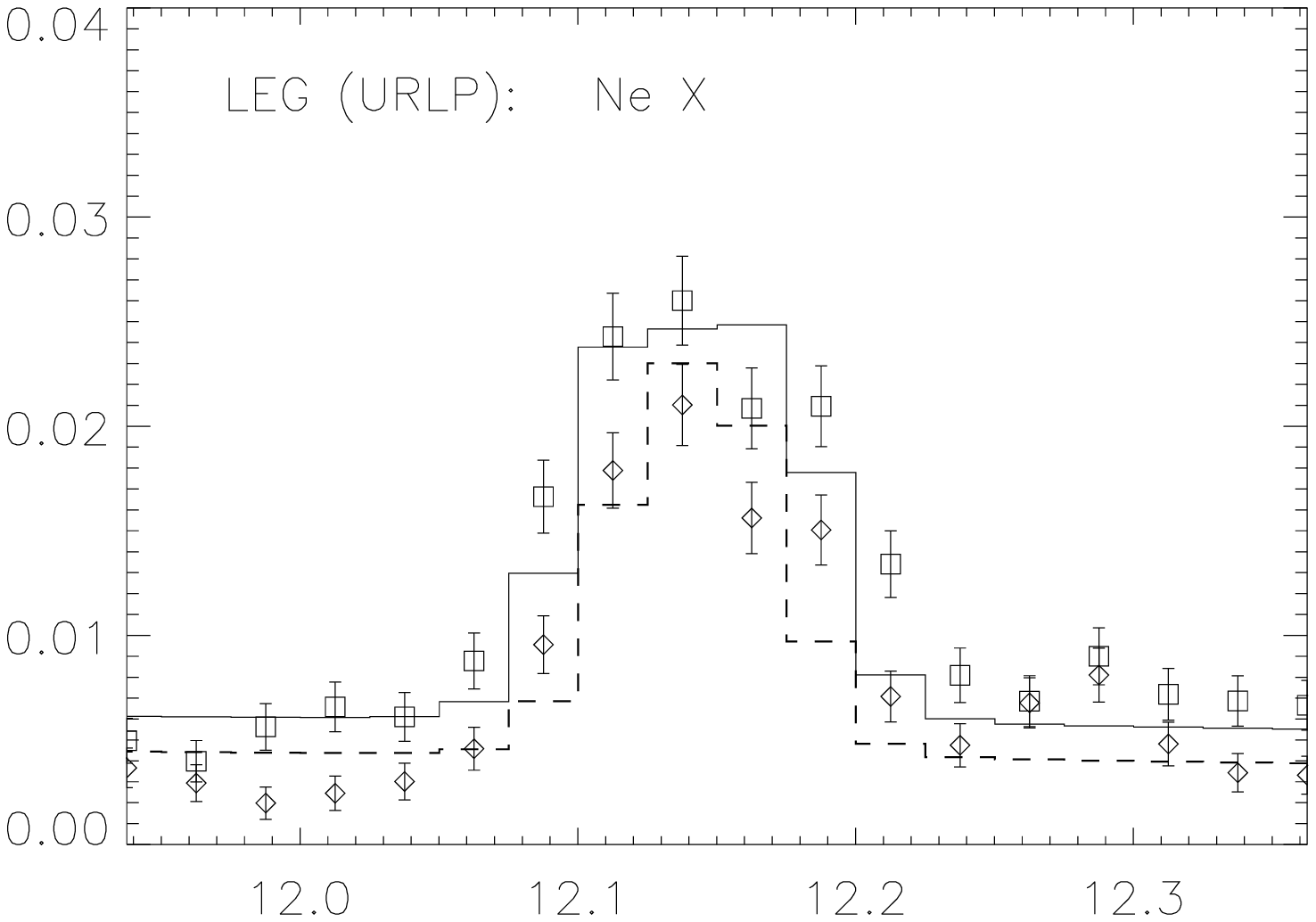}
\includegraphics[width=1.55in, height=1.25in]{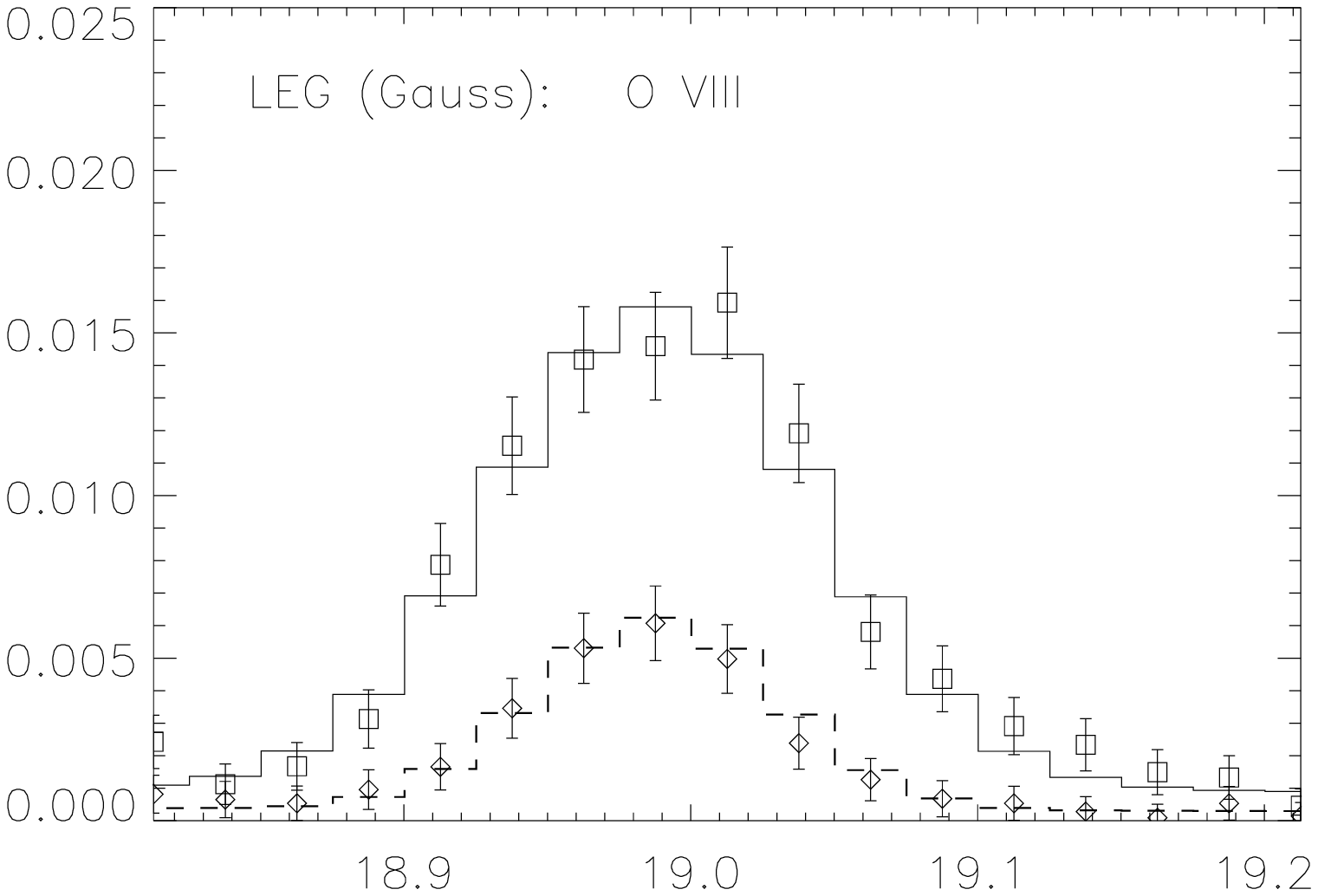}
\includegraphics[width=1.55in, height=1.25in]{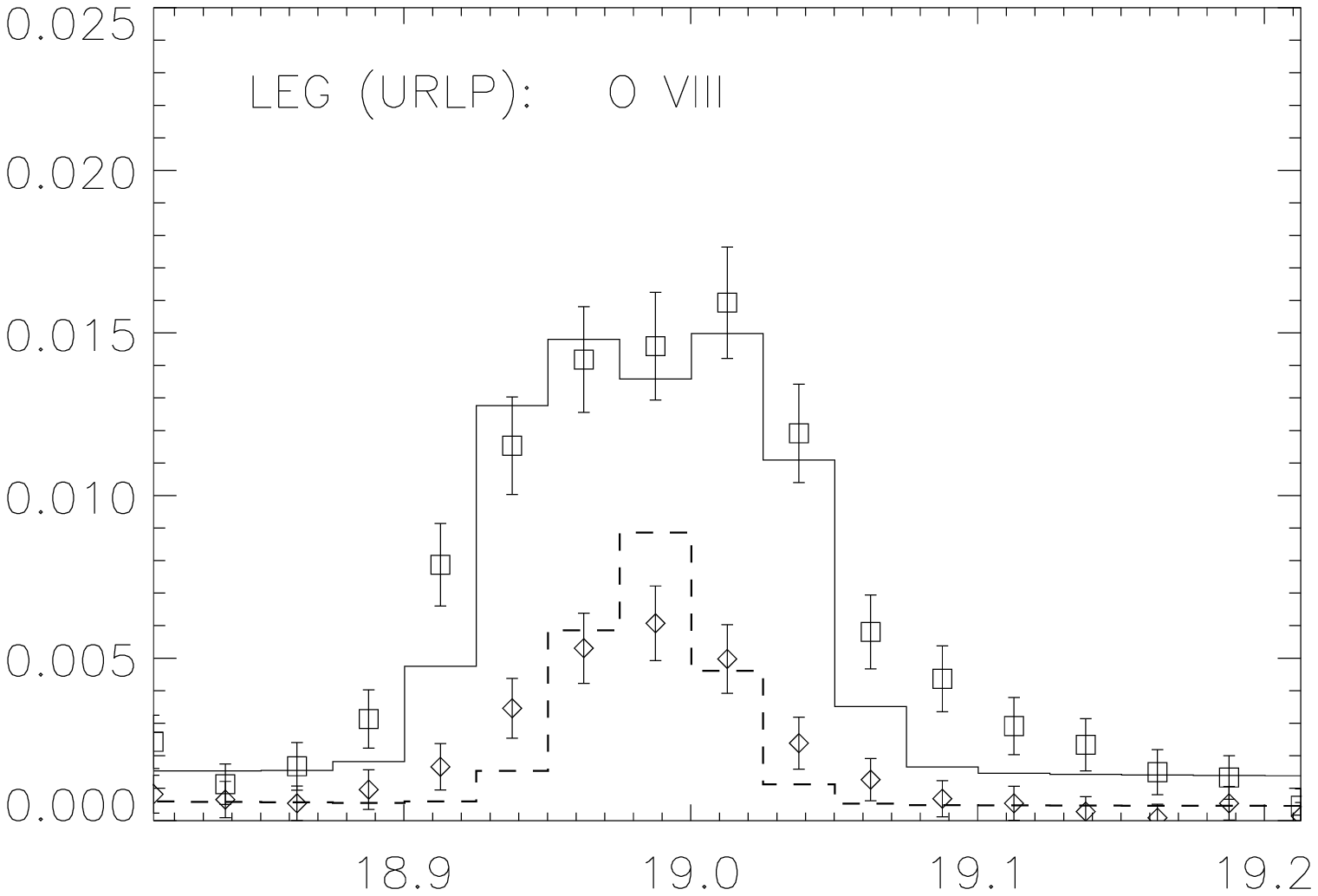}
\end{center}
\caption{Examples of fits to the line profile of some strong lines
in the X-ray spectra of \snr in 2007.
The first row presents the He-like triplets of Si XIII and Ne IX
for the LEG and MEG spectra, assuming that each component has
Gaussian profile.
The second row shows the H-like doublets of Ne X and O VIII as
detected in the LEG spectra. Both Gaussian and uniform-ring line
profile (URLP) models were used in the fits.
The broadened (LEG$+1$; MEG-1) and narrowed (LEG-1; MEG$+1$) lines
are shown by solid and dashed stepped lines, respectively.
Horizontal axes -- observed wavelength
(\AA); vertical axes -- flux density (photons s$^{-1}$ \AA$^{-1}$).
}

\label{fig:lines}
\end{figure}

For the second technique, we
constructed a custom model in XSPEC in which we assume that the
line is emitted from a gas with uniform density and bulk gas
velocity, occupying a ring-like structure.
We
call this model the 'Uniform-Ring Line Profile' (URLP) model.
Details are found in Appendix~\ref{app:urlp}.
Examples of fits with the both techniques to spectral lines in
the HETG 2007 and LEG 2007 spectra are given in Fig.~\ref{fig:lines}.
At wavelengths smaller than
$\approx 15$~\AA, we fitted the HEG and MEG data simultaneously;
otherwise we included only the MEG data.

\begin{figure}[ht]
\begin{center}
\includegraphics[width=3in, height=2.25in]{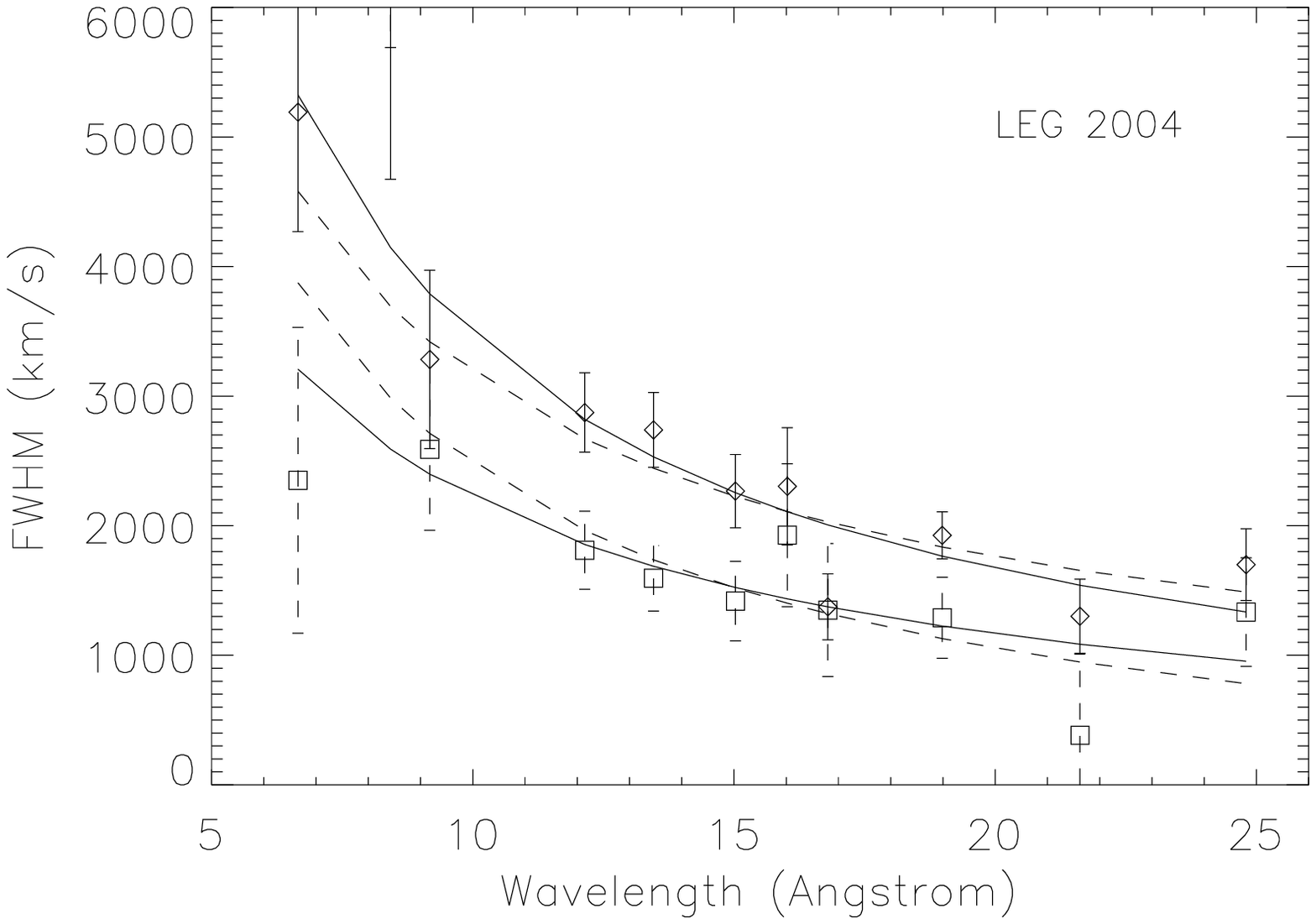}
\includegraphics[width=3in, height=2.25in]{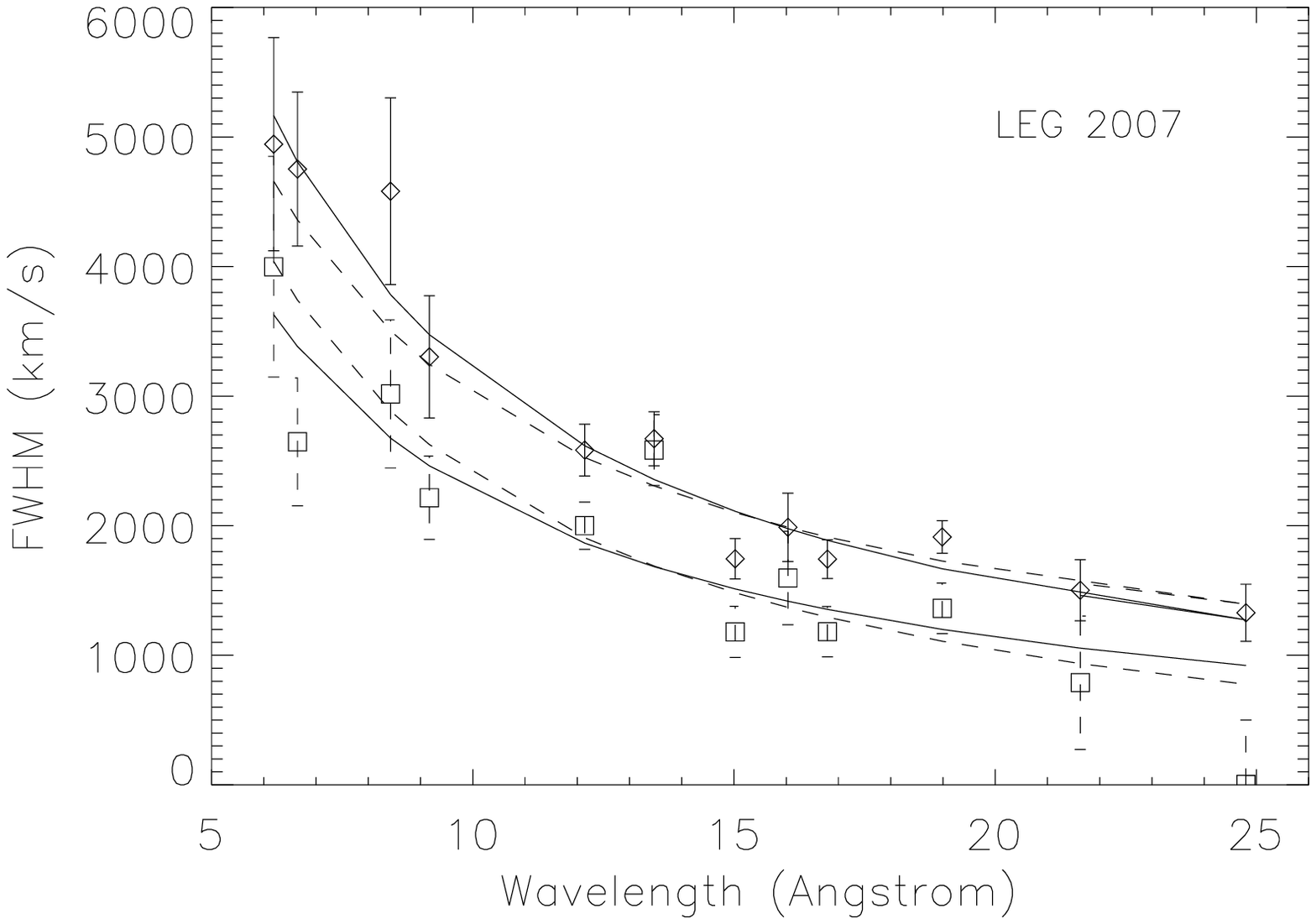}
\includegraphics[width=3in, height=2.25in]{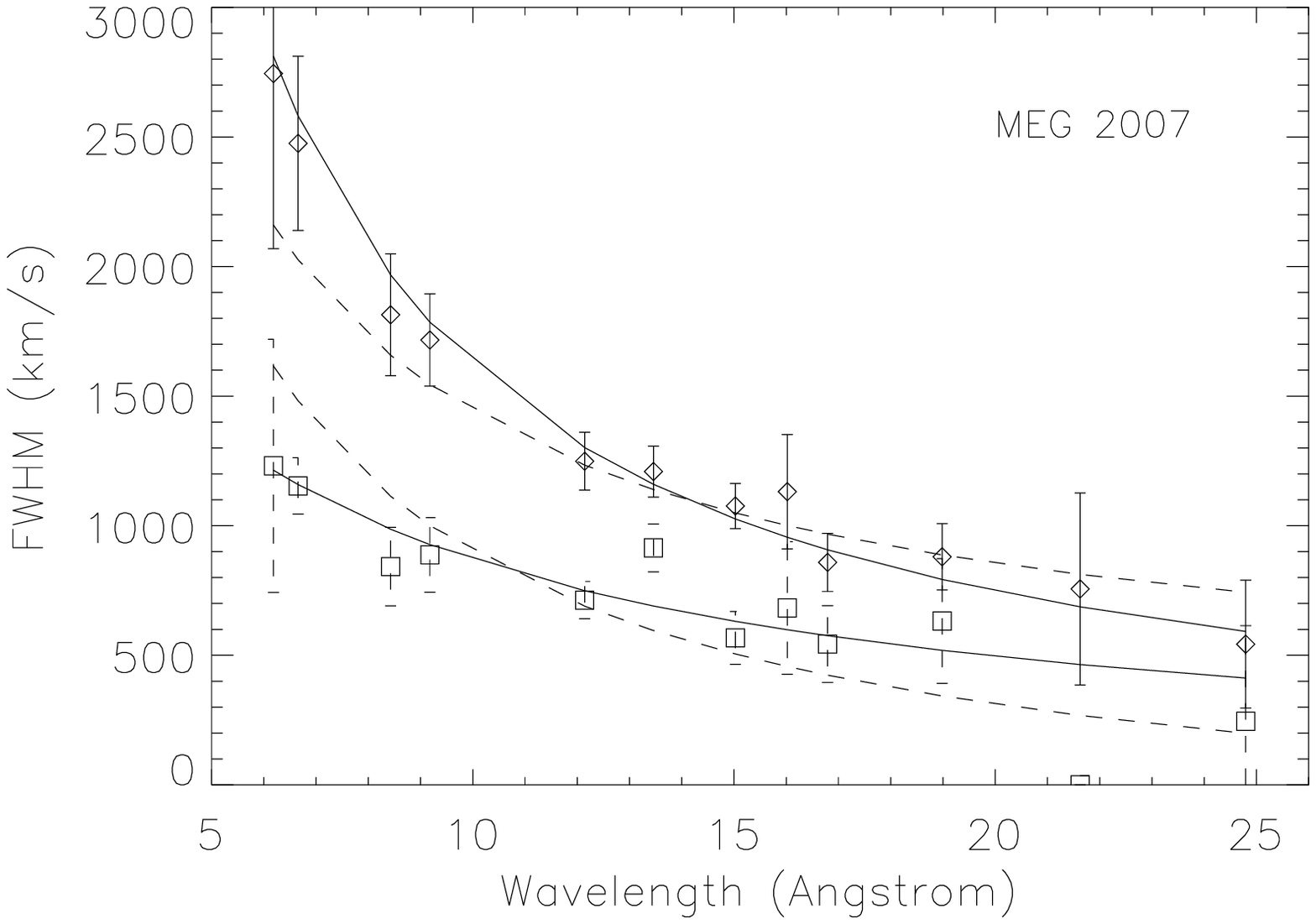}
\includegraphics[width=3in, height=2.25in]{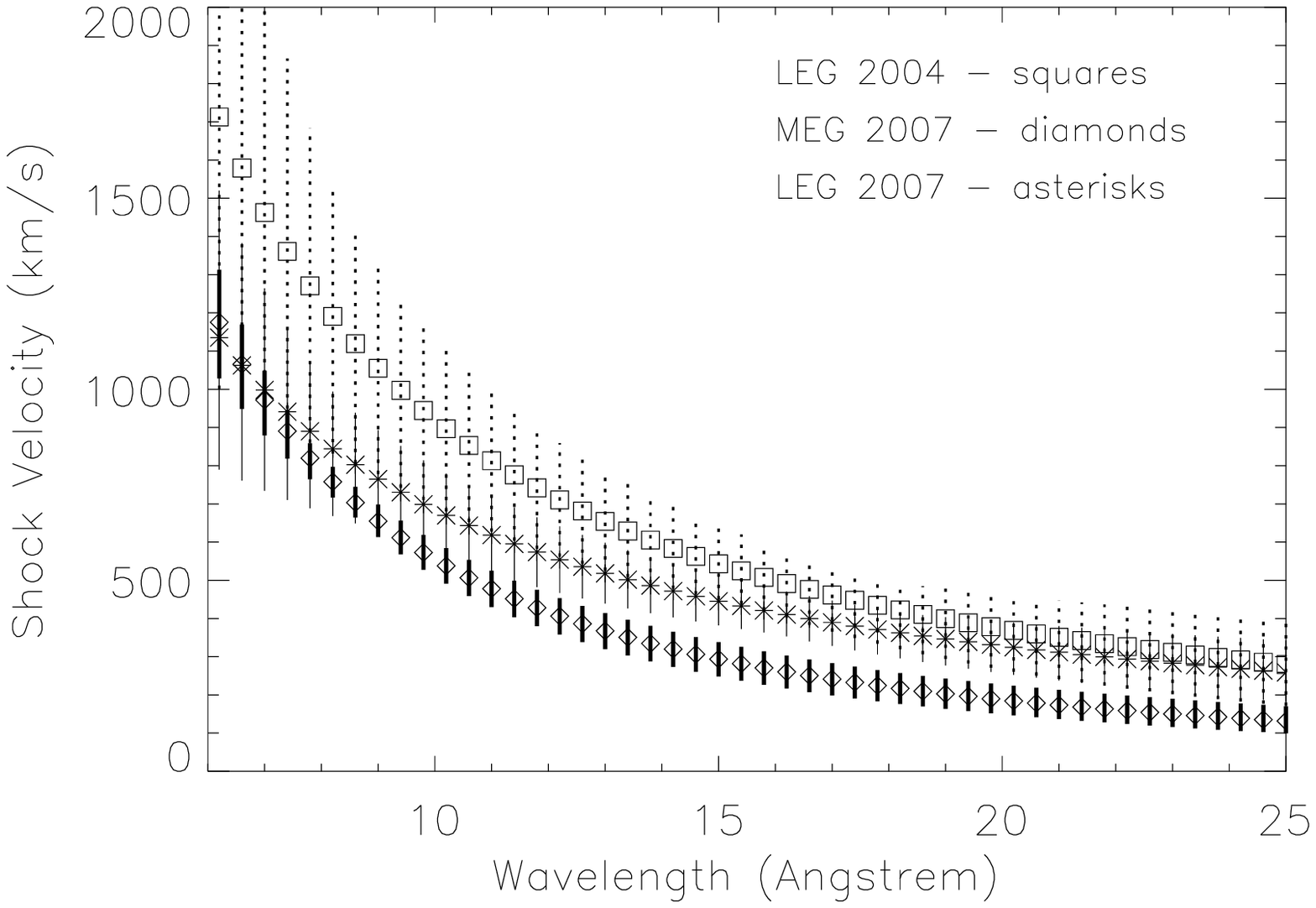}
\end{center}
\caption{
Measured line widths (FWHM) for the broadened ({\it diamonds}) and
narrowed ({\it squares}) spectral arms, respectively.
The solid and dashed curves correspond to the line broadening
parameter
in the cases with and without shock stratification
(see eq.~[\ref{eqn:stratified}]).
The lower right-hand side panel shows the
derived shock stratification.  The `error bars' indicate the
confidence
limits corresponding to the $\pm 1\sigma$~errors in the fits.
The LEG 2004 results are from Z05.
}
\label{fig:fwhm}
\end{figure}

Figure~\ref{fig:fwhm} shows the results of using the Gaussian
technique that renders a comparison with the results from
the LETG 2004 analysis.
The derived FWHMs of strong emission lines clearly illustrate the
spatial-spectral effects, namely, the line widths are broadened in
one of the spectrograph's arms and they are narrowed in the other
one. Also, the line widths are smaller in MEG spectra compared to
those in the LEG spectra. This is a direct result of the
different spectral scale (\AA~per arc-second) in the two
instruments with that of LEG being a factor $\sim2.5$~ larger.
Therefore, the spatial broadening is larger in LEG
(see the 1-st term in eq.~[\ref{eqn:stratified}]) which translates
to larger observed line widths.
In the shock-stratified case,
the best-fit values and the corresponding $1\sigma$~errors from
the fits are:
$\Delta \lambda_0 = 0.0207\pm0.0006 $~\AA, $z_0 =
0.00089\pm0.00005$, and $\alpha = -1.57\pm0.28$~(MEG 2007);
$\Delta \lambda_0 = 0.0454\pm0.0012 $~\AA, $z_0 =
0.00061\pm0.00096$, and $\alpha = -1.06\pm0.46$~(LEG 2007);
We note that given the grating dispersions of
0.056 \AA/arcsec (LEG) and 0.0226 \AA/arcsec (MEG), the derived
values for the source `half-size' ($\Delta \lambda_0$) are
consistent between MEG and LEG data: 0.92\arcsec and 0.81\arcsec,
respectively.
The lower right-hand side panel in Fig.~\ref{fig:fwhm} shows the
derived shock stratification.  The `error bars' indicate the
confidence
limits corresponding to the $\pm 1\sigma$~errors in the fits.
Two things are immediately evident: 
\begin{itemize} 
\item 
Despite the
relatively large errors, the fast shock velocities in
the distributions are slowing down (LEG 2004 vs. LEG 2007). This
is consistent with the temperature trend discussed in
\S~\ref{sec:global_fits}.  

\item 
Although they are qualitatively
consistent, the shock velocities inferred from the LEG data are
greater than those derived from the MEG spectra (LEG 2007 vs. MEG
2007).

\end{itemize}

We note that D08 already reported relatively low bulk gas
velocities in \snr from an analysis of the HETG 2007 spectra based
on a line-profile model that assumes a uniform gas motion and a
ring-like geometry.  Here, we arrive at the same conclusion by
employing a completely different modeling technique.

\begin{figure}[ht]
\begin{center}
\includegraphics[width=3in, height=2.25in]{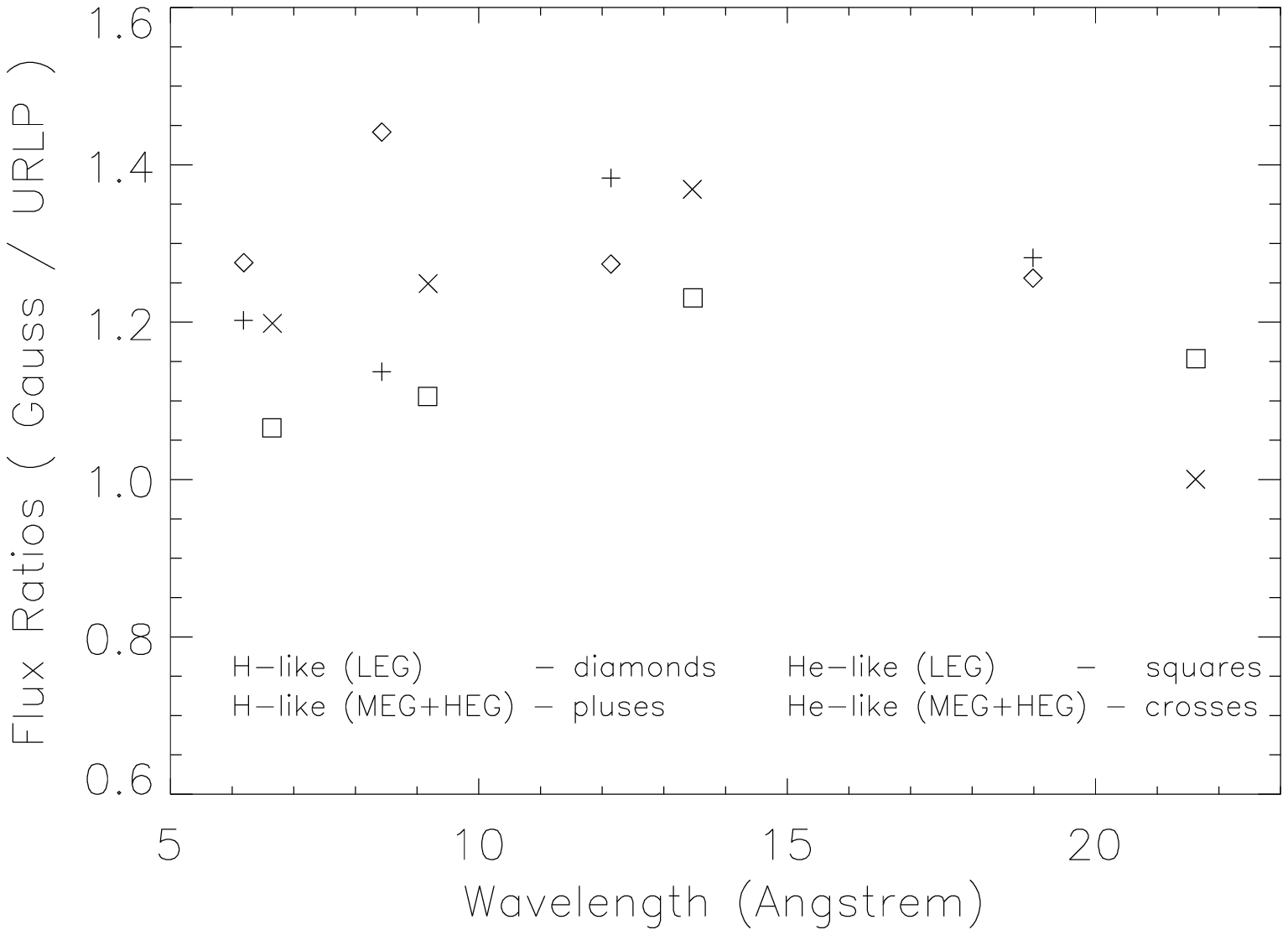}
\includegraphics[width=3in, height=2.25in]{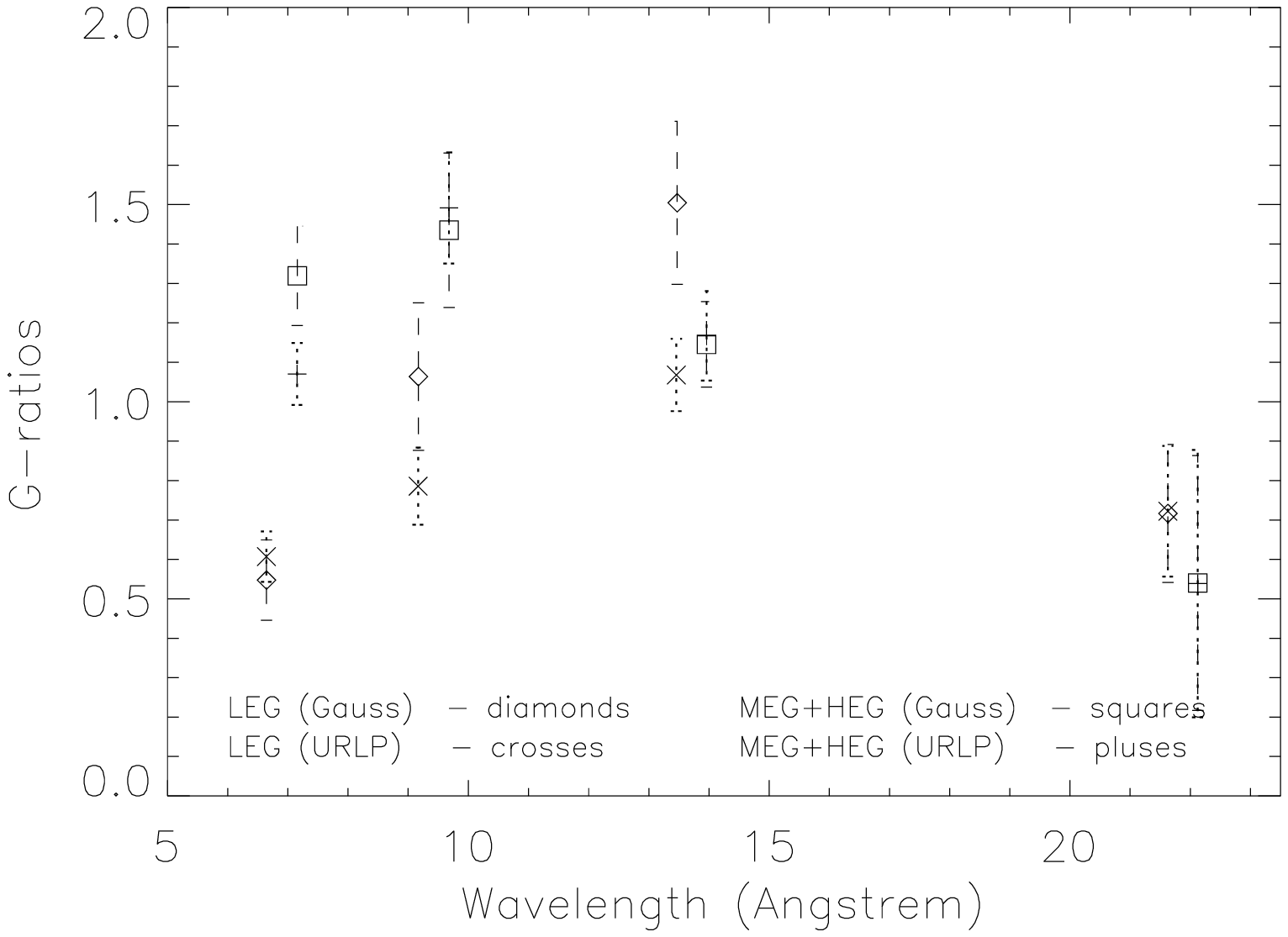}
\includegraphics[width=3in, height=2.25in]{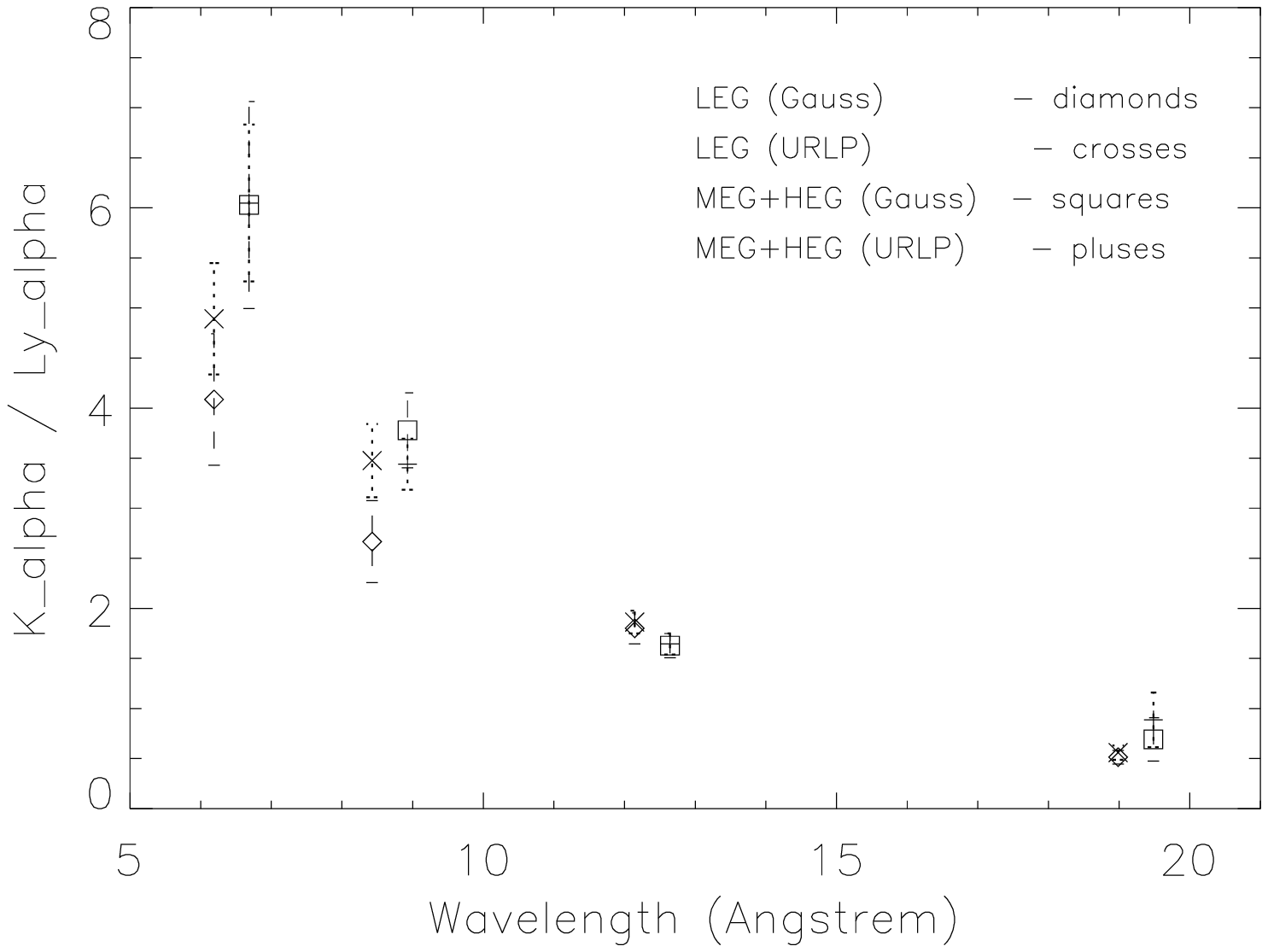}
\includegraphics[width=3in, height=2.25in]{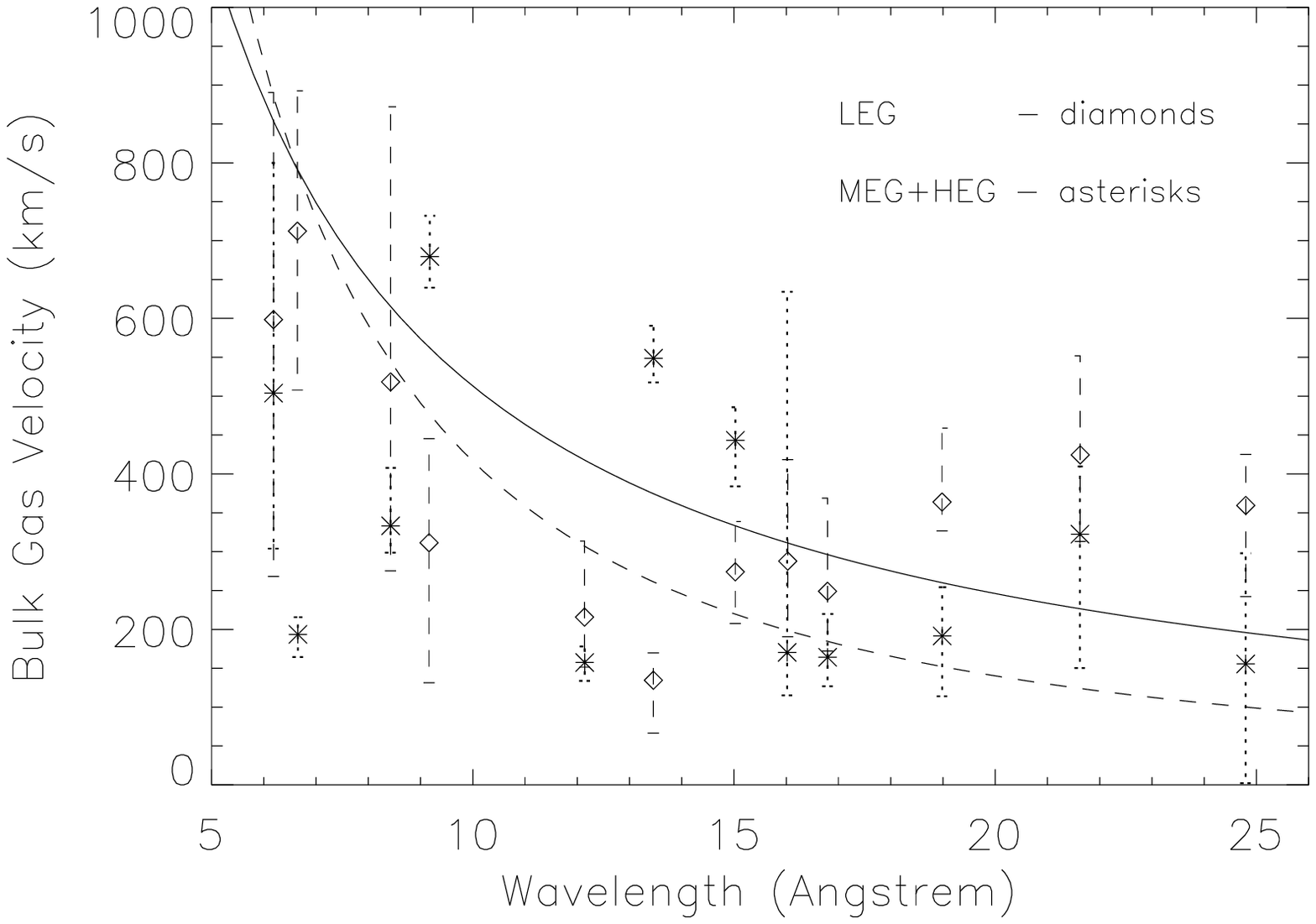}
\end{center}
\caption{
Comparisons between results from the Gaussian and URLP fits to the
line profiles of the `major' emission lines in the 2007 X-ray spectrum
of
\snr
(Si XIV 6.18\AA; Si XIII 6.65\AA; Mg XII 8.42\AA; Mg XI 9.17\AA;
Ne X 12.13\AA; Ne IX 13.45\AA; O VIII Ly$\alpha$ 18.97\AA; O VII
21.60\AA).
{\it Upper left:} flux ratios.
{\it Upper right:} the G-ratios (Si XIII; Mg XI; Ne IX; O VII);
the data for the MEG$+$HEG are shifted by 0.5\AA~for clarity.
{\it Lower left:} the K$_{\alpha}$/Ly$_{\alpha}$ ratios
([Si XIII/Si XII]; [Mg XI/Mg XII]; [Ne IX/Ne X]; [O VII/O VIII]);
the data for the MEG$+$HEG are shifted by 0.5\AA~for clarity.
{\it Lower right:} the derived bulk gas velocity from the
URLP
fits for all the strong lines in the spectra (additional to the listed
above `major' lines are: Fe XVII doublet 15.01\AA~and 15.26\AA;
O VIII Ly$\beta$ 16.01\AA; Fe XVII triplet 16.78\AA, 17.05\AA~and
17.10\AA; N VII 24.78\AA).
For comparison, the solid and dashed lines show the
shock stratification from  Fig.~\ref{fig:fwhm} (lower right panel)
for LEG 2007 and MEG 2007, respectively, with bulk gas velocities
equal to 3/4 of the shock velocities.
In all these figures the notation MEG$+$HEG means that the MEG
and HEG data for a spectral line were fitted simultaneously requiring
the same line flux, bulk gas velocity etc. This is done for all the
lines
with a wavelength smaller than 15.5\AA. The results for all other
HETG lines are based only on the MEG spectra. Error bars correspond to
the associated $\pm 1\sigma$ errors. No reddening correction has been
applied to the line fluxes.
}
\label{fig:compare}
\end{figure}

Figure~\ref{fig:compare} and Table~\ref{tab:flux} compare some
results from the application of
the two techniques to the grating spectra in 2007 (HETG and LETG).
The basic conclusions are as follows.

\begin{itemize}

\item
The line centroids show no apparent deviations from the redshift
($+286$\kms) of the Large Magellanic Cloud. Thus, there is no
evolution of this parameter between 2004 and 2007.

\item
The line fluxes determined from the Gaussian model are greater
than those determined from the URLP model, typically by 20-25\%.
This is probably due to the fact that the lines do not have wings
in the URLP model.  In fact, the fluxes derived here from the
HETG 2007 data generally bracket the D08 values.
This can be attributed to the fact that D08 also included additional
Gaussian line broadening in their ring profile model making it
intermediate between a pure Gaussian and the URLP model.

\item
In the HETG data, the `G-ratios' [$G=(f + i)/r$, where $f, i,$
and $r$ stand for forbidden, intercombination, and resonance line
fluxes, respectively of the He-like triplets] determined from
the Gaussian and URLP models agree very well.  But with the LETG
data, the two models give noticeably different G-ratios for some
triplets.  We believe that this fact is a consequence of the inferior 
spectral resolution of the LETG data.

\item
On the other hand, the two models agree very well in determining
ratios of K$_{\alpha}$/Ly$_{\alpha}$ ratios both for the HETG and
LETG spectra.  These ratios are determined by well-separated
emission lines for which spectral resolution is not a critical
factor.

\end{itemize}

The bulk gas velocities derived from HETG 2007 data are somewhat
lower than those derived from the LETG 2007 data although for some
lines the opposite is true. Since there are
no drastic changes in the spectra in the interval between these
observations (see also Fig.~\ref{fig:dem}), we doubt that this
difference is real.  Instead, we believe that the lower velocities
determined from the HETG 2007 data are more accurate  due to its
better spectral resolution. This is particularly true for the
He-like triplets at shorter wavelengths.
On the other hand, at longer wavelengths ($\sim \lambda > 18$~\AA)
the LETG results should be preferable due to the better quality
of the LETG spectra.
In general, we believe that taken together the results from the two
different techniques (lower right panel in Fig.~\ref{fig:compare})
bracket the bulk gas velocities in the X-ray emitting
region of \snr in 2007.
This is also supported by our
MARX\footnote{See http://space.mit.edu/CXC/MARX/} simulations.
Namely, we simulated line emission from a ring-like structure with
a uniform bulk gas velocity in the range (200 - 400)\kms at longer
wavelengths (12.13~\AA ~and 18.97~\AA) and in the range (400 -
800)\kms
at short wavelengths (6.18~\AA) for a \snr {\it Chandra} exposure of
300 ksec.  We extracted the MEG and LEG spectra following the
standard analysis procedure. We fitted them with our URLP model and
recovered the bulk gas velocities with an accuracy within 10\%
of the input values.

In the simple shock picture, the shock velocities should exceed
the bulk velocities by a factor 4/3.  But in fact, the post-shock
temperatures implied by the bulk velocities and equation~(\ref{eq:kT})
are far lower than those inferred from the spectral fits (see
\S~\ref{subsec:res}).
As we discuss in \S~\ref{sec:disc}, this emphasizes the importance
of reflected shocks.

\subsection{Line ratios and light curves}

The X-ray emission line spectrum of shock-heated plasmas is
sensitive to ionization age, $n_e t$, as well as
post-shock temperature. If $n_e t < 10^{13}$~cm$^{-3}$~s, the
shocked plasma will be under-ionized compared to a plasma in coronal
equilibrium at the same temperature.  Moreover, inner-shell
excitation and ionization boost the emission of the forbidden line
in the He-like triplets. These effects are discussed in detail by
\citet{L99}, \citet{M99}, and \citet{Bor01}.

Following Z05, we analyzed the K$_{\alpha}$ and Ly$_{\alpha}$
spectral lines of the H-like and He-like species of O, Ne, Mg and
Si to determine those regions in the parameter space of electron
temperature and ionization age for which the observed line ratios
correspond to the values that would result in a plane-parallel
shock.  As in our analysis of the LETG
2004 observations, we see again that no single combination of
electron temperature and ionization age is consistent with all the
observed ratios. Therefore, the new data confirm that the X-ray
emission of \snr comes from a distribution of shocks having a
range of ionization ages and postshock temperatures.
We also note the overall similarity between the LETG and HETG
results for the line ratios of O, Ne and Mg, and the apparent
differences for the G-ratios of the silicon He-like triplet
This emphasizes again the importance of spectral resolution
for deriving such quantities, e.g., the Si lines are well resolved
in MEG (also in HEG) while they are not resolved in LEG
(Fig.~\ref{fig:lines}).

\begin{figure}[ht]
\begin{center}
\includegraphics[width=6in, height=5.in]{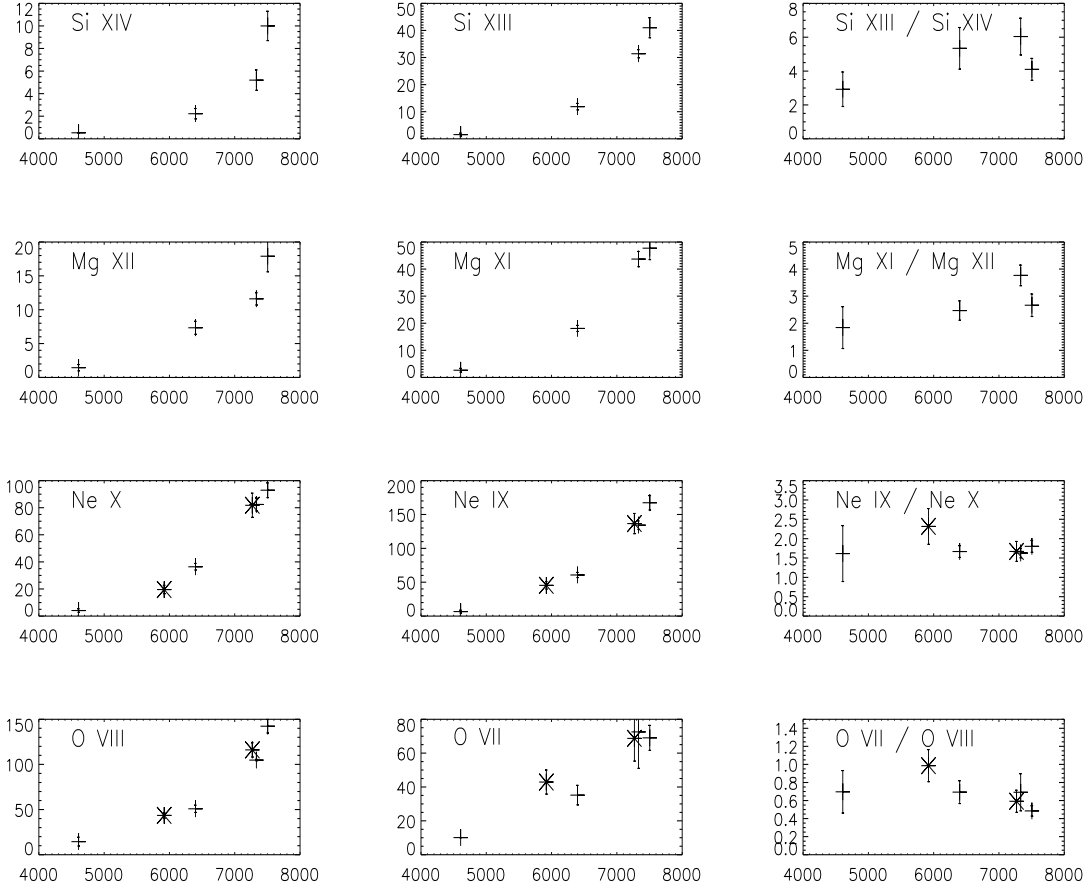}
\end{center}
\caption{
Light curves of some strong spectral lines in \snr.
Plots in each row show consecutively the H-like Ly$_{\alpha}$,
He-like K$_{\alpha}$ and their ratio. Horizontal axes are the days
after explosion; vertical axes (for the plots in the first and second
columns) are observed line fluxes ($10^{-6}$ photons cm$^{-2}$
s$^{-1}$). The data points shown
by pluses (and error bars) are for the {\it Chandra} gratings
observations and are taken from Michael et al. (2002;  for day
4608), Z05 (for day 6398) and Table~\ref{tab:flux}. The data
points shown with asterisks (and error bars) are for the
{\it XMM-Newton} gratings observations from Haberl et al. (2006; for
day 5920) and Heng et al (2008; for day 7269).
}
\label{fig:lc}
\end{figure}

Up to now, we have acquired dispersed spectra of \snr with {\it
Chandra} spanning four epochs (Michael et al. 2002; Z05, Z06; D08
and this work), while two additional dispersed spectra have been
acquired with {\it XMM-Newton} \citep{Ha06, He08}.
Figure~\ref{fig:lc} displays light curves of prominent Ly$_\alpha$
and K$_\alpha$ emission lines from these sources and, in the third
column, the ratio of K$_\alpha$/Ly$_\alpha$.\footnote{Spectral
lines with wavelength shorter than $\sim 12$~\AA~ were not discussed
by \citet{Ha06} and \citet{He08}. The 2007 {\it XMM-Newton} line
fluxes from \citet{He08} were corrected by a factor of $0.461$.
This correction factor is based on a private communication between
D.Dewey, R.McCray, F.Haberl and K.Heng.}

As seen in Fig.~\ref{fig:lc}, the light curves of the spectral
line fluxes are very similar to the light curve of the total X-ray
flux (0.5 - 2 keV) of \snr \citep{P06, P07}.  Namely, they show a
considerable acceleration in brightness during the past few years.
A slightly different behavior is evident
in the light curves of the line ratios (third column
in Fig.~\ref{fig:lc}). We see a gradual increase of the ratio of
K$_\alpha$/Ly$_\alpha$ for the He- and H- like ions of Si and Mg,
while the corresponding ratios remain approximately constant for
Ne and O.  Keeping in mind that the lines of Si and Mg are emitted
from hotter plasmas than those of Ne and O, we note that such a
behavior is consistent with our  findings from the global fits to
the X-ray spectra of \snr (see section \S~\ref{sec:global_fits}).
That is, the average temperature in the high-temperature tail of the
shock distribution is decreasing with time while that of the cooler
plasma remains approximately constant.  The same trend is seen in
fits to the pulse-height spectra obtained from the imaging
observations \citep{P06}.

\section{Two-dimensional results}
\label{sec:2D}
Since \snr is spatially resolved by {\it Chandra}, it is
possible to extract spatially resolved spectral information from
the grating observations carried out so far. We recall that for
each of our observations the instrument configuration was chosen to
maximize the outcome from the spatial-spectral effects. Thus,
the actual roll angles put the dispersion direction closely along
North-South on the sky, and the cross-dispersion direction in
East-West. Given that the spatial-spectral effects
influence the spectra (emission line profiles) mostly in the
North-South direction (along the dispersion axis), it is then
feasible to extract and analyze the X-ray spectra of the East and
West halves of \snr.

The X-ray spectra from East and West halves can be distinguished by
making use of column `tg-d' in the event file of a grating
obsevation. For the data sets in 2007 (MEG and LEG), we
used this parameter to obtain event files of the East and West
parts for each observation in that data set. Then, we extracted
the positive and negative first-order spectra
and combined them into a total East and
West spectra for the data set at hand.
For consistency, we applied the same procedure
to the data obtained in 2004 (LEG) although their
quality (photon statistics) is lower than that of the 2007 data.

\begin{figure}[ht]
\begin{center}
\includegraphics[width=3in, height=2.5in]{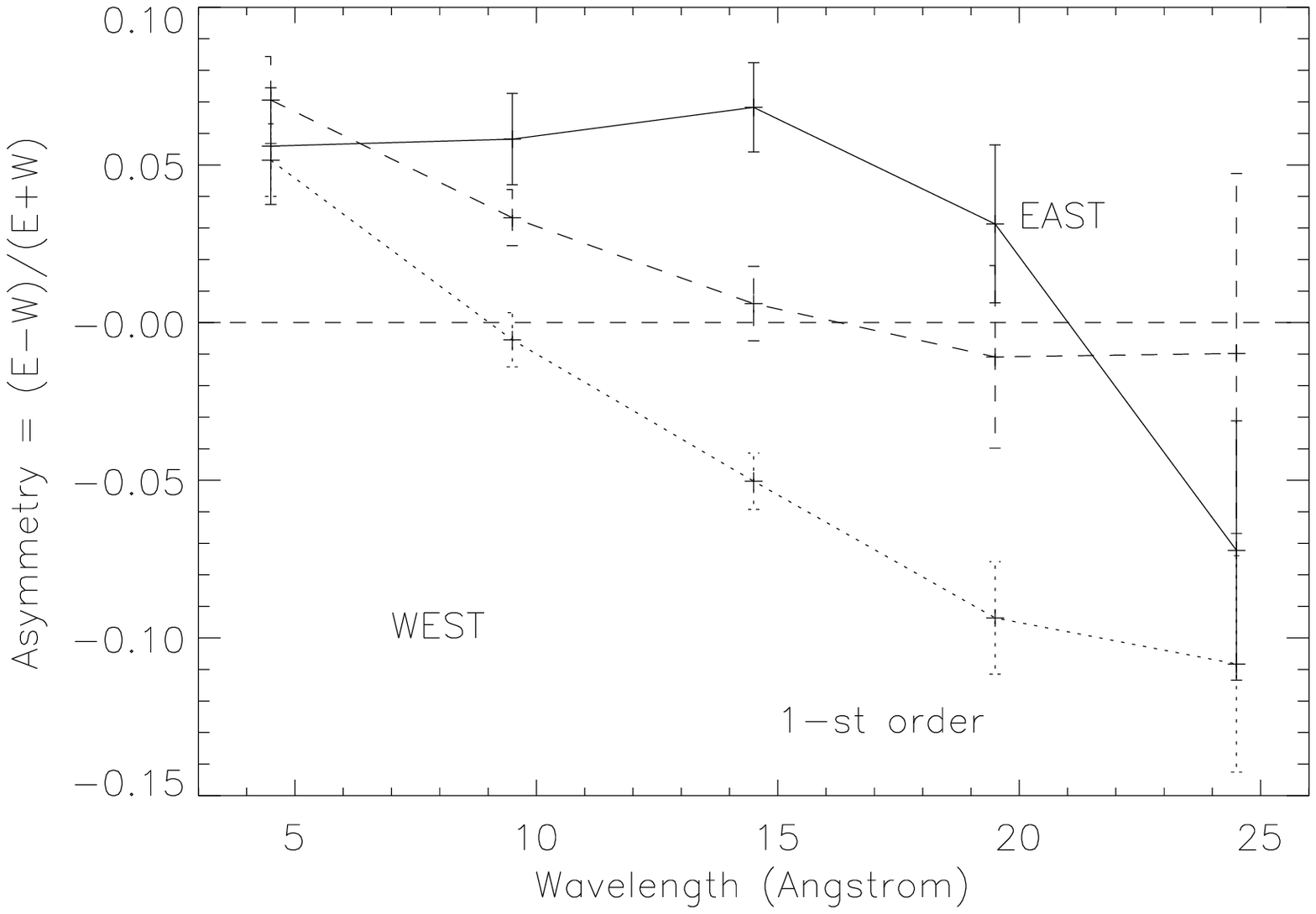}
\includegraphics[width=3in, height=2.5in]{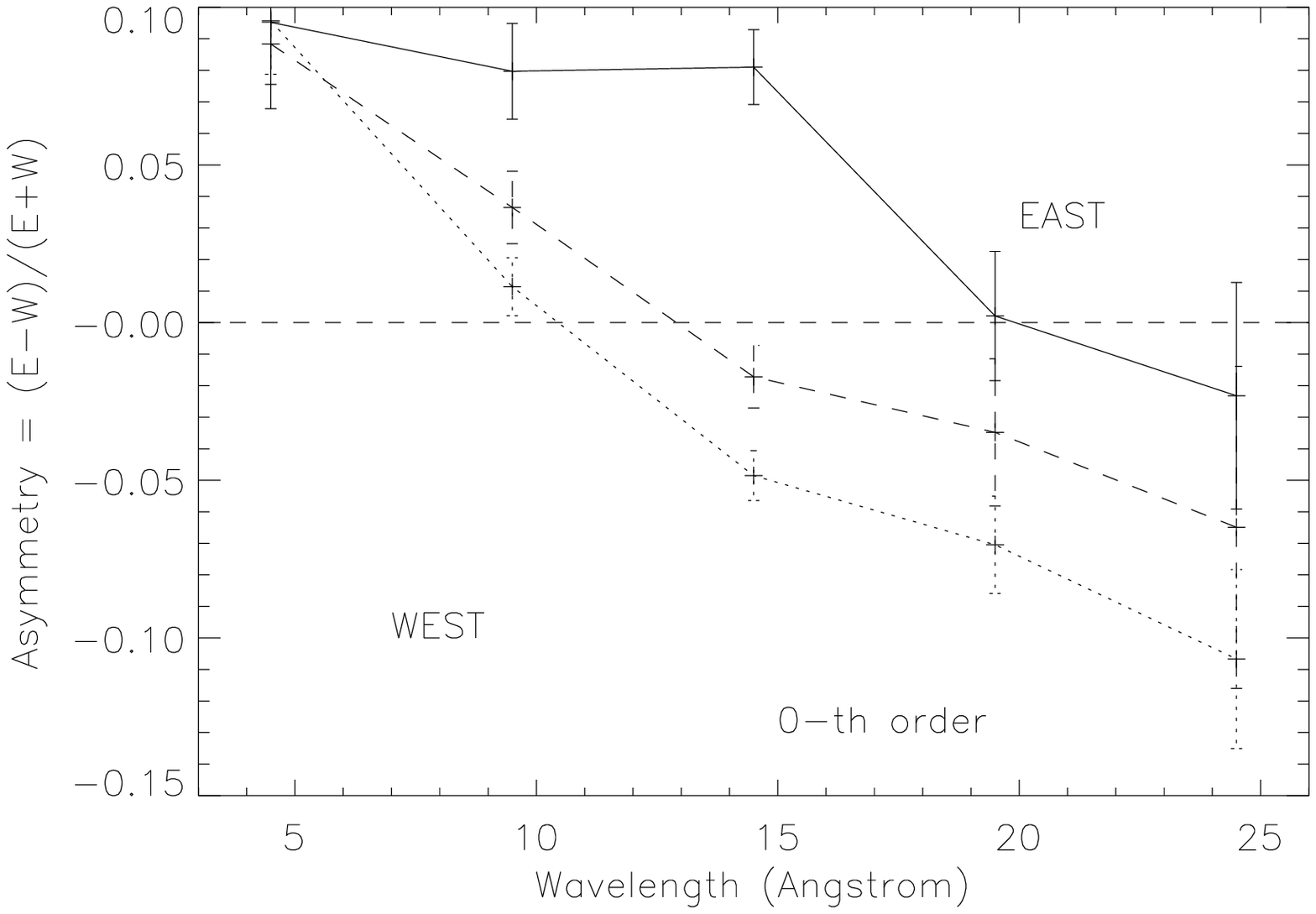}
\end{center}
\caption{East-West asymmetry in the grating spectra of \snr
measured in the wavelength (energy) intervals: (2-7)\AA,
(7-12)\AA, (12-17)\AA, (17-22)\AA~ and (22-27)\AA.
Results for LEG 2004, MEG 2007 and LEG 2007 are shown with solid,
dashed and dotted lines, respectively.
Results for the first-order spectra are shown in the left panel
and those for the zeroth-order spectra are in the right panel.
}
\label{fig:2dasym}
\end{figure}

To characterize the East/West asymmetry, we constructed the
parameter $A = (E-W)/(E+W)$, where $E$~ and $W$~ represent the
number of X-ray counts in the East and West spectra,
respectively. We split the entire spectrum in five wavelength
(energy) intervals and measured the asymmetry in each of them.
Figure~\ref{fig:2dasym} shows the results from this procedure.
Similarly, we measured the East/West asymmetry for the
zeroth-order spectra and we display the results in the right panel
of Fig.~\ref{fig:2dasym}.  A well established
trend  with wavelength (energy) is evident.
We emphasize that the actual values for the asymmetry can be affected
by the {\it Chandra} astrometry but this is not the case for the
observed trend. We therefore conclude that the
X-ray emission from the Eastern half dominates at short
wavelengths (high energies) and the opposite is true at long
wavelengths (low energies).

In general, high energy photons come predominantly from hotter
plasma (behind faster shocks) while  low energy ones are sign for
relatively cooler plasma (behind slower shocks). In other words,
the established East-West
asymmetry signifies an asymmetry in the shock distribution in \snr.
To check this, we fitted the East and West spectra for each data set
with the two-shock model discussed in \S~\ref{sec:global_fits}.
We held the X-ray absorption and abundances fixed to their best
values from our simultaneous global fits to the three data sets
under consideration (LEG 2004, MEG 2007, LEG 2007). The results
from the fits to the East and West spectra are:
(i) the emission measure corresponding to the faster shock is the
same for the
East and West parts in 2004 while the East half became dominant in
2007 (by 24\% in MEG; and by 10\% in LEG);
(ii) the emission measure in the slow shock evolved between 2004
and 2007 as the East part was dominant in 2004 (by 20\%) and
in 2007 this was true for the West half (by 37\% in MEG; and by
25\% in LEG);
(iii) the plasma temperatures for the cooler (slow shock) and
hotter (fast shock) components in East and West are within the
$1\sigma$~range of the values derived from the global fits (see
Table~\ref{tab:flux}).

Keeping in mind the uncertainties from the {\it Chandra}
astrometry, we also note that the East-West spectral extraction
are not strictly aligned between the three data sets since their
effective Nort-South directions fall within 15$^{\circ}$ ~from each
other.  Nevertheless, the two-dimensional results reported above makes
us confident to conclude that the X-ray emitting region in \snr does
show asymmetric shock distribution with its East part being richer
in fast shocks and its West part being richer in slow shocks.
This conclusion is consistent with the trend found
by \citet{P04},
who reported differential flux changes of the slow shock component
in \snr between 2000 and 2002.
On the other hand, the infrared images of the inner ring in \snr 
show that its Eastern half dominates the emission in this spectral
range \citep{Bou04, Bou06}. Since 
this is thermal emission from dust which is collisionally heated by the
hot gas \citep{Bou04, Bou06, Dwek08}, its spatial distribution traces 
that of the X-ray emitting plasma.
It is then conclusive that the fast shocks 
may also play an important role in heating the dust in the inner ring
which reemphasizes the importance of a more elaborated modeling of the
\snr emission in different spectral domains.

\section{Discussion} 
\label{sec:disc} 

The main results from our analysis of the new {\it Chandra}
grating spectra of \snr are: (1) it is possible to match the
overall X-ray spectrum in the framework of a simple shock picture;
but (2) the bulk gas velocities inferred from the line profiles
are much  far less than the velocity that we would expect of gas
shocked to the plasma temperature inferred from the spectra.  

A number of possible effects might account for a lower
difference between the widths of the broadened and narrowed line
profiles, which would result in measurements of effective bulk gas
velocities that may be less than the actual velocities.  For
example: (i) an emission line might be contaminated by weak lines
due to other species; (ii) the shock front might extend to a
greater latitude than the 10 degrees assumed in the present model.
Such a geometry will give a lower line width for the same bulk
velocity;  (iii) the line emissivity is not uniformly distributed 
with the azimuthal angle.  Moreover, a general caveat regarding our
simple picture is that we assume a single bulk gas velocity for
each line while in reality the lines likely form in multi-velocity
gas flows.

But, as
we have suggested (Z05 and D08), we think that the most likely
explanation for
the discrepancy is that most of the X-ray emitting gas has been
shocked
twice: first by the blast wave as it overtakes stationary gas; and
second, by a reflected shock that appears when the blast wave
suddenly encounters dense gas in the circumstellar ring.  The 
reflected shock will decelerate the gas that has been shocked by 
the blast wave and further raise its temperature and density.
If this is the
case, there will also be a transmitted shock that enters the dense
gas. The schematic of such a reflected shock structure (RSS) is 
shown in Fig.~\ref{fig:cartoon}.  
Possibly, the bimodal temperature 
distribution that we see
in the X-ray emitting gas results from a superposition of
high-temperature emission from the twice-shocked gas and
low-temperature emission from denser gas behind the transmitted
shock.  

In order to explore this possibility, we have constructed a
plane-parallel model for NIE X-ray emission from a blast wave, a
transmitted and a reflected shock.  Details of the model and its
implementation in XSPEC are given in appendix~\ref{app:sss}.  We
considered two basic cases for modeling the line profiles in the
resultant X-ray spectrum: (1) one that corresponds to
eq.~(\ref{eqn:stratified}) (which was originally adopted in Z06
and also used for our two-shock model in
\S~\ref{sec:global_fits}); and (2) a model in which the line
profiles correspond to the bulk gas velocities in the reflected
shock structure convolved with the spatial broadening due to
the size of the inner ring (the spatial-spectral effects for a
ring with an inclination angle $45^{\circ}$ and a small opening
angle $\beta = 10^{\circ}$~ are explicitly taken into 
account as described in appendix~\ref{app:urlp}).

The quality of the RSS model fits with a Gaussian line profile
(following
eq.~[\ref{eqn:stratified}]) is as good as in our two-shock model
and the derived abundances fall within the confidence limits given
in Table~\ref{tab:fit}. Therefore, in our actual fits we fixed the
abundances to their `best' values from the two-shock model in
order to decrease the number of free parameters. The simultaneous
fit ($\chi^2/dof=1834/2258$) to the entire data set (LEG 2004, MEG
2007 and LEG 2007) gave a column density of $N_H = 1.44\pm0.05
\times 10^{21}$~cm$^{-2}$~and the following parameters for
`average' RSS (transmitted shock, TS; direct shock, `blast' wave,
DS; reflected shock, RS): 

$kT_{TS} = 0.58\pm0.02$, $kT_{DS} = 2.43\pm0.18$, $kT_{RS} =
3.78\pm0.28$ (LEG 2004);

$kT_{TS} = 0.60\pm0.01$, $kT_{DS} = 2.06\pm0.09$, $kT_{RS} =
3.07\pm0.13$ (MEG 2007);

$kT_{TS} = 0.59\pm0.01$, $kT_{DS} = 1.74\pm0.07$, $kT_{RS} =
2.50\pm0.10$ (LEG 2007).

\noindent
Temperatures are in keV and the errors are $1\sigma$ values from
the fit. Emission measures in each of the shocks, relative to that
of the direct shock, in the same time sequence are:

$EM_{TS} = 4.7$, $EM_{DS} = 1.0$, $EM_{RS} = 0.4$ (LEG 2004);

$EM_{TS} = 4.6$, $EM_{DS} = 1.0$, $EM_{RS} = 0.6$ (MEG 2007);

$EM_{TS} = 6.5$, $EM_{DS} = 1.0$, $EM_{RS} = 1.2$ (MEG 2007).

\noindent
Thus, we see again that the temperature of the X-ray emitting
plasma is on the average decreasing with time, while an increasing
amount of plasma is affected by the reflected shocks.

On the other hand, the RSS model in conjunction with the ring
kinematics gives a best fit having much lower quality
($\chi^2/dof=2708/2260$). The model line profiles are narrower
than the observed profiles. We are not surprised: our
plane-parallel model is surely an oversimplification.  As we
have already shown (D08), we can fit the observed profiles of
individual lines in the HETG spectrum if we introduce a local
Gaussian Doppler broadening in addition to the term representing
the bulk velocity of the shocked gas. The complex hydrodynamics
(including normal and oblique shocks) that results from the
interaction of the blast wave with dense clumps in the ring
probably accounts for this extra broadening.  Of course we can
improve the quality of the fit to the RSS model if we introduce
additional variable parameters.  For example, if we add a Gaussian
broadening term with the same velocity dispersion for the entire
spectrum, the quality of fit improves to $\chi^2/dof=1960/2257$,
and the X-ray absorption and plasma parameters (shock temperatures
and abundances) reach values similar to those discussed above.  We
regard the addition of such a Gaussian broadening term as an {\it
ad hoc} way to allow for the complexity of the actual gas dynamics.

The parameters of the RSS model fit can be used to infer the
typical density contrast, $n_c/n_0$, between the clumps and their
surroundings (the smooth interclump medium)
in the inner ring of \snr (see
appendix~\ref{app:sss}). Namely, the deduced values are: $n_c/n_0
= 11.7, 8.8, {\rm and~} 6.9$~for LEG 2004, MEG 2007 and LEG 2007,
respectively. We believe that such a gradual decrease of the
inferred density contrast suggests the following picture: as the
blast wave penetrates deeper into the environment of the inner
circumstellar ring, it encounters interclump gas of steadily increasing
density and slows down while the clumps have approximately the same
high density.
The X-ray emission is increasingly
dominated by slow shocks that are transmitted into embedded dense
clumps, while shocks reflected from these clumps further compress
and heat the gas that was already shocked once by the blast wave.
Such a picture is in qualitative accord with the observed
temperature evolution of the X-ray emitting plasma (see
\S~\ref{sec:global_fits}).  It is also in accord with the
decelerating radial expansion of \snr as deduced from the X-ray
images (see Fig.2 in Park et al. 2007) and from the analysis of
the X-ray light curve (e.g. Park et al. 2006).

We emphasize that the X-ray spectra and images manifest
only part of the complex hydrodynamic interaction of the blast
wave with the inner circumstellar ring. The similar morphologies
of the optical and soft X-ray images (e.g., McCray 2007) suggest
that the shocks responsible for the optical hotspots are related
to those responsible for the X-ray emission.  However, they are
not the same shocks.  In their analysis of the optical spectrum of
Spot 1, \citet{Pun02} showed that the optical emission comes from 
regions of high densities ($n_e \sim 10^6$~cm$^{-3}$), which must
result from radiative shocks that can compress the preshock gas by
a factor $\geq 100$.  Such radiative cooling cannot take place in
the hot plasma seen in the X-ray spectra.  Taking the standard
cooling function of optically-thin plasma (Raymond, Cox \& Smith
1976) for plasma temperatures $10^5 \leq T \leq 4\times10^7$~K,
we estimate that the radiative cooling time is $t_{cool} \approx
270 n^{-1}_4 V^{3.2}_{1000}$~yr, where $n_4$ is the preshock
nucleon number density in units of 10$^4$~cm$^{-3}$, and
$V_{1000}$ is the shock velocity in 1000\kms - i.e., much
greater than the time since the interaction began.  Including
effects of non-equilibrium ionization may decrease this estimate
but not by an order of magnitude.

The spatial correlation of the relatively cool ($T \sim 10^4$ K)
optical spots and the hotter ($T \geq 10^6$ K) X-ray emitting
regions indicates that much denser clumps may co-exist with less
dense ones.  The hydrodynamics of such interaction may be quite
complex.  The optical emission comes from shocks transmitted into
the dense clumps by the impact of the blast wave, while the X-ray
emission can come both from shocks transmitted into less dense
clumps, or shocks reflected off the optically emitting clumps.
Alternatively, when the blast wave overtakes a clump of
intermediate density, the transmitted shock resulting from
encounter of the blast wave at nearly normal incidence may be fast
enough to heat the plasma to X-ray emitting temperatures, while
the transmitted shock resulting from passage of the blast wave
around the sides of the clump may be slow enough to undergo
radiative cooling and collapse, resulting in the observed optical
emission (see \S 4.1 in Pun et al. 2002; also Borkowski et al.
1997).  Clearly, we need continued multiwavelength spectroscopic
observations to sort out the real physical picture of the shock
interaction in \snr.

\section{Summary}
\label{sec:conclusions}

In this work, we presented spectral analysis of the entire data
set (LETG and HETG) for \snr obtained with {\it Chandra} in 2007.
The basic results and conclusions are as follows.

\begin{itemize}

\item
  Global fits of shock models to the spectra of \snr show that
the average temperature of the X-ray emitting plasma behind the
slower shocks has remained nearly stationary for the last few years,
while that behind the faster shocks has gradually decreased.

\item
  Light curves of strong X-ray emission lines and their ratios
confirm that the temperature of the high-temperature component of
the X-ray emitting gas has decreased.

\item
  Profiles of the strong emission lines were modeled in two
different ways.  The results are consistent and show that the
deduced bulk gas velocities of the X-ray emitting plasma are too
low to account for the observed temperatures in a simple shock
model.  This result suggests that much of the X-ray emission seen
in \snr comes from gas that has been shocked twice, first by the
blast wave and second by shocks reflected off dense clumps (Z05;
D08).

\item
  To explore the reflected-shock scenario in a more quantitative
manner, we have developed a new model that assumes a
plane-parallel geometry and takes into account the details of
non-equilibrium ionization. Global fits with this model to the
entire 2004-2007 gratings data set show that the contribution from
reflected shocks to the X-ray emission from \snr increases with time.
Despite its oversimplification, this model illustrates the general
physical picture.  It points to more elaborate and realistic
models that may result from fits to spectral data spanning
evolution of the supernova remnant for several more years.

\item
 The bulk gas velocities derived from the line profiles seen in
the LETG data tend to be larger than those from the HETG data. We
attribute this discrepancy to the inferior spectral resolution of
the LETG compared to the HETG, and therefore believe that the
velocities derived from the HETG data are more accurate,
especially, at shorter wavelengths.
At wavelengths $\sim \lambda > 18$~\AA, the LETG results might
be considered more reliable as result from a better photon statistics
in the LETG spectra.

\item
 Two-dimensional analysis of the X-ray grating observations
indicates an asymmetry in the shock distribution in \snr, namely,
the East half of the remnant is richer in fast shocks and the West
half is richer in slow shocks.

\end{itemize}

\acknowledgments
This work was supported by NASA through Chandra grant GO7-8062X to
the University of Colorado at Boulder, and by contract SV3-73016
to MIT for Support of the Chandra X-ray Center.
The authors appreciate the comments by an anonymous referee.

Facilities: \facility{CXO (HETG, LETG)}

\appendix

\section{Uniform Ring Model for Line Profiles}
\label{app:urlp}

The spatially resolved X-ray images of \snr (e.g., Park et al.
2006) show that the X-ray emission originates in a ring-like
structure. On the other hand, the first grating observations of
this object with good photon statistics (Z05, Z06) clearly showed
that this structure is `flat' (not just a hollow sphere). Thus, it
is natural to assume that the X-ray emission of this object
originates in a ring with parameters similar to those of the
optical inner ring in \snr. 

In our uniform ring model, we assume that the X-ray emitting
region is an equatorial section of a spherical shell with inner
and outer diameters of 1.55\arcsec ~and 1.7\arcsec, respectively,
inclination angle of 45$^{\circ}$ to the observer's line of sight
(l.o.s.), which extends in latitude to an angle $\pm \beta$ that
can be adjusted.  The X-ray emitting
gas is uniformly distributed in the ring and expands radially with
a constant velocity.  Such an extended object is spatially
resolved by {\it Chandra}, so that its spatial extent will be
manifested as equivalent line broadening in the dispersed
spectrum. Also, the blue- and red-shifted Doppler velocities in
different parts of the X-ray emitting region will influence the
emission line profile. These spatial-spectral effects\footnote{For
a general discussion of the spatial-spectral effects see \S~8.5.3
in the {\it Chandra} Proposer's Observatory Guide, ver. 10.}
distort the dispersed line images so that the line profiles will
be stretched in one of the spectrograph's arms and compressed in
the other one.  

In order to take these effects into account, we perform a 3D
integration to simulate the theoretical line profiles.  For each
parcel of gas, we calculate the line-of-sight projection of the
radial velocity (or Doppler shift) and the effective wavelength
shift (spatial shift) that corresponds to the position of this
parcel of gas in the geometrical ring. These two terms determine
the resultant profile of a given spectral line. Note that the
spatial broadening in terms of equivalent Doppler velocities
depends both on the wavelength of the emission line and on the
spectrometer at hand. The dispersion of the {\it Chandra} gratings
are 0.056 \AA/arcsec (LEG), 0.0226 \AA/arcsec (MEG) and 0.0113
\AA/arcsec (HEG); these map the spatial extent of the ring to
quite different equivalent gas velocities if, for example, a
spectral line at 6~\AA~ or 12~\AA~ is observed with each of these
instruments.

Figure~\ref{fig:urlp} illustrates the  effects of various model
parameters on the theoretical line profiles.  As expected, when
the X-ray emission region does not have a `flat' geometry  -- for
example, if the ring is actually a spherical shell ($\beta =
90^\circ$), the difference between the line profiles in the two 
arms of the spectrograph vanishes.

We constructed two custom XSPEC models to simulate the
spatial-spectral effects for \snr.  The first one explicitly
follows the description given above and gives the resultant
profiles.  The parameters are: the ring's inner and outer radii,
its inclination, and equatorial half opening angle, $\beta$; the
central wavelength of the spectral line, its total flux and the
bulk gas velocity. 
The second model, used to fit the global X-ray spectra, assumes
the same geometry but a combination of plane-parallel transmitted
and reflected shocks, as discussed in \S~\ref{sec:disc}.
Fits based on this model require substantially more computing time
in XSPEC than the standard spectral-line broadening models since a
3D integration must be performed for each wavelength bin in the
spectrum.

\section{Reflected Shock Structure} 
\label{app:sss}

We consider a simplified picture of a reflected shock structure
(RSS), in which a blast wave propagates through relatively
rarefied (nucleon number density $n_0$) intercloud gas at constant
velocity $V_0$ and encounters a static dense cloud (density
$n_c$).  After the encounter, a shock is transmitted into the
dense cloud and a reflected shock begins to propagate backwards,
further compressing the shocked gas behind the blast wave.  We
assume plane-parallel geometry, adiabatic shocks and equal
electron and ion postshock temperatures.

\begin{figure}[ht]
\begin{center}
\includegraphics[width=5in, height=4.in]{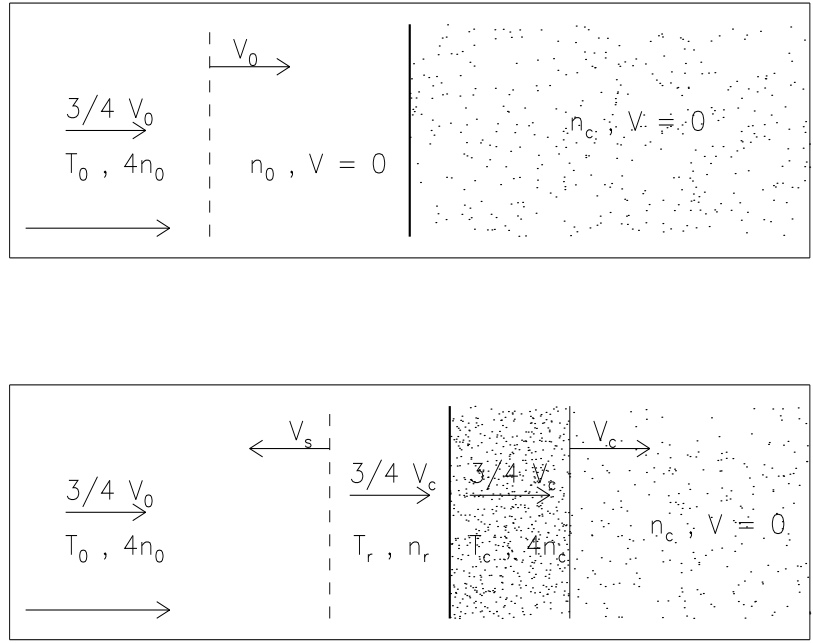}
\end{center}
\caption{
Schematic of the plane-parallel reflected shock structure (RSS).
The post-shock
velocities and densities have their values for strong shocks with
adiabatic index $\gamma = 5/3$.
{\it Upper} panel shows the situation before the interaction of
the `blast' wave (forward shock) with a dense `cloud'.
The dashed vertical line denotes the forward shock and the thick
solid line denotes the edge of the cloud.
{\it Lower} panel depicts the situation after the interaction as
a transmitted shock and a reflected shock have appeared.
Here , positions of the reflected and transmitted shocks are
denoted by a dashed and solid line, respectively, while the thick
solid line shows the position of the contact discontinuity.
The long arrow in the low left corner in each panel indicates the
positive direction of the x-axis.
}
\label{fig:cartoon}
\end{figure}

\begin{figure}[ht]
\begin{center}
\includegraphics[width=2in, height=1.5in]{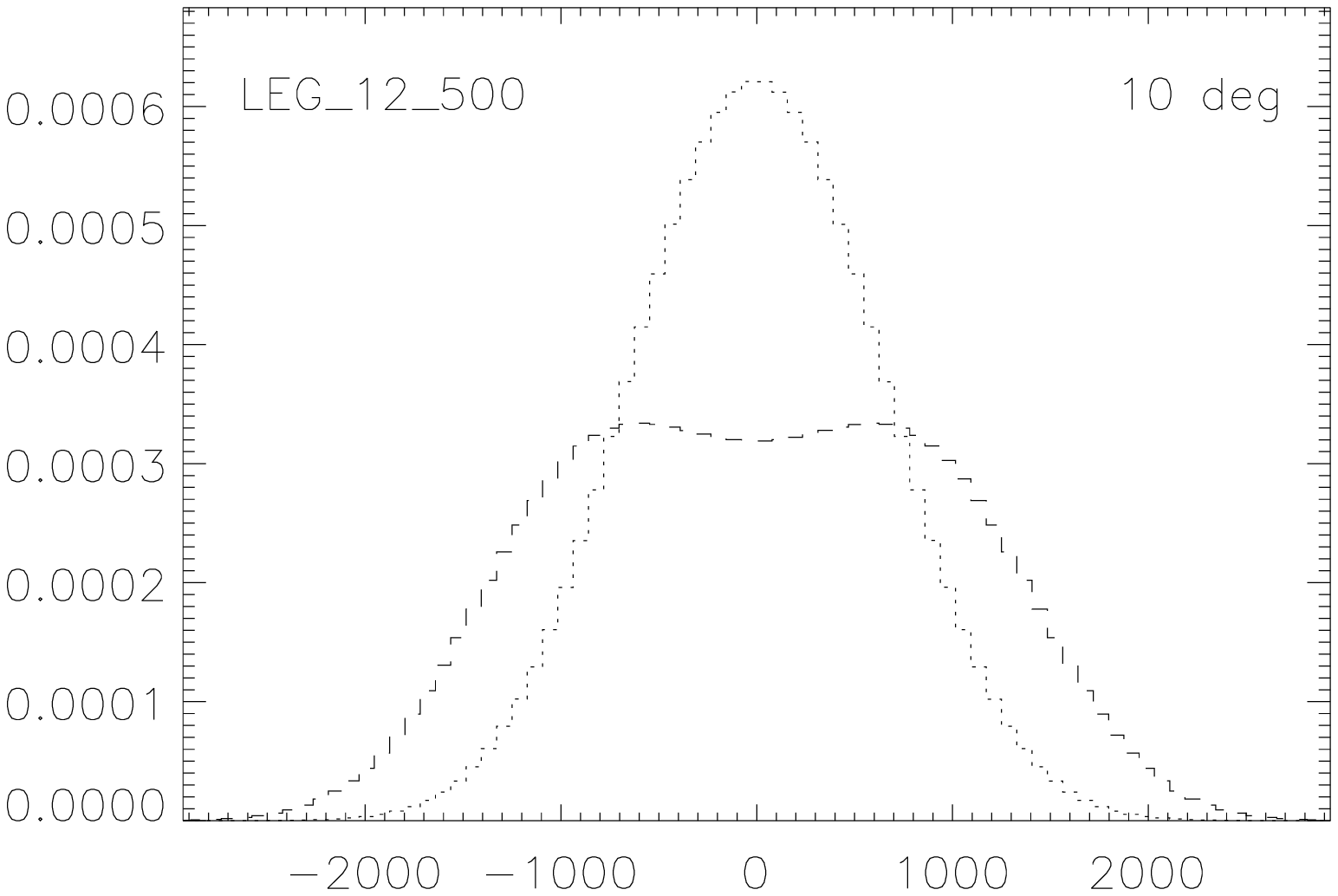}
\includegraphics[width=2in, height=1.5in]{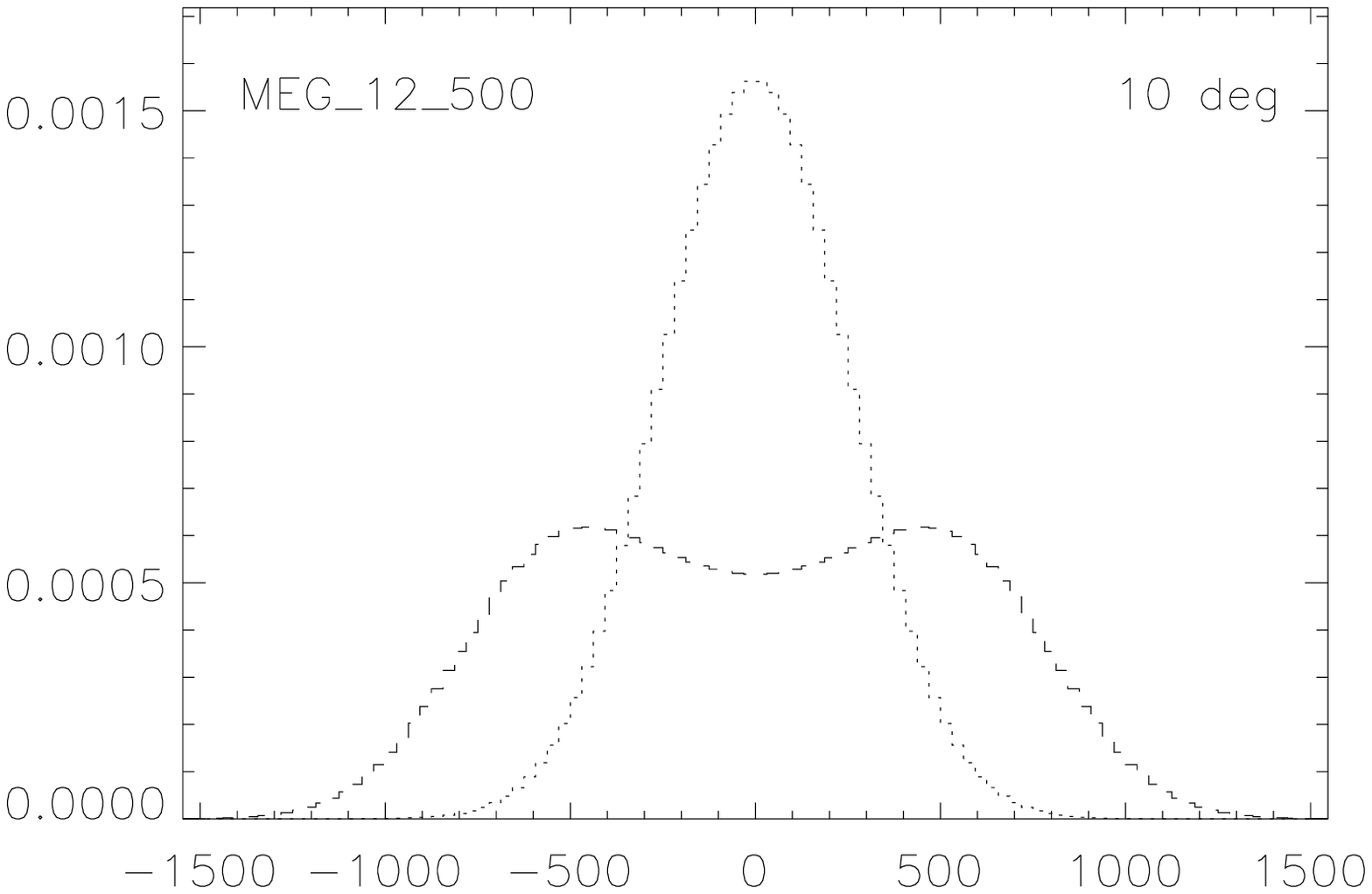}
\includegraphics[width=2in, height=1.5in]{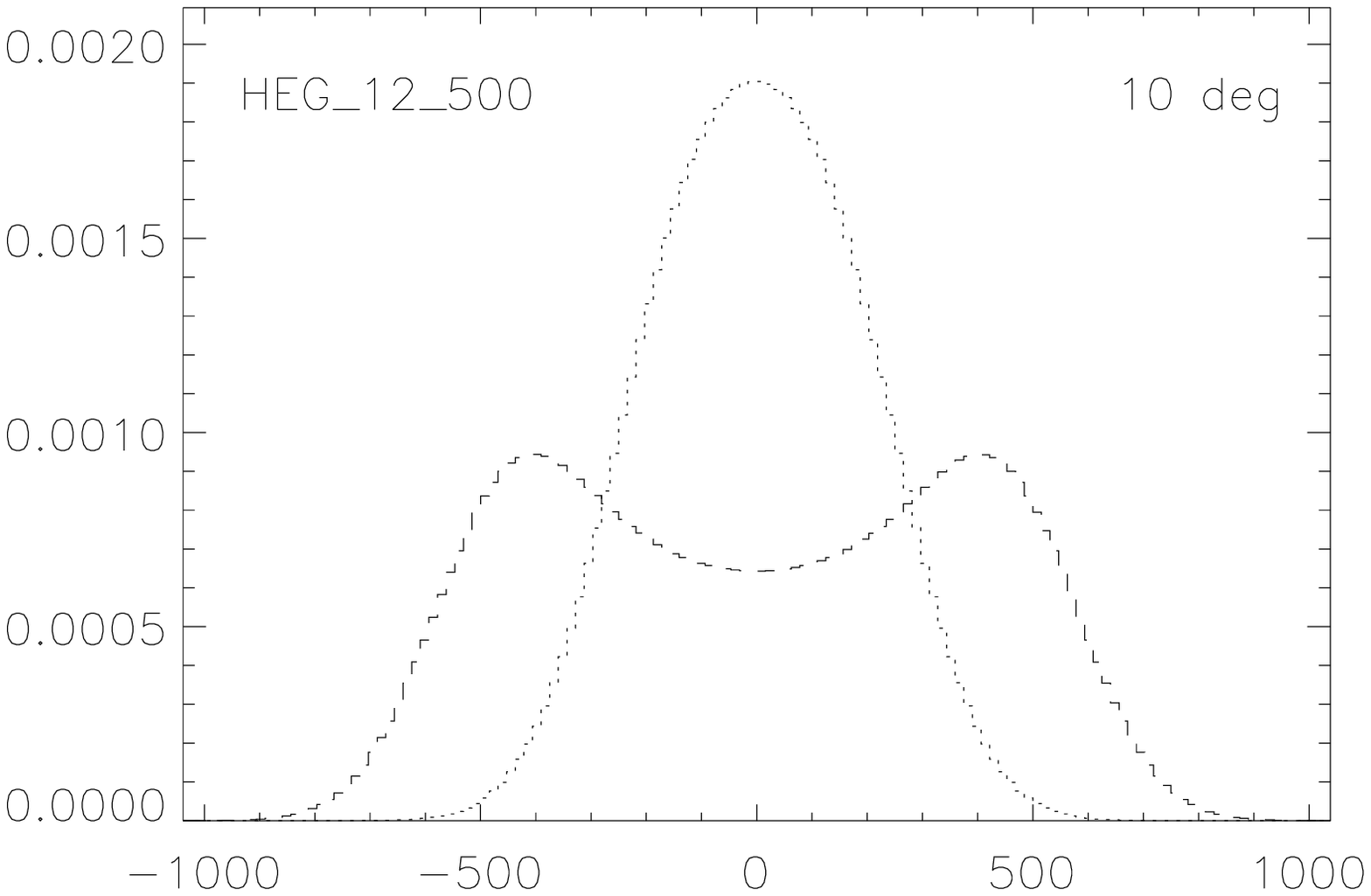}
\includegraphics[width=2in, height=1.5in]{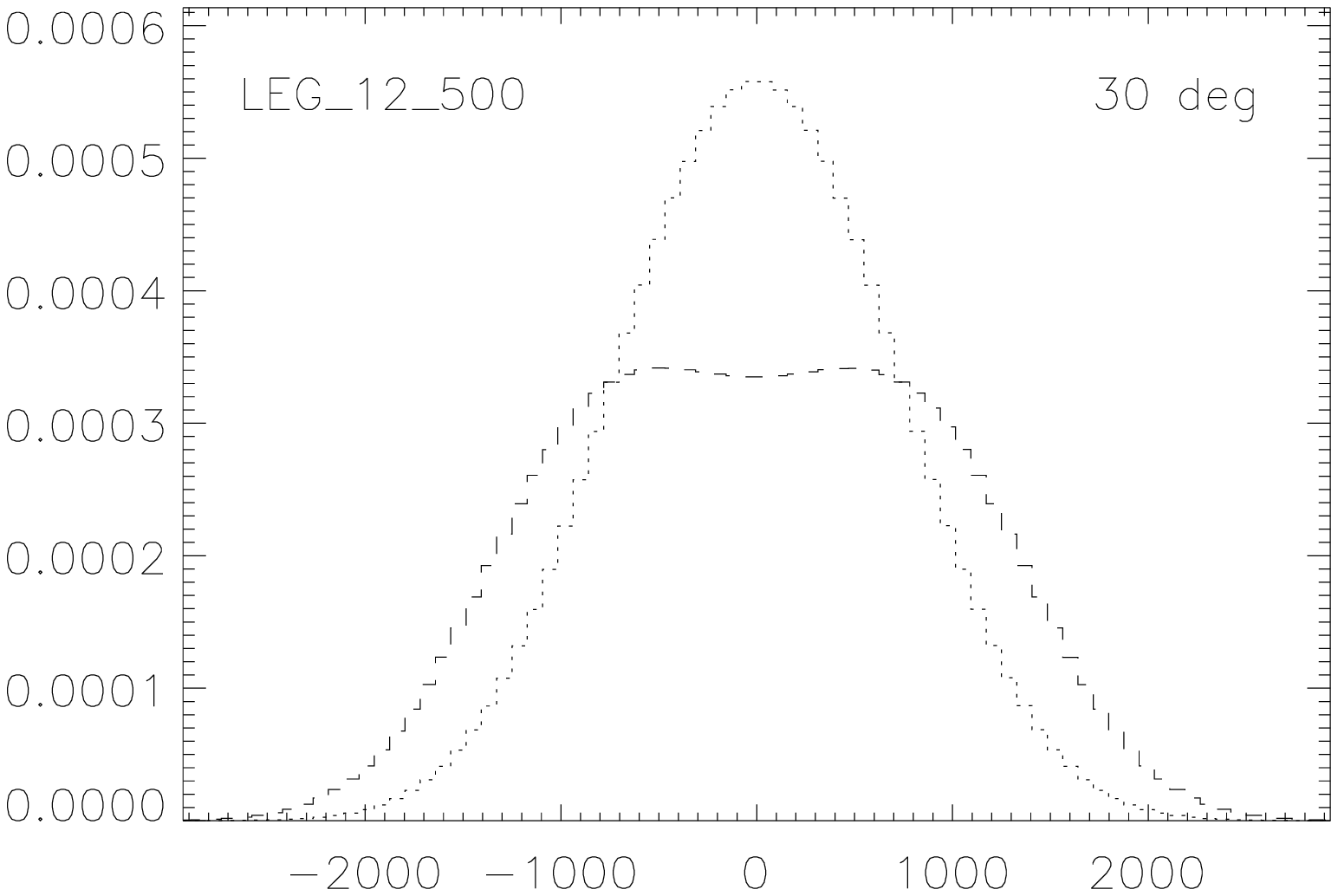}
\includegraphics[width=2in, height=1.5in]{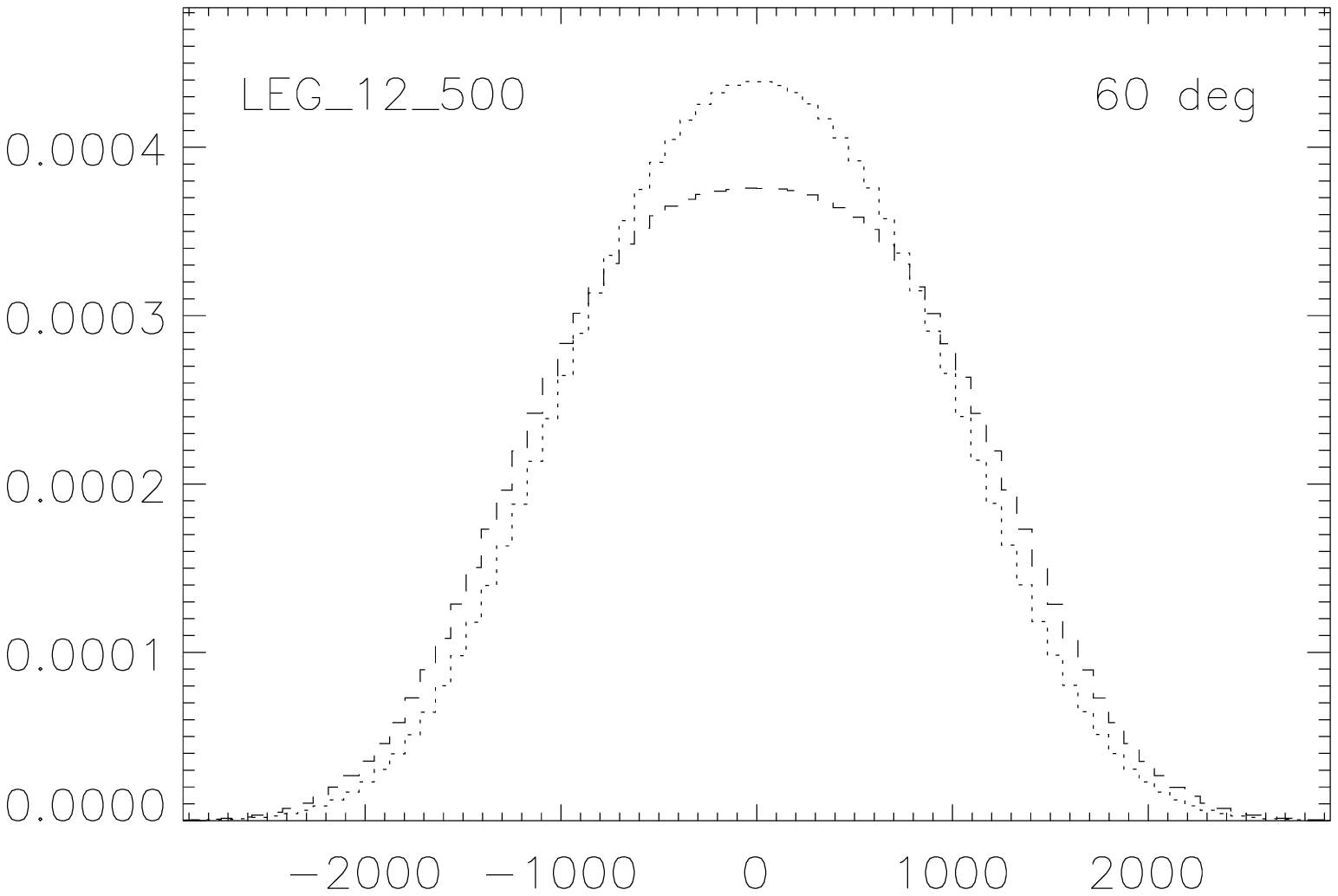}
\includegraphics[width=2in, height=1.5in]{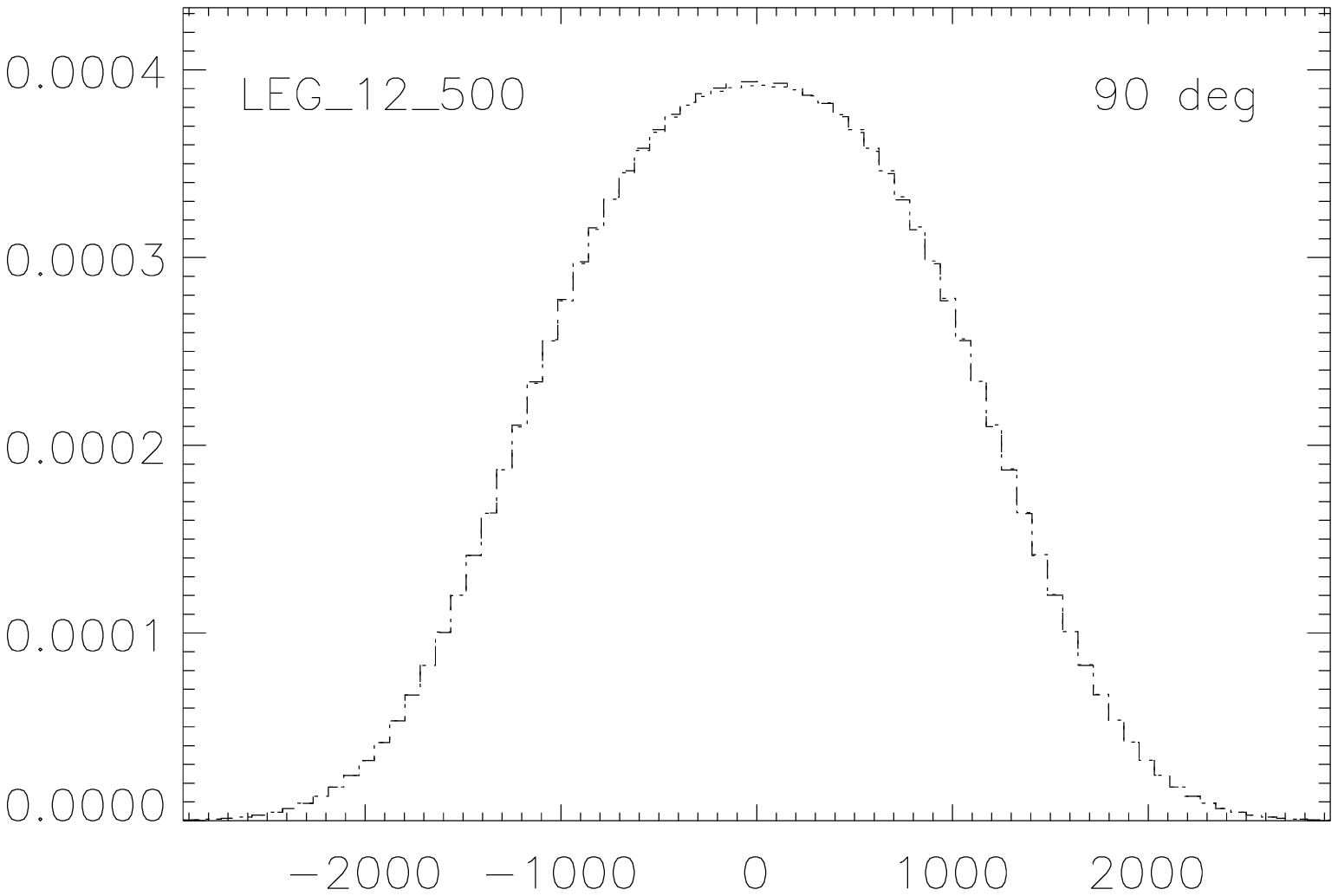}
\includegraphics[width=2in, height=1.5in]{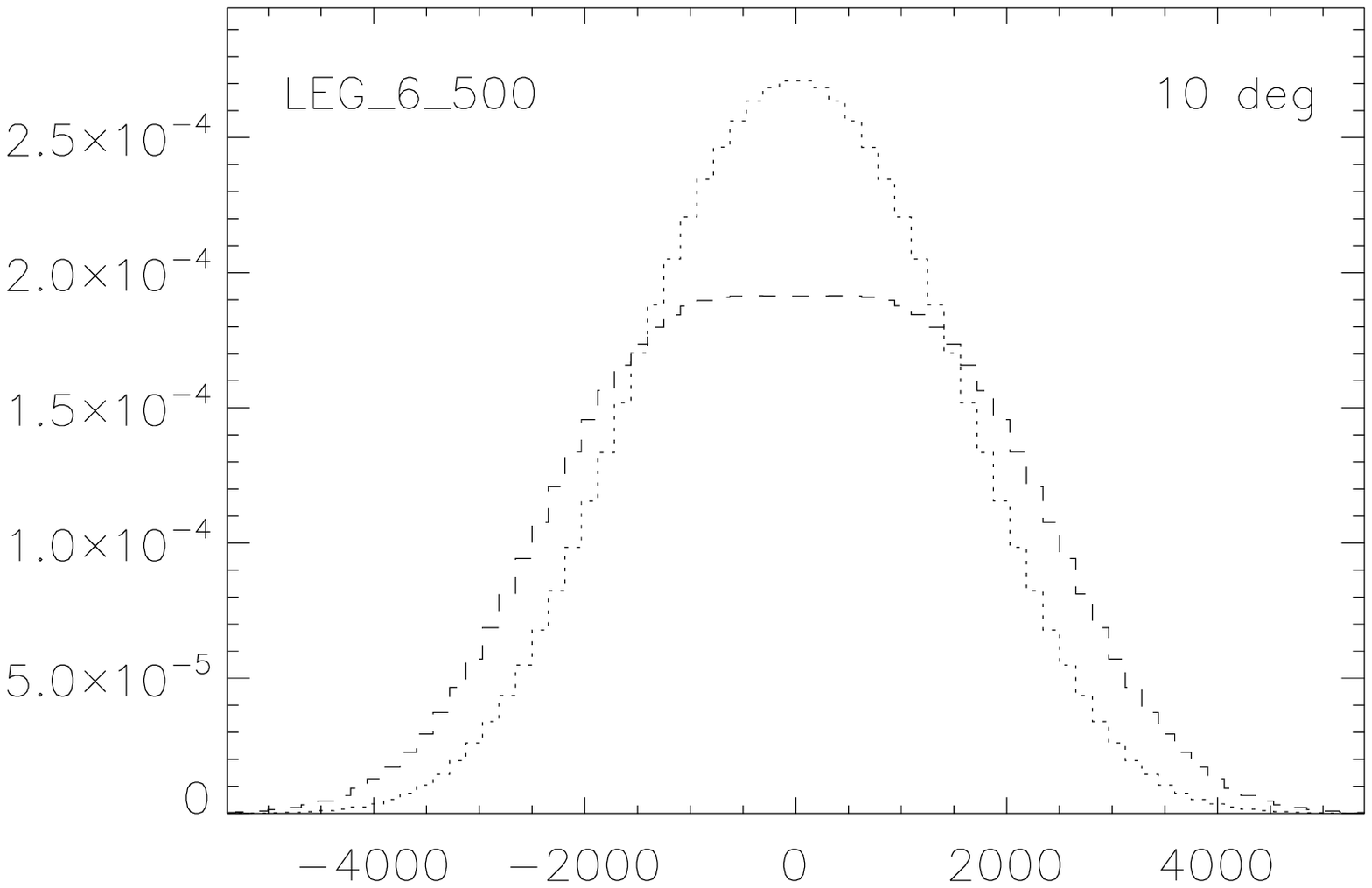}
\includegraphics[width=2in, height=1.5in]{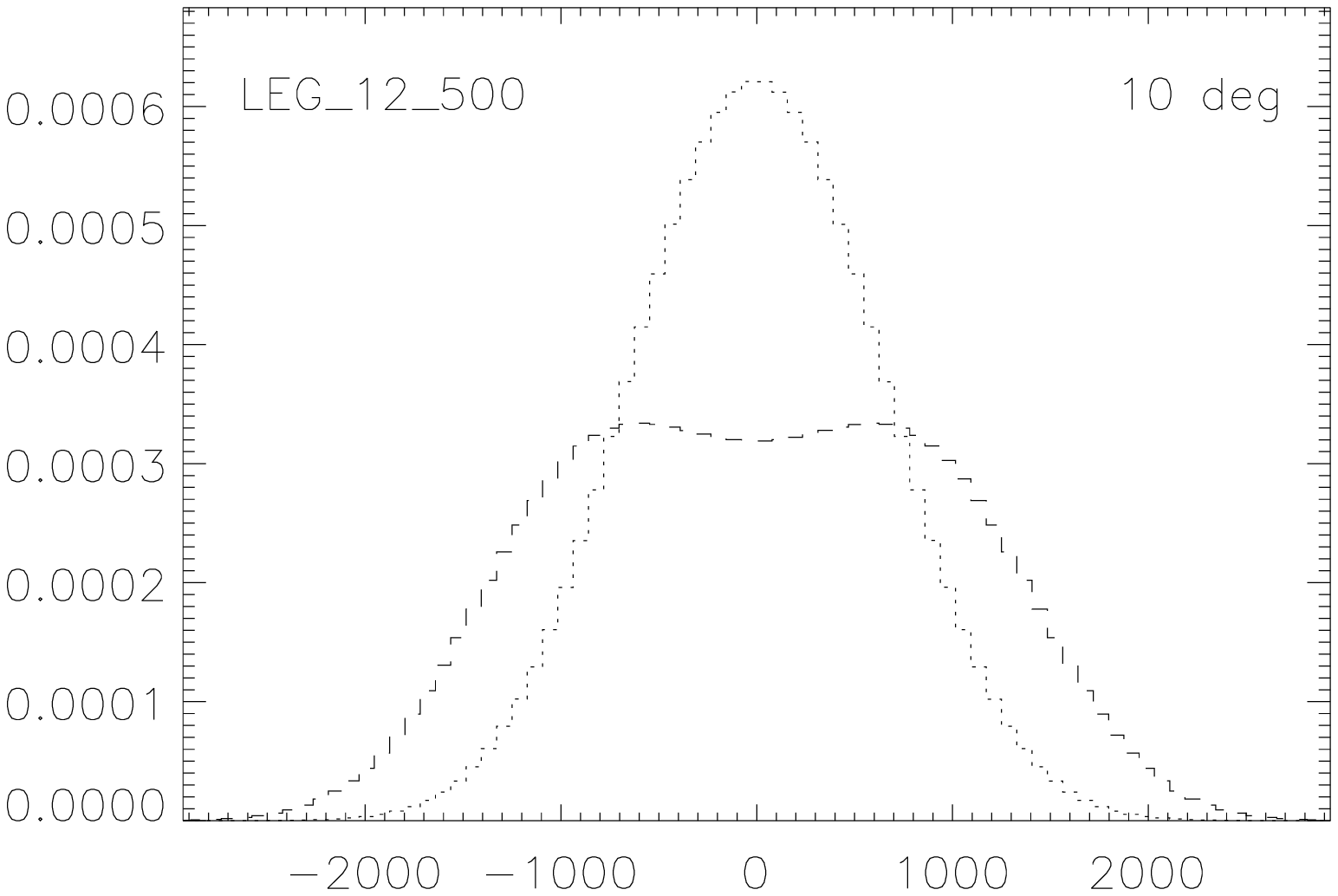}
\includegraphics[width=2in, height=1.5in]{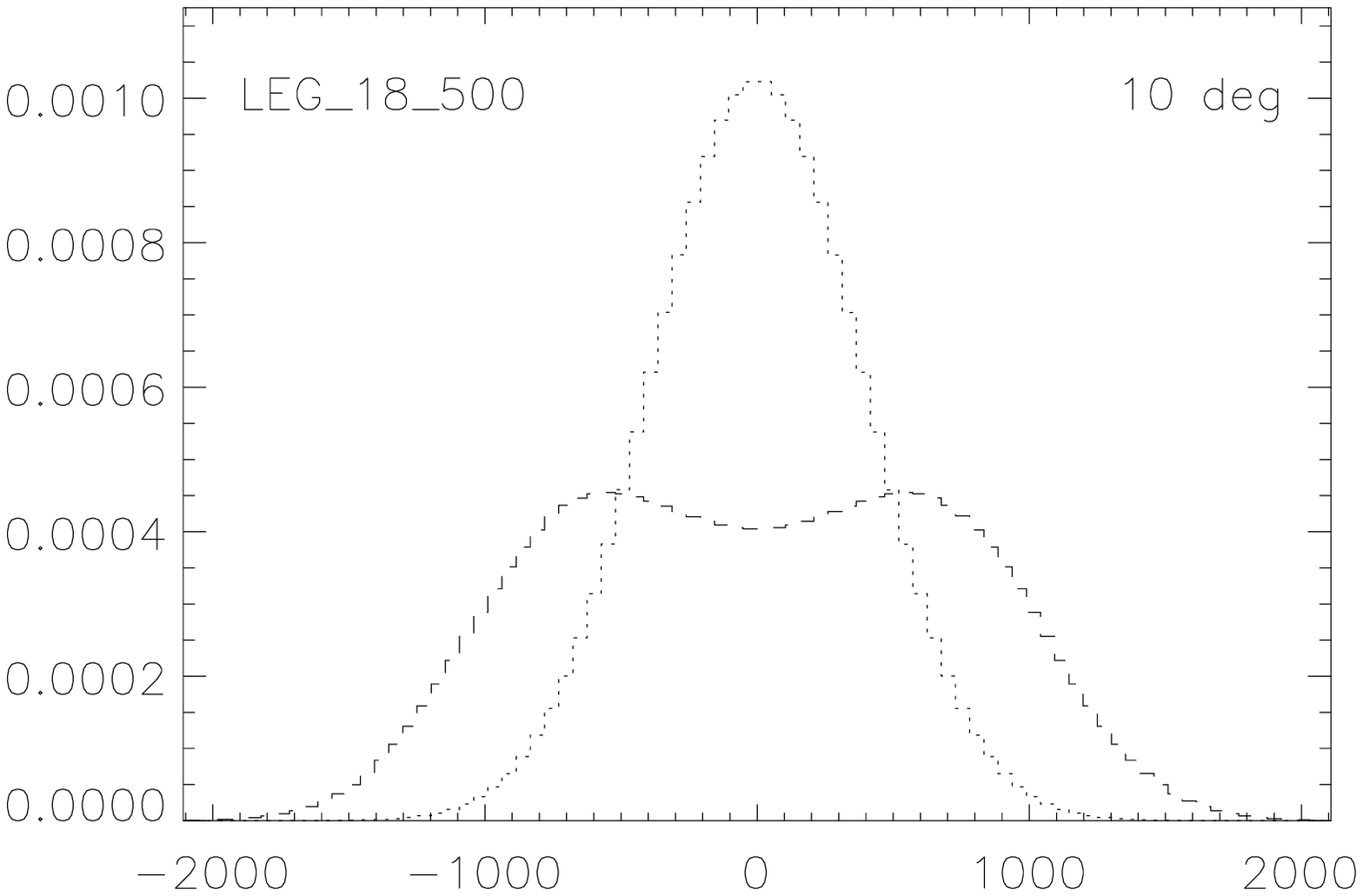}
\includegraphics[width=2in, height=1.5in]{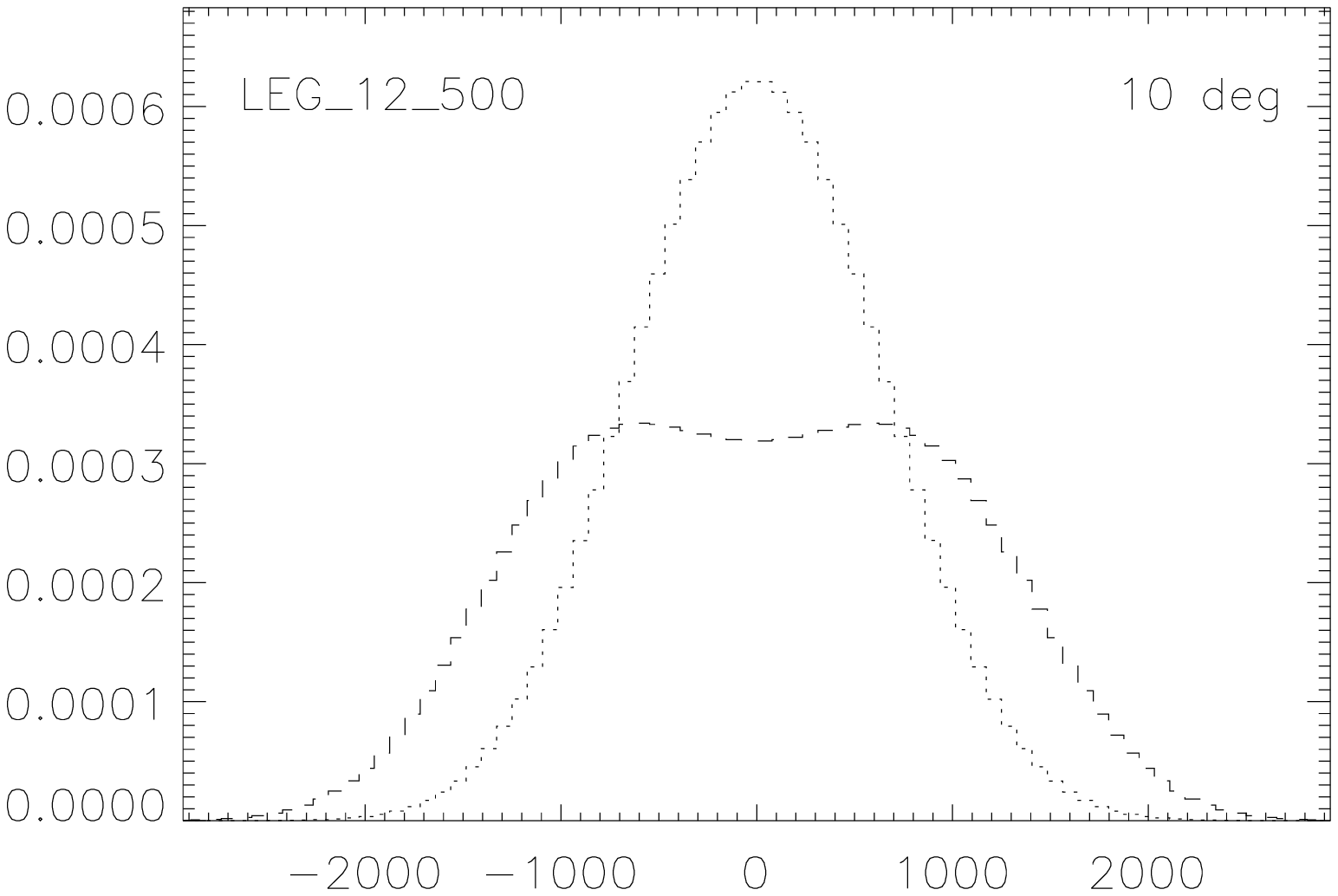}
\includegraphics[width=2in, height=1.5in]{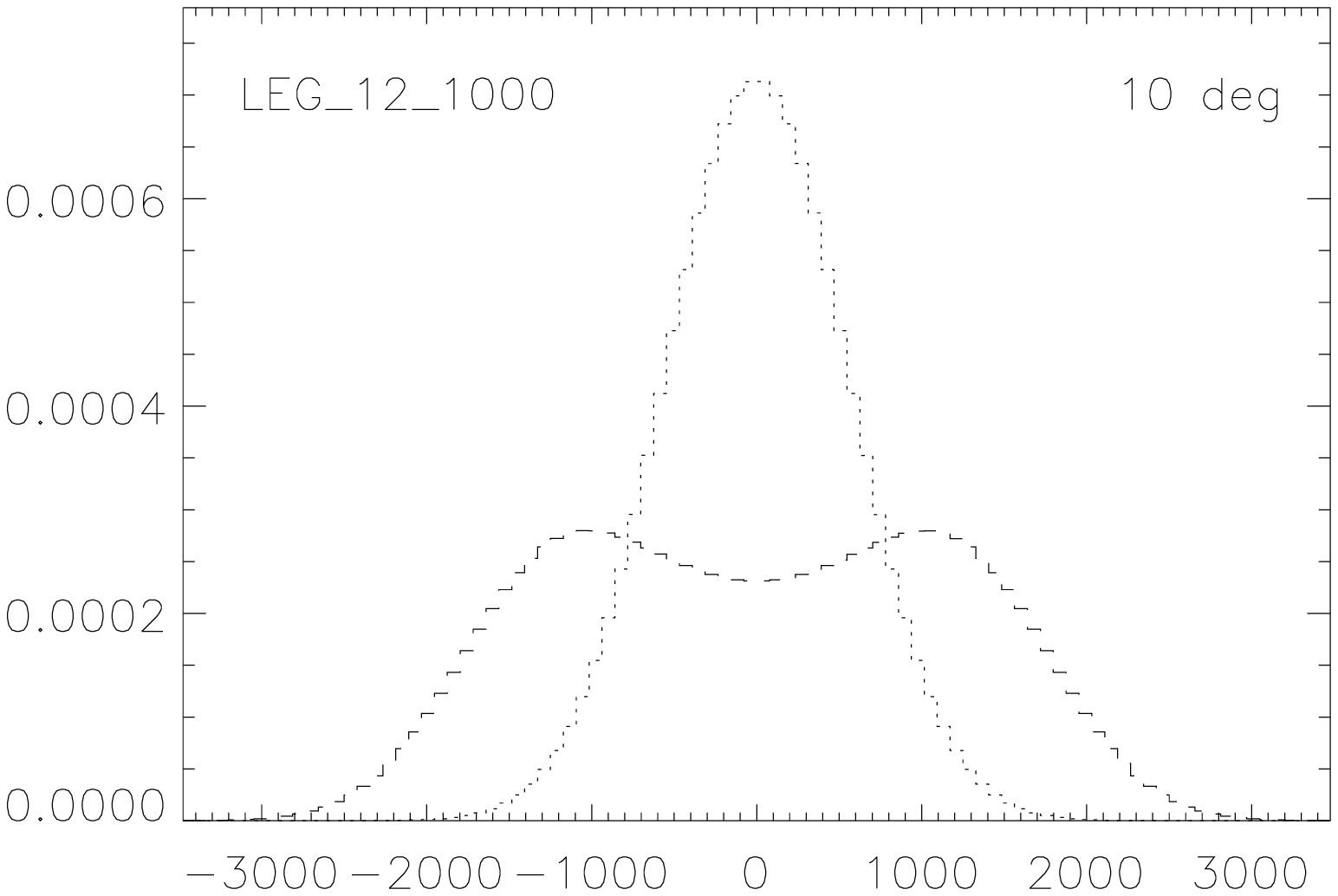}
\includegraphics[width=2in, height=1.5in]{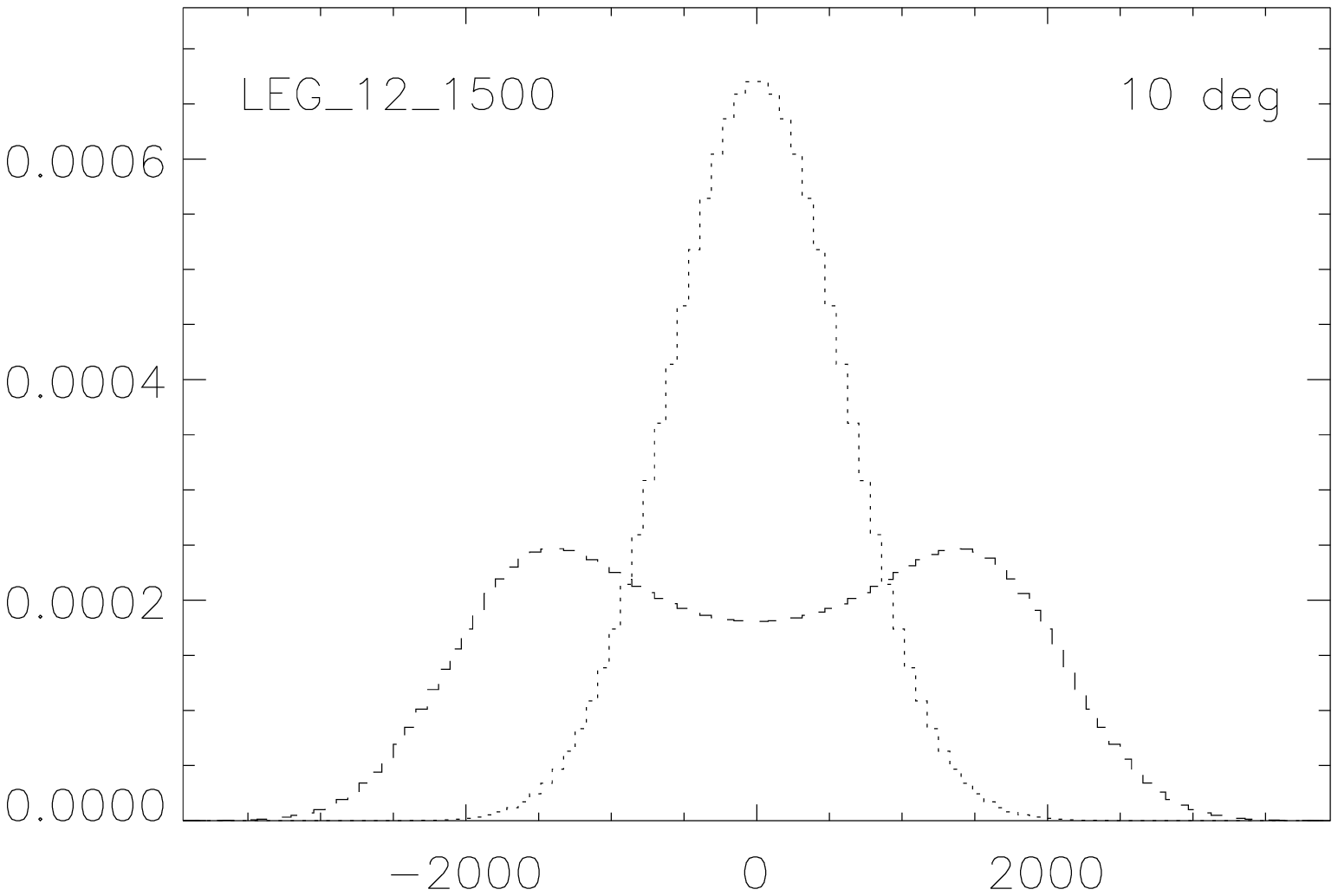}
\end{center}
\caption{
Theoretical line profiles from uniform equatorial ring with
inclination angle 45$^{\circ}$.
Labels in the upper left corner of each panel consecutivley
indicate the {\it Chandra} gratings,fiducial line wavelength in
\AA, and the bulk gas velocity of the emitting plasma in \kms.
Labels in the upper right corner of each panel indicate the half
the opening angle of the equatorial ring.
The line profiles are also convolved with a response
function (assumed to be a Gaussian) that corresponds to the gratings 
spectral resolution.
The stretched and narrowed line profiles are depicted by dashed
and dotted lines, respectively.
Horizontal axes are the Doppler velocities (\kms);
vertical axes are the flux densities per \kms.
}
\label{fig:urlp}
\end{figure}

The RSS that appears after the interaction has three
discontinuities separating plasmas with different characteristics.
The unshocked dense gas in the cloud is compressed, heated and set
into motion by the transmitted shock.  A contact discontinuity
separates this shocked cloud gas from intercloud gas that has been
shocked twice.  This gas will have the same bulk velocity and
pressure as the shocked cloud gas but lower density and higher
temperature.  Further downstream is the reflected shock, which
separates this twice-shocked intercloud gas from intercloud gas
that has been shocked once by the blast wave.
Figure~\ref{fig:cartoon} shows the RSS schematic as the post-shock
velocities and densities have their values for strong shocks with
adiabatic index $\gamma = 5/3$.  

The physical parameters in various parts of RSS follow from the
standard solutions (velocity, density and pressure) for a
plane-parallel shock: 

$V_2 = \frac{(\gamma - 1)M^2+2}{(\gamma + 1)M^2} V_1$
,\hspace{0.5cm}
$ \rho_2 = \frac{(\gamma + 1)M^2}{(\gamma - 1)M^2+2} \rho_1
$,\hspace{0.5cm}
$P_2 = \frac{2 \gamma M^2 - (\gamma - 1)}{\gamma (\gamma + 1)
M^2} \rho_1 V_1^2 \hspace{0.2cm} = \hspace{0.2cm} 
\frac{2 \gamma M^2 - (\gamma - 1)}{(\gamma + 1)} P_1  $

\noindent
where $M$ is the Mach number of the flow 
($ M^2 = \rho_1 V_1^2/\gamma P_1 $), and subscripts `1' and
`2' denote the flow parameters in front and behind the shock
front,  respectively.  Velocities are given in the rest frame of
the shock itself, thus, the parameter $V_1$ represents the actual
shock velocity.
Taking adiabatic index $\gamma = 5/3$, we find:
\begin{equation}
V_2 = \frac{M^2+3}{4 M^2} V_1 ,\hspace{0.5cm}
\rho_2 = \frac{4 M^2}{M^2+3} \rho_1 ,\hspace{0.5cm}
P_2 = 3\frac{5 M^2 - 1}{20 M^2} \rho_1 V_1^2 \hspace{0.1cm} = 
\hspace{0.1cm} \frac{5 M^2 - 1}{4} P_1 
\label{eqn:sh_sol}
\end{equation}
In the case of strong shocks ($M >> 1$), the well-known solution
$V_2 = V_1/4$, $\rho_2 = 4\rho_1$, and~$P_2 = 3/4\rho_1V_1^2$
follows.

Since our final goal is to simulate the RSS X-ray emission, in
what follows we use the plasma temperature instead of the gas
pressure: thus, $T_0$, $T_r$ ~and~ $T_c$ denote this parameter
behind the blast wave, reflected and transmitted shocks,
respectively.  We introduce the following dimensionless variables:
$x = V_s/V_0$ and $w = V_w/V_0$, where $V_s$ is the velocity of
the reflected shock in the observer's rest frame, and $V_w = 3/4
V_c$ is the plasma velocity behind the reflected and transmitted
shocks. Using the first formula in eq.~(\ref{eqn:sh_sol}), we find
that the velocity of the reflected shock is given by:
$$ V_w - V_s =  \frac{M^2+3}{4 M^2} \left(\frac{3}{4}V_0 -
V_s\right) ,\hspace{0.5cm} 
 w - x =  \frac{M^2+3}{4 M^2} \hspace{0.1cm} \frac{3- 4x}{4} 
$$
With the corresponding Mach number for the reflected shock, the
variable $x$ can be given as:
\begin{equation}
M^2 = \frac{3- 4x}{5} ,\hspace{0.5cm}
 x = \frac{(3 + 8w) - \sqrt{64w^2 - 96w + 81}}{12} 
\label{eqn:x}
\end{equation}
Thus, for any value of parameter $w \in [0, 3/4]$ we can use 
eqs.~(\ref{eqn:sh_sol}) and (\ref{eqn:x}) to derive the solution
for the reflected shock structure: 
$w \rightarrow x \rightarrow M$.

On the other hand, the condition of pressure equality across the
contact discontinuity and eqs.~(\ref{eqn:sh_sol}) give:
\begin{equation}
\frac{n_0}{n_c} = \frac{4}{5M^2 - 1}\hspace{0.1cm}
\frac{T_c}{T_0} ,\hspace{0.5cm}
8x^2 - 4x - 4 +(3 - 4x)\sqrt{\frac{n_0}{n_c}[(3 - 4x)^2-1]} = 0 
\label{eqn:x1}
\end{equation}
Now for any value of the density contrast $ n_0/n_c \in [0,1]$,
equations
~(\ref{eqn:sh_sol}) and (\ref{eqn:x1}) give the necessary RSS
solution: $n_0/n_c \rightarrow x \rightarrow M$.

\begin{figure}[ht]
\begin{center}
\includegraphics[width=3in, height=2.25in]{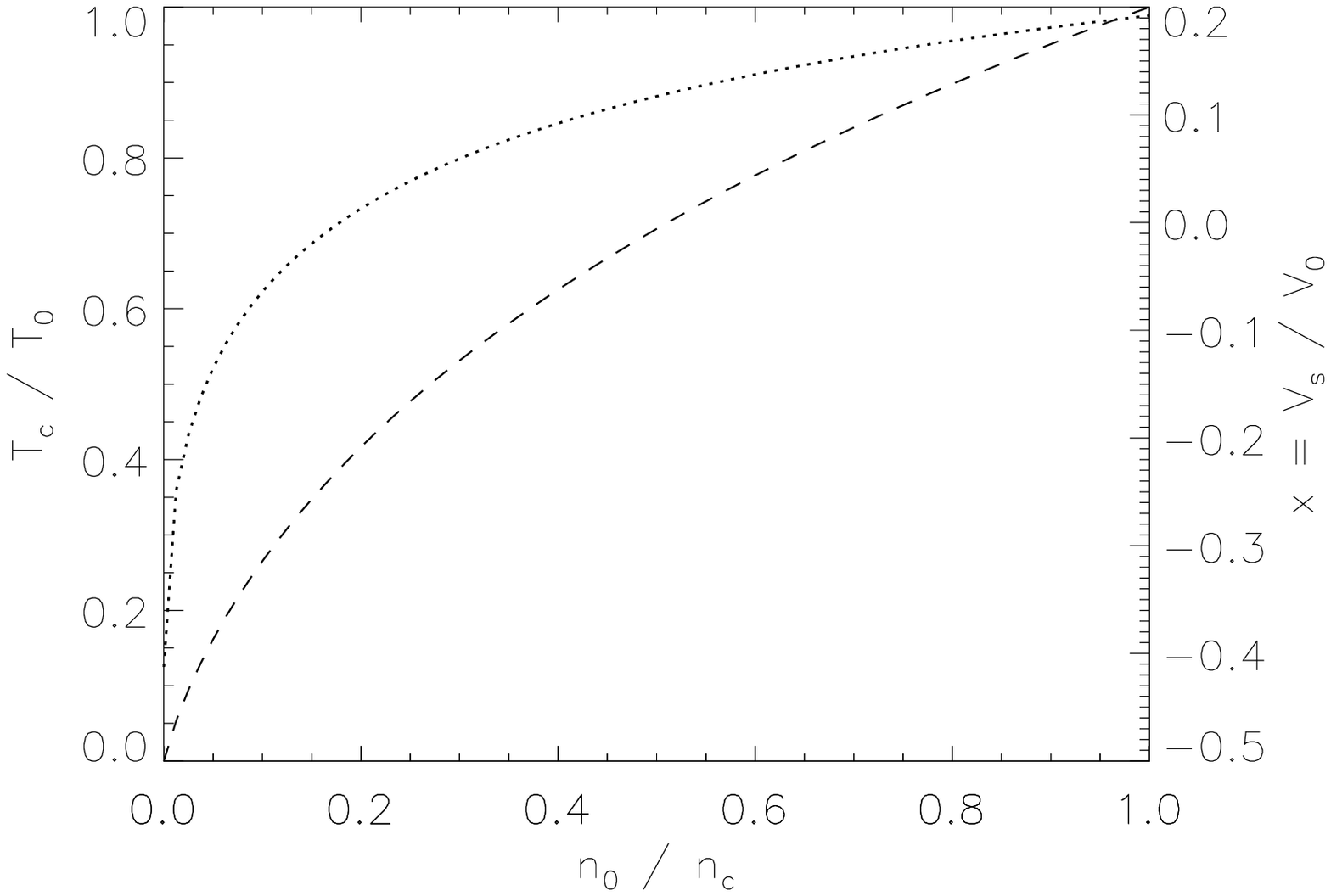}
\includegraphics[width=3in, height=2.25in]{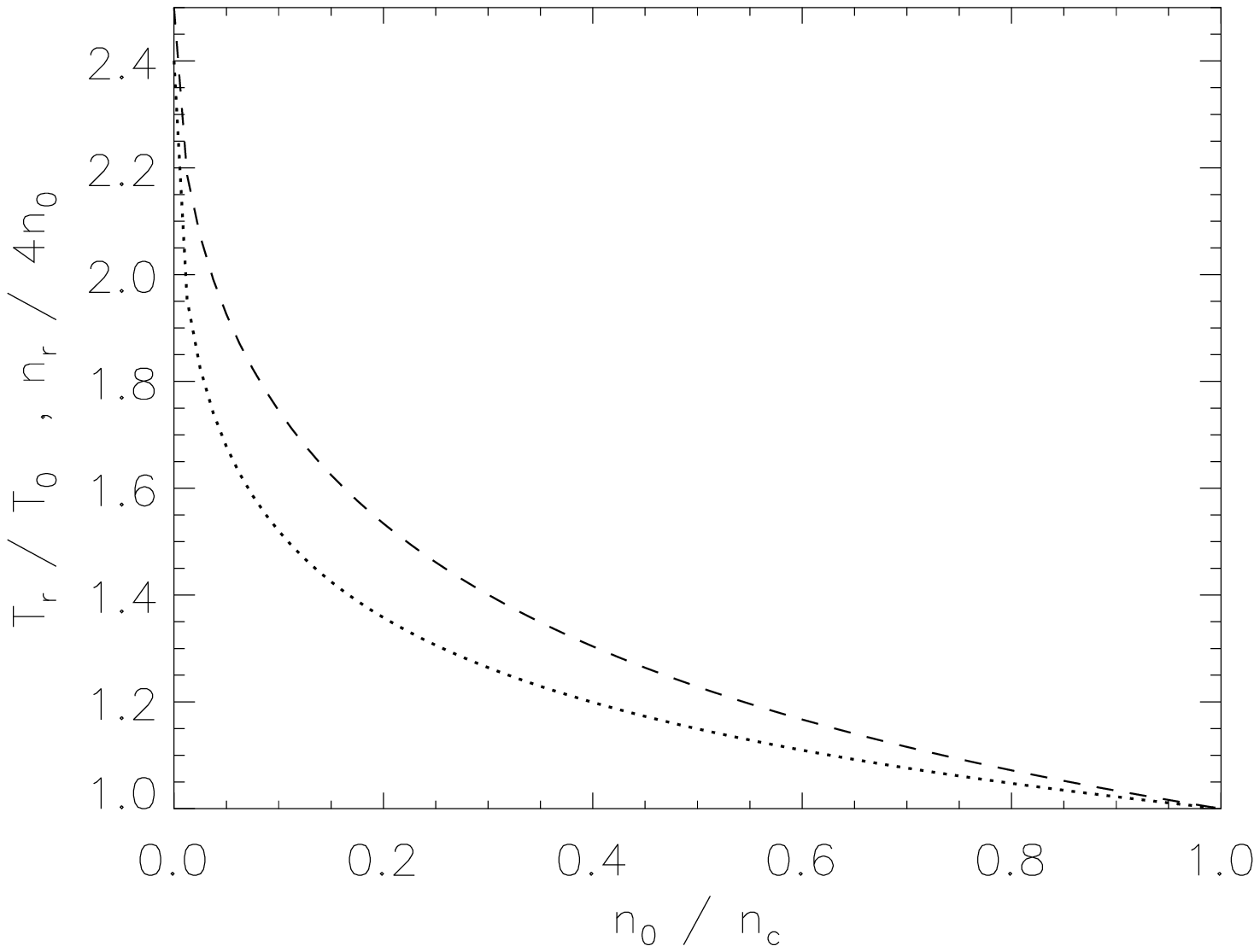}
\end{center}
\caption{
Shock parameters in the reflected shock structure (RSS) as
function of the density contrast $n_0/n_c$.
{\it Left} panel: the resultant contrast ($T_c/T_0$, dashed line)
between the temperatures behind the transmitted and forward shock,
as well as the velocity ($V_s/V_0$, dotted line) of the
reflected shock surface in space (the observer's coordinate system).
{\it Right} panel: the temperature ($T_r/T_0$, dotted line) and
density
($n_r/4 n_0$, dashed line) jumps behind the reflected shock,
having assymptotic ($n_c >> n_0$) values of 2.4 and 2.5, respectively.
}
\label{fig:rss}
\end{figure}

Figure~\ref{fig:rss} shows the RSS solution as function of the
density contrast between the cloud and its surroundings.  The two
limiting cases are worth noting. First, when the blast wave
interacts with a very dense cloud ($n_0/n_c = 0$), we find the
classical solution of the problem of a shock interaction with a
wall \citep{LL95}: $M = \sqrt{5}$, and density and temperature
jumps at the reflected shock front by factors 2.5 and 2.4,
respectively.  The reflected shock velocity has space velocity
$V_s = - V_0/2$, i.e., it is moving opposite to the original blast
wave with half the blast wave velocity.  At the other limit, with
no density discontinuity ahead of the blast wave ($n_0/n_c = 1$),
the derived Mach number is equal to unity, and no RSS appears. Thus,
the basic variable we used for deriving the RSS solution has an
range of possible values $ x \in [-0.5,\frac{3-\sqrt{5}}{4}]$,
which corresponds to the
Mach number of the reflected shock ranging from $ M \in [1,\sqrt{5}]$.

To calculate the X-ray emission from the RSS, we need to know the
temperature and emission measure of plasma in each zone of the
RSS.  We recall that for plane-parallel shock the emission measure
of the shocked gas is related to the ionization age of the shock,
$\tau = n_e t$, namely, $EM \propto n_2 V_2 \tau $. Given that the
transmitted and reflected shocks have equal age (they were `born'
at the same instant), we find:
\begin{eqnarray}
\frac{EM_c}{EM_d} & = & \frac{5M^2 - 1}{4}\hspace{0.1cm}
\frac{\tau_c}{\tau - \tau_l} \sqrt{\frac{T_0}{T_c}}\nonumber \\
\frac{EM_r}{EM_d} & = & d_{jump} \frac{\tau_l}{\tau -
\tau_l}\label{eqn:xspec}  \\
 \frac{\tau_l}{\tau_c} & = & \frac{4}{5M^2 - 1}\hspace{0.1cm}
\frac{T_c}{T_0}\hspace{0.1cm} (3 - 4x)\nonumber  \\
 \frac{\tau_r}{\tau_l} & = & \frac{d_{jump}}{3 - 4x}\nonumber \,,
\end{eqnarray}
where $\tau$ and $\tau_l$ denote the ionization ages of the blast 
wave and of the twice-shocked gas, respectively; $\tau_c$ and
$\tau_r$ are the ionization ages of the transmitted and reflected
shocks, respectively;  
$d_{jump} = n_r / 4n_0$ is the density enhancement behind the
reflected shock; and $EM_d$, $EM_r$ and $EM_c$ respectively denote
the emission measure of the hot plasma in the blast wave,
reflected shock and the transmitted shock.

The ionization age is an important parameter for simulating the
X-ray spectra of shocked gases, which are typically in the state
of non-equilibrium ionization (NEI).  The ionization ages are
slightly different in the three basic RSS regions. The plasma
behind the transmitted shock has a standard distribution for its
ionization age in $[0, \tau_c]$. The ionization age of the plasma
in the
blast wave that has not yet interacted with the reflected shock
ranges from $[\tau_l, \tau]$. In the both case, the ionization age
gradually increases downstream in the flow.  On the other hand,
the NEI effects are more complicated to characterize in the
twice-shocked gas.  The gas entering the reflected shock had been
shocked already by the blast wave and so its ionization state had
evolved from its initial condition.  But, the reflected shock
first interacts with the `youngest' gas behind the blast wave, and
this plasma becomes the `oldest' in the doubly-shocked region.

With all these considerations at hand, we developed a custom model
for XSPEC which simulates the X-ray emission from a reflected
shock structure and takes into account the NEI effects. The free
parameters of the model are: temperatures ($T_0$, $T_c$) and
ionization ages ($\tau$, $\tau_c$) of the plasma behind the blast
wave and the transmitted shock, respectively; chemical
composition; and the total emission measure.  We recall that the
$w$ parameter is defined for any given pair ($T_0$, $T_c$) of
temperatures by  
$w = 3/4 (V_c/V_0) = 3/4 \sqrt{T_c/T_0}$.
From this we obtain solutions of (eqs.~[\ref{eqn:x}]) as well as
the emission measures from different RSS regions and a lower limit
($\tau_l$) for the  ionization age of the plasma in the blast wave
(eqs.~[\ref{eqn:xspec}]). The NEI effects in the blast wave and
the transmitted shock are taken into account following the
\citet{Bor01} {\it vpshock} model in XSPEC.  We modify this
approach for the gas behind the reflected shock to account for the
initial NEI plasma evolution following the blast wave passage.

We created two variants of the RSS model in XSPEC. In the first,
we handled the line broadening in the same way as described by Z06
-- that is, we assumed a Gaussian profile with a variable width
following eq.~(\ref{eqn:stratified}). In the second variant we
consider line profiles produced by a uniform-disk (see
appendix~\ref{app:urlp}) having individual bulk gas velocities in
each RSS region (see Fig.~\ref{fig:cartoon}) that are
self-consistently derived in the model.




\clearpage
\begin{table}
\begin{center}
\caption{Two-Shock Model Results (First-Order Spectra)
\label{tab:fit}}
\begin{tabular}{llll}
\tableline
\tableline
 Parameters & LEG 2004 & MEG 2007  & LEG 2007  \\
\tableline
$\chi^2$/dof  & \dotfill & 1819/2248 & \dotfill \\
N$_H$(10$^{21}$ cm$^{-2}$) & \dotfill & 1.30 [1.18 - 1.46] &
                             \dotfill \\
kT$_1$ (keV)  & 0.53 [0.50 - 0.55] & 0.56 [0.53 - 0.59]& 
                0.54 [ 0.53 - 0.56] \\
kT$_2$ (keV)  & 2.70 [2.54 - 3.03] & 2.43 [1.94 - 2.72] & 
                1.90 [1.84 - 1.97] \\
tau$_1$$^{(a)}$  & 3.20 [2.63 - 4.30] & 3.63 [3.33 - 4.58] & 
                   4.78 [4.06 - 5.88] \\
tau$_2$$^{(a)}$  & 1.58 [1.23 - 2.13] & 2.23 [1.81 - 2.76] & 
                   2.67 [2.23 - 3.25]\\
EM$_1$$^{(b)}$   & 4.12 & 9.12 & 10.9  \\
EM$_2$$^{(b)}$   & 1.71 & 4.22 & 5.63 \\
H    & 1 & 1 & 1 \\
He(2.57)  & 2.57 & 2.57 & 2.57  \\
C~(0.09)   & 0.09 & 0.09 & 0.09  \\
N~(1.63)   & \dotfill & 0.56 [0.50 - 0.65] & \dotfill \\
O~(0.18)   & \dotfill & 0.081 [0.074 - 0.092] & \dotfill \\
Ne(0.29)   & \dotfill & 0.29 [0.27 - 0.31] & \dotfill \\
Mg(0.32)   & \dotfill & 0.28 [0.26 - 0.29] & \dotfill \\
Si(0.31)   & \dotfill & 0.33 [0.32 - 0.35] & \dotfill \\
S~(0.36)   & \dotfill & 0.30 [0.24 - 0.36] & \dotfill \\
Ar   & 0.537 & 0.537 & 0.537 \\
Ca   & 0.339 & 0.339 & 0.339 \\
Fe(0.22)   & \dotfill & 0.19 [0.19 - 0.21] & \dotfill \\
Ni   & 0.618 & 0.618 & 0.618 \\
F$^{(c)}_X$(0.5-2 keV) & 1.48 [1.37 - 1.56] & 3.27 [3.14 - 3.40]
&
                         3.84 [3.70 - 3.96] \\
F$^{(c)}_X$(0.5-6 keV) & 1.84 [1.68 - 1.94] & 4.08 [3.80 - 4.21]
&
                         4.65 [4.48 - 4.78] \\
\tableline
\end{tabular}

\tablecomments{
Note --  Resutlts from the {\it simultaneous} fit to the LETG 2004, 
MEG 2007 and LEG 2007 spectra.
90\%-confidence intervals are given in brackets, and all
abundances are expressed as ratios to their solar values 
\citep{AG89}.
For comparison, the inner-ring abundances of He, C, N, and O
\citep{LF96}; those of Ne, Mg, Si, S and Fe
typical for LMC \citep{RD92}
are given in the first column in parentheses.}
\tablenotetext{a}{ ($n_e t$) in units of $10^{11}$ cm$^{-3}$~s.}
\tablenotetext{b}{ $EM = \int n_e n_H dV$ in units of
$10^{58}$~cm$^{-3}$
if a distance of 50 kpc is adopted.}
\tablenotetext{c}{ The observed X-ray flux in units of
$10^{-12}$ ergs cm$^{-2}$ s$^{-1}$.}
\end{center}
\end{table}

\clearpage

\begin{table}
\begin{center}
\caption{SNR 1987A: Line Fluxes\label{tab:flux}}
\begin{tabular}{lrrrrr}
\tableline\tableline
\multicolumn{1}{c}{ } & 
\multicolumn{1}{c}{$\lambda_{lab}$$^{a}$} &
\multicolumn{2}{c}{LETG 2007 $^{b}$} & 
\multicolumn{2}{c}{HETG 2007 $^{b}$} \\ 
\multicolumn{1}{c}{Line} & 
\multicolumn{1}{c}{(\AA)} &
\multicolumn{1}{c}{Gauss} & 
\multicolumn{1}{c}{URLP} & 
\multicolumn{1}{c}{Gauss} & 
\multicolumn{1}{c}{URLP} \\ 
\tableline
Si XIV L$_{\alpha} $ & 6.1804  & 10.0 $\pm$ 1.3 & 7.9
$\pm$ 0.8 & 5.2 $\pm$ 0.9 & 4.3 $\pm$ 0.5 \\
Si XIII K$_{\alpha} $ & 6.6479  & 41.0 $\pm$ 3.7 & 38.5
$\pm$ 2.2 & 31.4 $\pm$ 1.5 & 26.2 $\pm$ 1.0 \\
Mg XII L$_{\alpha} $ & 8.4192  & 17.9 $\pm$ 2.3 & 12.4
$\pm$ 1.1 & 11.6 $\pm$ 0.9 & 10.2 $\pm$ 0.6 \\
Mg XI K$_{\alpha}$   & 9.1687  & 47.7 $\pm$ 4.2 & 43.2
$\pm$ 2.7 & 43.7 $\pm$ 2.8 & 35.0 $\pm$ 1.6 \\
Ne X  L$_{\alpha} $  & 12.1321  & 92.9 $\pm$ 5.4 & 72.9
$\pm$ 5.4 & 82.4 $\pm$ 4.8 & 60.0 $\pm$ 2.6 \\
Ne IX K$_{\alpha}$   & 13.4473  & 167.3 $\pm$ 10.6  & 135.9
$\pm$ 5.8 & 134.2 $\pm$ 6.3  & 98.0 $\pm$ 4.7 \\
Fe XVII$^{c}$ & 15.0140  & 93.7 $\pm$ 5.5 & 86.9 $\pm$
3.6 & 88.5 $\pm$ 4.7 & 78.4 $\pm$ 3.8 \\ 
O VIII L$_{\beta} $ & 16.0055  & 49.5 $\pm$ 6.6 & 36.3
$\pm$ 3.3 & 36.5 $\pm$ 7.6 & 28.8 $\pm$ 3.8 \\
Fe XVII$^{c}$ & 16.7800  & 135.0 $\pm$ 10.3 & 123.4 $\pm$
9.3 & 111.2 $\pm$ 9.3 & 98.4 $\pm$ 9.0 \\
O VIII L$_{\alpha} $ & 18.9671  & 142.3 $\pm$ 7.5 & 113.3
$\pm$ 5.8 & 104.8 $\pm$ 10.8 & 81.8 $\pm$ 8.6 \\
O VII K$_{\alpha}$  & 21.6015  & 73.2 $\pm$ 8.9  & 63.4
$\pm$ 7.4 & 72.5 $\pm$ 21.5  & 72.5 $\pm$ 21.0 \\
N VII L$_{\alpha} $ & 24.7792  & 69.0 $\pm$ 7.4 & 62.2
$\pm$ 6.5 & 73.9 $\pm$ 17.8 & 65.3 $\pm$ 14.7 \\
\tableline
\end{tabular}
\tablenotetext{a}{The laboratory wavelength of the main component.}
\tablenotetext{b}{The observed total line/multiplet flux in units
of $10^{-6}$ photons cm$^{-2}$ s$^{-1}$ and the associated 
$1\sigma$ errors for two different line profiles: Gaussian and
URLP.}
\tablenotetext{c}{The total flux from the strongest component
$\lambda\lambda 15.01, 15.26$~\AA~and~
$\lambda\lambda\lambda 16.78, 17.05, 17.10$~\AA, respectively.}
\end{center}
\end{table}

\clearpage

\clearpage

\clearpage

\clearpage


\clearpage

\clearpage

\clearpage

\clearpage

\end{document}